%% file: TOPQ_2012_03.tex
\pdfoutput=1
\documentclass[11pt,a4paper,pdfa=true]{article}
\usepackage{jheppub} 
\usepackage{atlasphysics}
\usepackage{subfigure}
\usepackage{mathrsfs}
\usepackage{url} 
\usepackage{xspace}
\usepackage{dcolumn}
\usepackage{booktabs}
\usepackage{afterpage}
\usepackage{rotating}
\usepackage{multirow}
\usepackage{esvect}
\usepackage{array}
\usepackage{preprintcover}

\newcolumntype{C}[1]{>{\centering\arraybackslash}p{#1}}

\graphicspath{{figures/}}

\usepackage[T1]{fontenc} 

\input{macros}

\input{AAAcoverpage}

\RequirePackage{ifpdf}
\begin{document}
\author{The ATLAS Collaboration}
\title{Measurement of the \ttbar\ production cross-section as a function of jet multiplicity and jet transverse momentum in 7~\TeV\ proton--proton collisions with the ATLAS detector}

\abstract{
The \ttbar\ production cross-section dependence on jet multiplicity  and jet transverse momentum is reported  for  proton--proton
collisions at a centre-of-mass energy of 7~\TeV\ in the single-lepton channel.
The data were collected with the ATLAS detector at the CERN Large Hadron Collider and comprise the full 2011 data sample corresponding to an integrated luminosity of 4.6~\ifb.  Differential cross-sections are presented 
as a function of the jet multiplicity for up to eight jets using jet transverse momentum thresholds of 25, 40, 60, and 80~\GeV, and as
a function of jet transverse momentum up to the fifth jet.  The results are shown after background subtraction and corrections for all detector effects, within a kinematic range closely matched to the experimental acceptance.
Several QCD-based Monte Carlo models are compared with the results.  Sensitivity to the parton shower modelling  is found at the higher jet multiplicities, at high transverse momentum of the leading jet and in the transverse momentum spectrum of the fifth leading jet.  The MC@NLO+HERWIG MC is found to predict too few events at higher jet multiplicities.  }



\maketitle
\input{introduction}

\input{detector-description}

\input{data-sample}
\label{sec:recoselection}

\input{object-selection-new}

\input{event-selection-new}

\input{monte-carlo}

\input{systematic-uncertainties}

\input{results}

\input{corrections}

\input{propagation}

\input{results-particlelevel}

\input{conclusions}

\input{acknowledgements}

\clearpage
\newpage

\appendix
\input{appendixRecoLevelResults}

\clearpage
\input{appendixCorrectionFactors}
\clearpage
\input{appendixSystematicTables}

\clearpage
\newpage

\afterpage{\clearpage}
\bibliographystyle{JHEP}
\bibliography{TOPQ_2012_03}

\clearpage
\newpage

\include{atlas_authlist}


\end{document}

%% file: macros.tex
\newcommand{\wjets}{\ensuremath{W\!+\!\textrm{jets}}\xspace}
\newcommand{\wTwojets}{\ensuremath{W\!+\!2\;\textrm{jets}}\xspace}
\newcommand{\wTwojet}{\ensuremath{W\!+\!2\;\textrm{jet}}\xspace}

\newcommand{\zjets}{\ensuremath{Z/\gamma^{*}\!+\!\textrm{jets}}\xspace}

\def\mtw{\ensuremath{m_\mathrm{T}(W)}}
\def\ejets{\ensuremath{e+\textrm{jets}}}
\def\mujets{\ensuremath{\mu+\textrm{jets}}}

\def\JVF{\ensuremath{\textrm{JVF}}}

\def\Nparti{\ensuremath{N_\textrm{part}^\textrm{i}}}
\def\fpartrecoi{\ensuremath{f_\textrm{part!reco}^\textrm{i}}}
\def\frecopartj{\ensuremath{f_\textrm{reco!part}^\textrm{j}}}
\def\mrecopartij{\ensuremath{M_\textrm{reco,j}^\textrm{part,i}}}
\def\facceptj{\ensuremath{f_\textrm{accpt}^\textrm{j}}}
\def\Nrecoj{\ensuremath{N_\textrm{reco}}^\textrm{j}}
\def\Nbgndj{\ensuremath{N_\textrm{bgnd}}^\textrm{j}}
\def\fmisassignj{\ensuremath{f^\mathrm{j}_\mathrm{misassign}}}

%% file: AAAcoverpage.tex
\PreprintCoverPaperTitle{Measurement of the \ttbar\ production cross-section as a function of jet multiplicity and jet transverse momentum in 7~\TeV\ proton--proton collisions with the ATLAS detector}

\PreprintIdNumber{CERN-PH-EP-2014-114}  

\PreprintCoverAbstract{
The \ttbar\ production cross-section dependence on jet multiplicity  and jet transverse momentum is reported  for  proton--proton
collisions at a centre-of-mass energy of 7~\TeV\ in the single-lepton channel.
The data were collected with the ATLAS detector at the CERN Large Hadron Collider and comprise the full 2011 data sample corresponding to an integrated luminosity of 4.6~\ifb.  Differential cross-sections are presented
as a function of the jet multiplicity for up to eight jets using jet transverse momentum thresholds of 25, 40, 60, and 80~\GeV, and as
a function of jet transverse momentum up to the fifth jet.  The results are shown after background subtraction and corrections for all detector effects, within a kinematic range closely matched to the experimental acceptance.
Several QCD-based Monte Carlo models are compared with the results.  Sensitivity to the parton shower modelling  is found at the higher jet multiplicities, at high transverse momentum of the leading jet and in the transverse momentum spectrum of the fifth leading jet.  The MC@NLO+HERWIG MC is found to predict too few events at higher jet multiplicities.
}  

\PreprintJournalName{JHEP}  
\RequirePackage{ifpdf}

%% file: introduction.tex
\section{Introduction}

Final states of proton--proton ($pp$) collisions at the Large Hadron Collider (LHC)~\cite{Evans:2008zzb} often include jets arising from
QCD bremsstrahlung due to the strongly interacting
partons in the initial state and the high centre-of-mass energy of the scattering process that allows for radiation in a large kinematic phase space. 
In this paper, an inclusive measurement of jets in 
 top--antitop (\ttbar) final states is presented, which is sensitive to the production mechanism of additional jets in these events.  The events studied have a high  partonic-system centre-of-mass energy and are complex final states consisting of several coloured partons, with sensitivity to various hard scales.

The production of additional jets in \ttbar\ events is sensitive to higher-order perturbative QCD effects.  The uncertainties associated with these processes are a significant source of uncertainty in precision measurements, such as the measurement of the top-quark mass~\cite{ATLAStopmass} or the inclusive \ttbar\ production cross-section at the LHC~\cite{Aad:2012qf}. Several theoretical approaches are available to model \ttbar\ processes, including NLO QCD calculations, parton-shower models and methods matching fixed-order QCD with the parton shower. The aim of this paper   is to test these theoretical approaches by making a direct measurement of jet activity in \ttbar\ events.  
Furthermore, \ttbar\ production with additional jets is a dominant background  
in certain Higgs boson production processes and decay modes
 and to many searches for new physics phenomena~\cite{searches, Mangano:2008ha}. 

Tests similar to those presented in this paper have been performed at lower energies, using measurements of jets associated with colour-singlet vector-boson production at the LHC~\cite{atlaswplusjets, Aad:2011qv} and at the Tevatron~\cite{Abazov:2009av,Abazov:2013gpa, CDFZpt,Aaltonen:2007ip}.  The CMS collaboration recently 
measured the cross-section of additional jets normalised to the inclusive \ttbar\ production cross-section~\cite{Chatrchyan:2014gma}. The present measurement is complementary to the measurement of \ttbar\ production with a veto on additional jet activity~\cite{ATLAS:2012al},
which is mostly sensitive to the first perturbative QCD emission.

In the Standard Model (SM), a top-quark decays almost exclusively to a $W$ boson and a $b$ quark.
The $W$ boson decays into a pair of leptons ($e\overline{\nu}_{e}$, $\mu\overline{\nu}_{\mu}$, $\tau\overline{\nu}_{\tau}$) or into a pair of  quark-jets.  $\tau$ leptons produced by $W$ boson decays can also decay into leptons ($e\overline{\nu}_{e}\nu_{\tau}$, $\mu\overline{\nu}_{\mu}\nu_{\tau}$).  Selected events are classified by the decay of one or both of the $W$ bosons into leptons, as either single-lepton or dilepton channel, respectively.  

In this paper,  the \ttbar\ production cross-section is measured differentially in jet multiplicity and in jet transverse momentum (\pt)   in the single-lepton channel, without explicit 
separation between jets related to \ttbar\ decays and  additional jets.  The jet multiplicity is measured  for several different jet \pt\ thresholds in order to probe the \pt\ dependence of the hard  emission.
The jet multiplicity, especially for values greater than four, is closely related  to the number of hard emissions in QCD bremsstrahlung processes.  

In addition, the  differential cross-section with respect to the jet \pt\  is presented separately for the 
 five highest \pT\ jets.  These differential cross-sections  are particularly sensitive to the modelling of higher-order QCD effects in Monte Carlo (MC) generators~\cite{mangano2007,Corke:2010zj}.
Therefore, a precise measurement can be used to discriminate between different models and to determine their free parameters.  Furthermore, a precise measurement of the leading jet \pT\ could  be used to 
determine the \pT\ of the \ttbar\ system above approximately 130~\GeV, since for large transverse momenta the leading jet \pT\ is correlated with the \pT\ of  the \ttbar\ system as illustrated in figure~\ref{fig:alpgenpt}. 
Therefore, measurements of the leading jet \pT\ provide complementary information with respect to existing differential production cross-section measurements of the top-quark~\cite{Aad:2012hg,Chatrchyan:2012saa}.

The present analysis uses $pp$ data collected during 2011 corresponding to an integrated luminosity of $4.59\pm 0.08$~\ifb~\cite{Aad:2013ucp}. 
The  measurements are corrected for all known detector effects and are presented in the form of differential cross-sections, defined within the detector acceptance (``fiducial'' cross-sections) in order to avoid model-dependent extrapolations and to facilitate comparisons with theoretical predictions. 
The fiducial volume definition follows  previous kinematic definitions of cross-section measurements involving top quarks~\cite{Aad2012244}.  In addition, the objects used to define the fiducial volume  at particle level  were reconstructed   such that they closely match  the reconstructed objects in data.

\begin{figure}[htbp]
\centering
\includegraphics[width=0.59\textwidth]{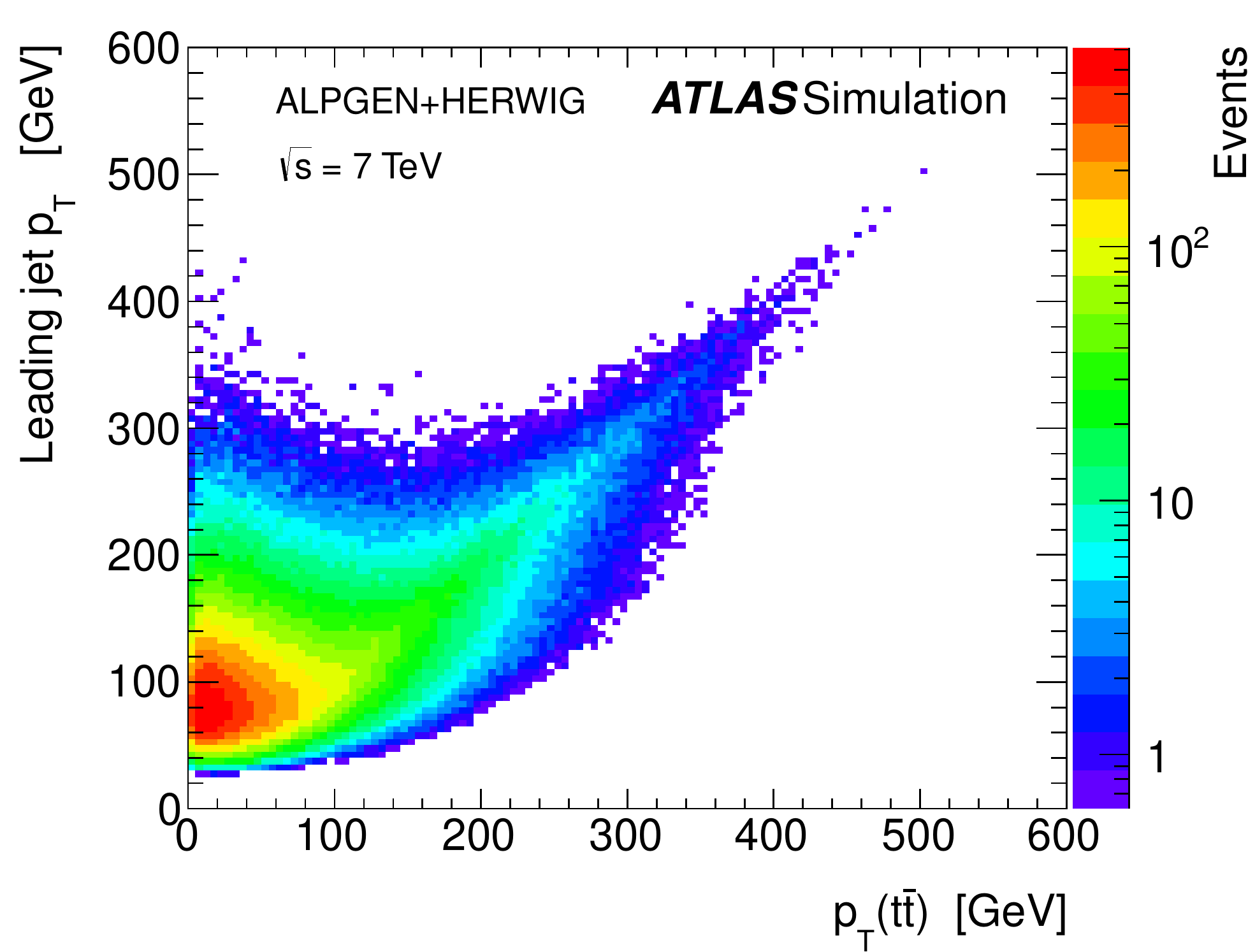}
\caption{The relationship between the \pt\ of the \ttbar\ system in the single-lepton channel and the \pT\ of the highest \pT\ jet in \ttbar\ events generated with {\sc ALPGEN}+{\sc HERWIG}.    The \pt\ of the \ttbar\ system is taken at parton level and the leading  jet is constructed at particle level. 
\label{fig:alpgenpt} }
\end{figure}

%% file: detector-description.tex
\section{The ATLAS detector}

The ATLAS detector~\cite{:2008zzm} covers nearly the entire solid angle around the LHC-beam collision point.  Due to the complexity of the final state in the selected events, the present analysis relies on all main ATLAS detector  subsystems.

The ATLAS reference system is a Cartesian 
right-handed coordinate system, where the nominal collision point is at the origin. 
The anti-clockwise beam direction defines the positive $z$-axis, while the positive 
$x$-axis is defined as pointing from the collision point to the centre of the LHC 
ring and the positive $y$-axis  points upwards. The azimuthal angle $\phi$ is measured around the beam axis, and the polar angle $\theta$ is measured with respect to~the $z$-axis. The pseudorapidity is defined as $\eta =  -\ln \tan (\theta/2) $.

 The ATLAS detector consists of an inner tracking detector (ID), comprising a silicon pixel detector, a silicon microstrip detector (SCT), and a transition radiation tracker (TRT).  The ID is surrounded by a superconducting solenoid that provides a 2~T magnetic field. 
The ID is used for reconstruction of tracks and primary vertices and plays a crucial role in 
$b$-quark jet identification. It is surrounded by high-granularity liquid-argon (LAr) electromagnetic (EM) sampling calorimeters with lead absorbers.  An iron absorber and scintillating
tile calorimeter provides hadronic energy measurements in 
the central pseudorapidity range of $|\eta| < 1.7$. The end-cap and forward regions are instrumented with LAr 
calorimeters for both electromagnetic and hadronic energy measurements up to $|\eta| = 4.9$. 
The calorimeter system is surrounded by a muon spectrometer (MS) that incorporates a system of air-core superconducting toroid magnets arranged with an eight-fold azimuthal coil symmetry around the calorimeters, and a system of three stations of chambers 
for triggering and for precise track measurements. 

The online event selection relies on a three-level trigger system.  A hardware-based first-level trigger is used to initially reduce the event rate by $\mathcal{O}(300)$.  The detector readout is available for two stages of software-based (higher-level) triggers.  In the second level, partial object reconstruction is carried out to improve the selection and reduce the rate of soft $pp$ interactions recorded.  At the
last level, the event filter, the full online event reconstruction is used, which reduced the rate to approximately 300~Hz during the 2011 run period.

%% file: data-sample.tex
\section{Data sample and event selection}
Data were selected from the full
2011 data-taking period using the $pp$ LHC running periods during which all ATLAS sub-detectors
were fully operational, corresponding to an integrated luminosity of
$4.59\pm0.08$~\ifb.

During this data-taking period, the peak luminosity delivered by the LHC was high enough to produce multiple $pp$ collisions from one $pp$ bunch crossing.  The LHC bunch structure and high luminosity also produced $pp$ collisions in immediately adjacent $pp$ bunch crossings.  The average number of $pp$ collisions, over all bunch crossings and all data analysed, was measured and is referred to as $\langle \mu \rangle$.
At the beginning of the data-taking period  $\langle \mu \rangle$ was around five, whereas by the end of period it was approximately eighteen.  The effects of 
particles created in additional collisions are mitigated by the object and event selections used in this analysis.

%% file: object-selection-new.tex
\subsection{Object reconstruction}
\label{object-selection}

Primary vertices were reconstructed from tracks within the ID.  The selected primary vertex was required to have at least
five tracks and to be consistent with the beam-collision
region in the $x$\,-\,$y$ plane.  If more than one primary vertex candidate was found, then the vertex with the
highest $\sum \pT^{2}$ of associated tracks was chosen to be associated with the hard scattering process.



Electron candidates were identified~\cite{Aad:2014fxa} as energy deposits (clusters) in
the electromagnetic calorimeters, with a matching reconstructed track in the 
ID.  These electrons were selected within the pseudorapidity range
$|\eta|<2.47$, excluding the barrel/end-cap transition
region of $1.37<|\eta|<1.52$. 
The energy cluster in the calorimeter was required to be isolated.  The isolation requirement was formed by 
calculating the total transverse energy within a cone of size $\Delta R = 0.2$ around the electron direction, where $\Delta R = \sqrt{(\Delta\phi)^2+(\Delta \eta )^2}$ and $\Delta \phi$ and $\Delta \eta$ are the difference of azimuthal angle and pseudorapidity, respectively. This calculation was performed after the exclusion of calorimeter cells associated with the electron cluster. The electron was considered isolated if this energy sum was below 
 10\% of the electron energy.  Similarly,  the summed \pT\ of additional tracks within a cone of
size $\Delta R =0.3$ around the electron direction was required to be below 10\% of the electron candidate track \pt.   
  The electron 
was required to have a longitudinal impact parameter with respect to the selected primary vertex of less than 2~mm.
The reconstructed \pT\ of electrons used in
the event selection was required to be greater than 25~\GeV, but
electrons with $\pT >15$~\GeV\ were considered when removing jets that
overlap with electrons and when applying a veto on events with additional
leptons.   

%

Muon candidates were required to have a reconstructed track in the MS matched with a track reconstructed in the ID, a reconstructed $\pT > 25$~\GeV\ and $|\eta|<2.5$ \cite{Aad:2014zya}.  The selected muons were required to be isolated in the calorimeter and tracking volume.  The calorimeter isolation was constructed from the sum of transverse energy components within a cone of  $\Delta R= 0.2$ around the direction of the muon and was required to be less than 4~\GeV.  The isolation within the ID was formed using a \pT\ sum of additional tracks within a cone of $\Delta R = 0.3$ around the direction of the muon and was required to be less than 2.5~\GeV.  To reduce the effects of additional primary vertices, the muon was required to have a longitudinal impact parameter with respect to the selected primary vertex of less than 2~mm.  In the same manner as the electron selection, muons with \pT\ as low as 15~\GeV\ were used to veto events with additional leptons.


Topological clusters~\cite{Lampl:1099735} were formed from calorimeter energy deposits.
These clusters were used as input to
the anti-$k_t$~\cite{Cacciari:2008gp} jet algorithm, which was run with a radius
parameter of 0.4. The jets were calibrated using the EM+JES scheme described in~\cite{Aad:2014bia,Aad:2011he}
to correct the jet energy, which was calibrated for electromagnetic particles to the response for hadrons, based on the jet energy and $\eta$.  In a first step, the calibration procedure corrected the jet energy relative to jets built from stable particles in MC simulations (see section~\ref{sec:measurement-def} for details). In a second step,
differences between data and MC simulation were evaluated
using in situ techniques exploiting the \pT\ balance between
high-\pT\ jets and  well measured physics objects.
 The calibrated jets are required to have $\pT > 25$~\GeV\  and $|\eta| < 2.5$. 
To suppress jets from additional
$pp$ interactions, the sum of the \pT\ of the tracks originating from the selected primary vertex and associated with the jet was required to be at least 75\% of the \pT\ sum of all tracks associated with the jet.  This quantity is referred to as the jet vertex fraction (JVF).  Jets with no associated tracks were also accepted.

The identification of the electron, muon and jet objects was performed independently of other object identifications, using clusters and tracks.  In particular, no distinction was made between clusters arising from electron energy deposits or from hadrons within a jet.  In order to optimise the object identification for the event selection of this analysis and to avoid double counting of energy deposits, the overlap between these identified objects was resolved as described below. 

In order to remove jets that were reconstructed from energy deposits associated with prompt electrons, jets 
 were removed from an event if they were within $\Delta R = 0.2$ of an electron with $\pT>15$~\GeV. 
To remove residual muons from heavy-flavour decays,  muons that were within $\Delta R = 0.4$ of any jet  were removed.  To apply a similar constraint on the  electrons,  electrons that were within $\Delta R = 0.4$ of any 
jet were removed from the events.  For this condition, the only jets considered were those remaining after the removal of jets associated with electrons as previously described.

 The missing transverse
momentum azimuthal angle and magnitude ($\met$) were reconstructed from the vector sum of the transverse momenta of the reconstructed objects (electrons, muons, jets) as well as the transverse-energy deposited in calorimeter cells not associated with these objects, within the range $|\eta| < 4.9$.  The object classification scheme for the electrons, muons and jets used to calculate \met\ was chosen to be the same as the definitions given above. 
Calorimeter cells not associated with an object were calibrated at the electromagnetic (EM) scale before being added to \met.  
This calibration scheme is similar to the one described in~\cite{Aad:2012re}.

Jets were identified as ``$b$-jets'' by detecting $b$-hadron decays within the jet.  These $b$-jets were identified using the MV1 algorithm~\cite{ATLAS-CONF-2011-102}, which 
combines several tagging algorithms into a single neural-network-based discriminant, taking into account jet \pt\ and \eta\ distributions.
The selection efficiency is approximately 70\% for $\pT>20$~\GeV\ in simulated \ttbar\ events.  The 
rejection factor for jets initiated by light quarks was found to be approximately 130.

%% file: event-selection-new.tex
\subsection{Event selection}
\label{sec:event-selection}

Data used in this measurement were 
collected by triggering on either a high-\pT\ electron, based on calorimeter energy deposits, shower shape and track 
quality constraints; or a high-\pT\ muon,
comprising a reconstructed track in the MS matched with a
reconstructed track in the ID.
The \pT\ threshold for the muon trigger was 18~\GeV, whereas
the electron trigger threshold was 20~\GeV\ or 22~\GeV\ according to the data-taking period.  The reconstructed lepton object was required to be within $\Delta R < 0.15$ of the lepton reconstructed by the high-level trigger.

The selected events were required to contain at least one reconstructed primary
vertex.  To avoid events with bad detector components or reconstruction performance, events were rejected that contained any jet with $\pT >20$~\GeV\ that was
identified as arising from calorimeter noise or out-of-time activity with respect to the
primary $pp$ collision~\cite{Aad:2014bia}. 
Furthermore, events in which an electron and a muon shared the same track were removed.

Events were selected if they contain exactly one reconstructed electron ($e$) or muon ($\mu$) and at least three jets with 
 $\pT > 25$~\GeV\ and $|\eta| < 2.5$. 
One of the jets was required to be $b$-tagged. In addition, $\met > 30$~\GeV\ and a transverse $W$ mass\footnote{The variable \mtw\ is defined as
 $\sqrt{2 \pT^{\ell} \pT^{\nu} (1 - \cos(\phi^{\ell} - \phi^{\nu}))}$, where $\ell$ and $\nu$ refer to the charged lepton ($e$ or $\mu$) and \met\ respectively.}   
 $\mtw > 35$~\GeV\ were required.
To reduce the contribution of dilepton \ttbar\ final states, events with additional leptons (electrons or muons) with $\pt > 15$~\GeV\ were excluded.  Events with jet-jet pairs  with $\Delta R < 0.5$ were excluded to reduce jet \pT\ migrations between particle and reconstructed jets.

In addition to this event selection, events for the  jet \pT\ measurement were required to have a leading jet with $\pT > 50$~\GeV\ and a $2^\mathrm{nd}$-leading jet \pT$ > 35$~\GeV.
 Measurements of the jet multiplicity were also performed by selecting events with the jet \pT\
threshold raised from 25~\GeV\ to 40~\GeV, 60~\GeV\ and 80~\GeV\ in both channels, where the rest of the event selection was as described before.

The numbers of selected events are shown in tables~\ref{tab:elyield} and \ref{tab:muyield} for the electron and muon channel, respectively.

\subsection{Estimation of backgrounds}
\label{sec:background}
\label{sec:wplusjets}

The dominant background in this measurement is the
associated production of $W$ bosons with jets (including those arising from charm and
bottom quarks), followed by  single-top-quark production and multijet production.  Smaller backgrounds arise from \zjets\ and
diboson production ($WW$, $WZ$, $ZZ$). 

The  normalisation of the \wjets\ contribution was extracted from a lepton charge
asymmetry measurement from data. The method uses the fact that 
the production of $W$ bosons at the LHC is charge asymmetric, and the
theoretical prediction of the ratio $r_\mathrm{MC} \equiv \frac{\sigma(pp\rightarrow W^+)}{\sigma(pp\rightarrow W^-)}$ has an uncertainty of only a few percent.  Most processes other than $W$ production are either mostly or completely charge symmetric.  The number of events in data with a positively (negatively) charged-lepton was measured and is referred to as $D^+$ ($D^-$).  Therefore, $N_{W^+}-N_{W^-} \approx D^+-D^-$, where $N_{W^+}$ ($N_{W^-}$) is the number of $W^+$ ($W^-$) events.
The \wjets\ estimate then comes from:
\begin{equation}
N_{W^+}+N_{W^-}=\frac{r_\mathrm{MC}+1}{r_\mathrm{MC}-1}(D^+-D^-)
\end{equation}
The normalisation was determined in \wjets\ events before
any $b$-tagging requirement, separately for the $W\!+\!3$~jet,
  $W\!+\!4$~jet and $W\!+\!\ge 5$~jet events.

The flavour composition was derived from a \wTwojets\ measurement
from data.  The number of \wTwojet\ events before and after $b$-tagging was
measured using the charge-asymmetry technique. The number of \wTwojet\ events after $b$-tagging can be expressed in terms of the number of \wTwojet\  events before $b$-tagging, the flavour fractions and $b$-tagging probabilities. The flavour fractions were adjusted to ensure that the 
 derived number of \wTwojet\ events after $b$-tagging matched the
 data.  The overall charge-asymmetry normalisation was fixed, and a fit procedure was used to extract the normalisation of the bottom and charm-quark
fractions ($Wb\bar{b}\!+\!\mathrm{jets}$,
$Wc\bar{c}\!+\!\mathrm{jets}$, and  $Wc\!+\!\mathrm{jets}$).  The heavy-flavour components were then extrapolated to events with higher jet multiplicities.

In the \ejets\ channel, either jets or electrons originating from photon
conversions can mimic an isolated electron from a $W$ boson decay and are referred to as the multijet background.
In the \mujets\ channel, the multijet background arises mostly from leptonic decays of heavy-flavour quarks. 
The shape and normalisation of the multijet background in the \ejets\
channel was obtained using a matrix method~\cite{Aad:2010ey}
with looser electron
identification cuts and no isolation requirement.  The $\met<20$~\GeV\ region was used as the control region for this method. 
The multijet background in the
\mujets\ channel was determined using the mean of two matrix
method estimates, which differ in their choice of normalisation region.  The first method uses a low-\mtw\ region, whereas the
second method uses a region where the selected muon
has a large impact parameter with respect to the primary vertex.  The low-\mtw\ region includes events that do not contain $W$ bosons, whereas the high impact parameter region includes muons from heavy-flavour decays.

Contributions from single-top-quark, \zjets, and diboson production were
evaluated using the corresponding MC samples and theoretical
cross-sections for these processes.

\input{TTbar_PowHeg_Pythia_P2011C_eljets}
\input{TTbar_PowHeg_Pythia_P2011C_mujets}

%% file: TTbar_PowHeg_Pythia_P2011C_eljets.tex
\begin{table}[ht]
\begin{center}
\begin{tabular}{|c|c|c|c|c|c|c|c|c|}\hline
       &       & \multicolumn{6}{c|}{Reconstructed jet multiplicity} \\ \hline
Source & Yield & 3 & 4 & 5 & 6 & 7 & $\geq$8 \\ \hline
$\ttbar$ & 25660 & 10060 & 9068 & 4335 & 1567 & 472 & 158\\
$\wjets$ & 7238 & 5257 & 1525 & 367 & 70 & 13 & 6\\
Multijet & 2150 & 1409 & 498 & 166 & 58 & 12 & 7\\
Single-top-quark & 2935 & 1904 & 760 & 215 & 45 & 9 & 1\\
$\zjets$ & 925 & 578 & 239 & 85 & 18 & 5 & 1\\
Diboson & 180 & 140 & 32 & 6 & 1 & 0 & 0\\
\hline
Expectation & 39087 & 19347 & 12123 & 5174 & 1759 & 512 & 172\\ \hline
Data ($4.59\pm 0.08$~\ifb) & 38318 & 19471 & 11791 & 4964 & 1544 & 424 & 124\\
\hline
\end{tabular}
\end{center}
\caption{The numbers of selected data, MC simulation and background events in the electron channel, for the 25~\gev\ jet \pt\ threshold.  The yield column shows the total number of events passing the full event selection, which requires three or more selected jets.  The POWHEG+PYTHIA MC simulation sample was used for the \ttbar\ prediction.  The numbers of \ttbar, single-top-quark, \zjets\ and diboson events were normalised to the integrated luminosity of the data.  The other yields were determined from fits to data distributions.}
\label{tab:elyield}
\end{table}

%% file: TTbar_PowHeg_Pythia_P2011C_mujets.tex
\begin{table}[ht]
\begin{center}
\begin{tabular}{|c|c|c|c|c|c|c|c|c|}\hline
       &       & \multicolumn{6}{c|}{Reconstructed jet multiplicity} \\ \hline
Source & Yield & 3 & 4 & 5 & 6 & 7 & $\geq$8 \\ \hline
$\ttbar$ & 30741 & 11953 & 10884 & 5220 & 1903 & 580 & 200\\
$\wjets$ & 10424 & 7514 & 2261 & 510 & 104 & 28 & 7\\
Multijet & 1063 & 737 & 227 & 68 & 23 & 7 & 3\\
Single-top-quark & 3498 & 2274 & 901 & 252 & 57 & 11 & 3\\
$\zjets$ & 546 & 368 & 126 & 40 & 10 & 1 & 0\\
Diboson & 211 & 166 & 38 & 7 & 1 & 0 & 0\\
\hline
Expectation & 46482 & 23013 & 14436 & 6096 & 2098 & 627 & 213\\ \hline
Data ($4.59\pm 0.08$~\ifb) & 46192 & 23447 & 14170 & 5851 & 1977 & 568 & 179\\
\hline
\end{tabular}
\end{center}
\caption{The numbers of selected data, MC simulation and background events in the muon channel, for the 25~\gev\ jet \pt\ threshold.  The yield column shows the total number of events passing the full event selection, which requires three or more selected jets.  The POWHEG+PYTHIA MC simulation sample was used for the \ttbar\ prediction.  The numbers of \ttbar, single-top-quark, \zjets\ and diboson events were normalised to the integrated luminosity of the data.  The other yields were determined from fits to data distributions.}
\label{tab:muyield}
\end{table}

%% file: monte-carlo.tex
\section{Monte Carlo simulation} \label{sec:monte-carlo}

MC simulations were used to correct the measurement for detector effects and to estimate some of the 
background contributions.

To derive corrections for detector effects, a good description of the \ttbar\ signal process is important.
Signal predictions rely on matrix-element calculations for short distance physics processes and on parton shower,
fragmentation and proton remnant modelling for long-range effects.
The potential bias of the final result due to a particular model chosen was estimated by generating MC samples using alternative models for each of these components. 

In modern MC generators, there are mainly two different approaches 
 used to provide predictions of \ttbar\ final states and their multijet topology.
The first approach focuses on a precise prediction using merged leading-order (LO) matrix elements for a given number of hard partons supplemented with parton-shower emissions in the soft-collinear region.  The second approach focuses on the most accurate prediction of the inclusive rates of  \ttbar\ production  by calculating the matrix elements at next-to-leading order (NLO).  Programs implementing this approach also provide an accurate description at leading order of the $\ttbar\!+\!1\;\textrm{jet}$ final state, and leading-logarithmic accuracy for additional jet production.
In this analysis, the first approach was used in the form of the {\sc ALPGEN}~\cite{MAN-0301} MC generator.
 This sample was compared with the alternative approach implemented in the {\sc MC@NLO}~\cite{Frixione:2002ik} and {\sc   POWHEG}~\cite{Alioli:2010xd} MC generators.  In both cases, 
the matrix-element calculation was matched to separate programs for the simulation of the long-range effects.

The {\sc  ALPGEN} sample was generated using version 2.13, with the
CTEQ6L1  parton distribution functions (PDFs)  and the associated value of the strong coupling constant $\alphas(m_\mathrm{Z})=0.129$~\cite{Stump:2003yu}.  The factorisation and renormalisation scales were set to the default values of the program, i.e. $\mu_\mathrm{F}^2 = \mu_\mathrm{R}^2 = \sum{\left(m^2 + \pT^2\right)}$, where the sum was calculated over top, heavy quarks and light quarks with mass $m$ and transverse momentum \pt. ALPGEN  was used to calculate 
LO matrix elements for up to  five hard partons. 
  Parton showering and fragmentation were performed using {\sc HERWIG}~\cite{COR-0001} v6.520 
together with {\sc JIMMY}~\cite{JButterworth:1996zw} for the multiple-parton interaction model using the AUET1 tune~\cite{PUB-2010-014}. 
The MLM parton-jet matching
scheme~\cite{MAN-0301} was applied,\footnote{using matching scale $ETCLUS$ of 20~\GeV\ and a matching radius of 0.7.} to avoid double counting
configurations generated by both the parton shower and the
matrix-element calculation.  This resulted in samples with up to four hard partons exclusively and five hard partons inclusively, where the inclusive five parton sample includes jets produced by the parton shower. The processes $\ttbar + b\bar{b}$ and $\ttbar + c\bar{c}$ were generated separately using the same programs and algorithm as described above. The exclusive heavy-flavour samples were combined with the $\ttbar$  inclusive samples, after the removal of overlapping events.  The overlapping events were rejected if  the \pT\ of the $b$ or  $c$-quarks was above 25~\GeV\ and they were matched to jets within a cone of $\Delta R = 0.4$.  This sample is referred to as ``{\sc ALPGEN}+{\sc HERWIG}'' in the following discussion.

 Further \ttbar\ samples were generated following the alternative approach with NLO perturbative QCD calculations. 
 A {\sc MC@NLO} sample was
produced with the CT10~\cite{CT10Lai2010} PDF set and using the default values of the program for renormalisation and factorisation scales, i.e.  
$\mu_\mathrm{F}^2 = \mu_\mathrm{R}^2 = (p_\mathrm{T,t}^2 +p_\mathrm{T,\bar t}^2)/2 + m_\mathrm{t}^2$, where $p_\mathrm{T,t}$  ($p_\mathrm{T,\bar t}$) refers to the \pT\ of the top (antitop) quark and $m_\mathrm{t}$ is the top mass. 
MC@NLO was  also interfaced to  {\sc
  HERWIG/JIMMY} with the AUET1 tune.  {\sc POWHEG} ({\sc POWHEG}-hvq, patch4)  samples were produced
with the CT10 PDF set, using the default setting of the hard-process scales $\mu_\mathrm{F}^2 = \mu_\mathrm{R}^2  = \pT^2  + m_\mathrm{t}^2$, where \pT\ corresponds to the parton-level top-quark transverse momentum.  POWHEG  was used to produce the matrix-element calculation and top-quark decay.  To assess the effect of different fragmentation, multi-parton interaction and parton-shower models, the same POWHEG sample was matched to  two different multi-purpose generators. 
   One  sample was produced by matching with {\sc
  PYTHIA6}~\cite{Sjostrand:2006za}, using the  ``C'' variant of the Perugia 2011 tune family~\cite{Skands:2010ak} that uses the 
CTEQ6L1 PDF.
  Another  sample was produced by matching to {\sc HERWIG}+{\sc JIMMY} with the AUET1 tune.   These samples are referred to as ``{\sc POWHEG+PYTHIA}'' and ``{\sc POWHEG+HERWIG}'', respectively, in the following text.
The {\sc POWHEG+PYTHIA} sample was used as the nominal \ttbar\ sample for the correction of detector effects.

The uncertainty on the predictions due to modelling of initial-state radiation (ISR) and final-state radiation (FSR) was estimated  using {\sc ALPGEN} v2.14 with the {\sc PYTHIA6} parton-shower, the CTEQ5L
PDF~\cite{CTEQ5}, and the Perugia 2011 family of tunes. For these variations,  the same $\alphas (m_\mathrm{Z})$ value was used for the  calculation of the matrix elements and for the parton shower as suggested in ref.~\cite{Cooper:2011gk}. For the {\sc ALPGEN+PYTHIA} central sample, the Perugia 2011 central tune which employs $\lambda_\mathrm{QCD}=0.26$ was used. Uncertainties due to ISR/FSR-modelling choices were estimated by varying the {\sc ALPGEN}
renormalisation scale associated with \alphas\ up and down at each local vertex in
the matrix element  relative to the original scale.  A factor of 2.0  (0.5) was applied, resulting in lower (higher) \alphas\ values,  respectively.
The effective \alphas\ value in the parton shower was varied by the same factors as  the matrix-element calculation and the corresponding  {\sc PYTHIA6} tunes ``Perugia 2011 radHi'' and ``Perugia 2011 radLo''~\cite{Skands:2010ak} were used.
  In this paper, these samples are referred to as ``\alphas\ down'' and ``\alphas\ up''.
These settings were
shown to produce variations that are similar to the uncertainty bands on the distributions of the additional jet-veto variables $f(Q_\mathrm{0})$ and $f(Q_\mathrm{sum})$ that are described in ref.~\cite{ATLAS:PUBtop}. 

To estimate radiation uncertainties in the {\sc POWHEG} predictions, the model parameter $h_\mathrm{damp}$, which effectively regulates the high-\pT\ radiation in  {\sc POWHEG}, was set to 172.5~\GeV\ (value used for $m_\mathrm{t}$) following a similar strategy as in ref.~\cite{ATLAS:PUBtop2} while  all other {\sc POWHEG} samples used the default value  $h_\mathrm{damp}\sim \infty$.  This sample was generated using POWHEG-BOX (revision 2330, version 1.0) and is referred to as ``{\sc POWHEG($h_\mathrm{damp}$)+PYTHIA}'' in the following discussion.

The effect of colour reconnection was estimated by generating a {\sc POWHEG}+{\sc PYTHIA6} sample in which no colour reconnection was allowed within {\sc PYTHIA6}, using the ``noCR''   Perugia 2011 tune~\cite{Skands:2010ak}.

The \ttbar\ cross-section for $pp$ collisions at a centre-of-mass energy of $\sqrt{s} = 7 \tev$ was calculated to be $\sigma_\mathrm{t\bar{t}}= 177^{+10}_{-11}$~pb for $m_\mathrm{t} = 172.5$~\GeV.  This calculation was carried out at next-to-next-to-leading order (NNLO) in QCD including resummation of next-to-next-to-leading logarithmic (NNLL) soft gluon terms~\cite{Czakon:2013goa,Czakon:2012pz, Czakon:2012zr,Baernreuther:2012ws,Cacciari:2011hy} with Top++2.0~\cite{Czakon:2011xx}.  The PDF and \alphas\ uncertainties were calculated using the PDF4LHC prescription~\cite{Botje:2011sn} with the MSTW2008 68CL NNLO~\cite{Martin:2009iq,Martin:2009bu}, CT10 NNLO~\cite{Lai:2010vv,Gao:2013xoa} and NNPDF2.3 5f FFN~\cite{Ball:2012cx} PDF sets, and added in quadrature to the scale uncertainty.  The NNLO+NNLL value is about 3\% larger than the exact NNLO prediction, as implemented in HATHOR 1.5~\cite{Aliev:2010zk}.  All \ttbar-MC samples were generated with $m_\mathrm{t} = 172.5$~\GeV\ and were normalised to
the NNLO+NNLL theoretical cross-section.
 
For the simulation of the background processes, samples of $W$ and $Z$ bosons with additional jets  were generated using {\sc ALPGEN}
v2.13, with the CTEQ6L1 PDF, {\sc HERWIG}  and {\sc JIMMY} with the AUET1 tune.  Separate configurations were used for each partonic final-state with one  to four associated partons.  Parton multiplicities of five or more were generated inclusively.  Since this analysis selects events based on identified $b$-jets, specific predictions of  
$Wb\bar{b}\!+\!\mathrm{jets}$, $Wc\bar{c}\!+\!\mathrm{jets}$, $Wc\!+\!\mathrm{jets}$ and $Zb\bar{b}\!+\!\mathrm{jets}$
 events are necessary. Therefore, these processes were generated using LO matrix-element  calculations and the overlap between these samples
and the respective inclusive jet-flavour samples was removed using the same method as previously
described for the \ttbar\ samples.
 In the case of \wjets, the 
normalisation was determined from  data as described in section~\ref{sec:wplusjets}, whereas the MC
simulation was used to provide the information on the  shape of the multiplicity spectrum. 

The $t$-channel single-top-quark sample was generated with the {\sc
  AcerMC} generator~\cite{KER-0401}, whereas {\sc MC@NLO} was used to
generate the $Wt$ and $s$-channel single-top-quark production processes.  The single-top-quark samples were
each normalised according to a calculation of the inclusive production cross-section at NLO accuracy complemented with  an approximate NNLO calculation
for the $t$-channel~\cite{singletop_tchan}, $s$-channel~\cite{singletop_schan} and $Wt$-channel~\cite{singletop_Wtchan}.  
Diboson events ($WW$, $WZ$, $ZZ$) 
were produced using {\sc HERWIG} normalised to the cross-section obtained from a NLO
calculation with MCFM~\cite{Campbell:1999ah} using the MSTW2008NLO PDF. 

To properly simulate the LHC environment, additional inelastic $pp$ interactions were generated with
 {\sc PYTHIA6} using the AMBT1 tune and then overlaid on top of the hard-processes.
The MC events were re-weighted such that the predicted $\langle \mu \rangle$ distribution matched that of the data run period.
The particles from additional interactions were added before the detector simulation, but were not used within the particle-level  definition described in section~\ref{sec:measurement-def}.

The {\sc POWHEG}+{\sc PYTHIA},  {\sc ALPGEN}+{\sc HERWIG}, {\sc MC@NLO}+{\sc HERWIG} and the central {\sc ALPGEN}+{\sc PYTHIA} MC samples were passed through a full {\sc Geant4}~\cite{AGO-0301} simulation of the ATLAS
detector~\cite{:2010wqa}.  The ISR/FSR variations, colour reconnection and {\sc POWHEG}+{\sc HERWIG} MC samples were passed through a parameterised simulation of the detector response~\cite{:2010wqa}.

%% file: systematic-uncertainties.tex
\section{Systematic uncertainties}
\label{sec:syst-reco}
 This section describes the sources of systematic uncertainties and how they were estimated for the signal and background 
yields.  The sources of these uncertainties include the object reconstruction and
identification, the jet energy scale (JES) calibration, the jet energy
resolution (JER), the $b$-tagging calibration, the multijet-background normalisation, and MC generator modelling.  Uncertainties relating to MC simulation modelling were evaluated for both signal and background MC samples.

\paragraph{Jet energy scale}
The JES uncertainty was evaluated using 21 nuisance parameters, which describe the \pT\ and $\eta$ dependence of the JES uncertainty.  The nuisance parameters include eleven parameters  for the effective  uncertainties of in situ measurements covering detector and modelling related uncertainties and statistical uncertainties, two parameters to model $\langle \mu \rangle$ 
dependence, one parameter for close-by jets, i.e. jet-jet pairs with a separation of $\Delta R < 1.0$, one parameter for the calibration of $b$-jets and two parameters for $\eta$-intercalibration, i.e. the uncertainty of the $\eta$ dependence of the calibration.  Uncertainties due to  different detector-simulation configurations used in the analysis and in the calibration were added as one additional uncertainty parameter ("relative non-closure").

Since details of   the fragmentation differ  between jets initiated by quarks and those initiated by gluons~\cite{Aad:2014bia}, 
the respective jet energy scale also differs slightly.  However, the in situ techniques mainly rely on processes that produce jets initiated by quarks.  Therefore, an additional uncertainty was assigned to cover potential differences  resulting  from the different quark/gluon flavour composition of the analysed sample and the jet response dependence on the jet flavour.
The quark and gluon fractions in the analysed sample were evaluated as a function of jet multiplicity, jet \pt\ and jet $\eta$,   using
the ALPGEN+HERWIG and MC@NLO \ttbar\ signal samples.  Depending on the jet multiplicity,  gluon fractions between 10\% and 60\% were predicted  within the 
acceptance of this measurement.  The predictions of the two MC models were found to agree within 10\% over the majority of the acceptance range.  The uncertainty on the predicted gluon-fractions was taken as the difference between the two 
MC models,  where 10\% was assigned as a conservative estimate when the difference between the two models was less than this.  For events with more than seven
jets, the uncertainty estimate for seven jet events was used.  The gluon-fraction and its associated uncertainty, 
together with the quark and gluon-response uncertainties, were used to  determine the resulting JES uncertainty, which was found to vary in the range
1.5--8\% depending on jet
\pT, \eta, and the jet multiplicity in the event.
An additional \pT-dependent uncertainty of up to
$2.5$\% was applied to jets matched to $b$-hadrons, to account for neutrino and muon energy losses. This was added in
quadrature to the inclusive JES uncertainty.

\paragraph{Jet energy resolution} The measurements of the jet energy resolution  from MC simulation and data were found to agree within their uncertainties~\cite{Aad:2011he}.  The resulting uncertainties on the measurement were evaluated by additionally smearing the jet energies by the systematic uncertainties on the jet energy measurement.  This resulted in an uncertainty of 2--20\%, depending on \pT\ and $\eta$.

\paragraph{Jet reconstruction efficiency} The jet reconstruction efficiency was derived from MC simulation and the uncertainty on the  efficiency was estimated in situ with  jets reconstructed from tracks in the ID 
that were matched to a jet reconstructed using calorimeter information. 
Data and MC simulation were found to agree within the uncertainties of the in situ method.  Therefore, a 2\% uncertainty was assigned to jets with $\pt<30$~\GeV~\cite{Aad:2011he}. The uncertainty at higher jet \pt\ is negligible.

\paragraph{$b$-tagging} The efficiency of the $b$-tagging algorithm was evaluated using MC samples.  The differences 
between the efficiency in data and MC simulation were evaluated using jets containing a muon 
within a multijet 
sample.   The \pT\ of the muon relative  to the jet axis, $\pT^\mathrm{rel}$, is in general harder for muons originating from $b$-hadron decays than from muons in $c$-jets and light-flavour jets. The $b$-tagging efficiency was extracted using template fits to the $\pT^\mathrm{rel}$ spectrum. 
The difference between data and MC simulation efficiencies 
 was expressed as a function of \pT\ and $\eta$ and was applied to the
MC simulation events used in this analysis. The uncertainties on this difference were derived from
the statistical and systematic uncertainties on the efficiency measurements~\cite{ATLAS-CONF-2012-043}. 

The mis-tag scale factors for $c$-jets and light-flavour jets were
measured using a vertex-mass method~\cite{ATLAS-CONF-2012-040}.  The vertex-mass was defined as the invariant mass of the charged particles associated with the
 secondary vertex.  Templates were derived from simulations and fitted to the vertex-mass distribution obtained from data after applying $b$-tagging.
Finally, the $c$-jet efficiencies were measured with an analysis based on $D^*$
mesons.

\paragraph{Jet vertex fraction} The efficiency of the $\JVF{}>0.75$ requirement was measured
using $Z \rightarrow \ell^+\ell^-$  events, with exactly  one additional jet after the suppression of
jets from additional primary interactions.  This suppression was achieved by
selecting events where the jet was produced with \pT\ balancing the $Z$ boson and an azimuthal opening angle close to $\pi$.  The systematic uncertainty due to
the JVF requirement was estimated by varying the selection parameters used to define the $Z\!+\!1\;\mathrm{jet}$ region and 
by applying the results from $Z \rightarrow \ell^+\ell^-$  events on events with \ttbar-decay topology.

\paragraph{Leptons} The mis-modelling of lepton trigger, reconstruction and selection
efficiencies in the simulation were corrected for by calculating data/MC correction factors
derived from measurements of these efficiencies in data.  $Z$ boson and $W$ boson decays ($Z \rightarrow \mu\mu$, $Z \rightarrow ee$, and $W
\rightarrow e\nu$) were used to obtain data/MC correction factors as functions of
the lepton kinematic distributions.  The uncertainties were evaluated by varying each of the lepton trigger, reconstruction and selection efficiencies within their associated one standard deviation errors, where each contribution was evaluated separately.  

The energy scale and resolution of reconstructed electromagnetic energy clusters were calibrated from resonance decays such as $Z \rightarrow ee$, $J/\psi \rightarrow ee$, or from studies of the energy/momentum ratio using isolated electrons from $W \rightarrow e\nu$.  Uncertainties on the scale and resolution were independently evaluated by fluctuating the scale or resolution correction applied to the MC events by the associated calibration factor uncertainty.
In a similar manner, the scale and resolution of the reconstructed \pT\ of muons were calibrated from $Z \rightarrow \mu\mu$ and $J/\psi \rightarrow \mu\mu$ decays.  The uncertainties on these calibrations were independently evaluated by smearing the correction applied to MC events by the associated calibration factor uncertainty.

\paragraph{Missing transverse momentum} Energy scale and \pT\ resolution corrections for $e$, $\mu$ and jets were included in the \met\ calculation.  For the calorimeter
cells not associated with a reconstructed electron or jet with \pT\
greater than 20~\GeV, an uncertainty dependent on
the total transverse energy in the calorimeter ($\Sigma \et$) was
assigned to their energy.  This is referred to as the ``Cell Out 
uncertainty''.  The uncertainty on \met\ due to additional $pp$ interactions was
estimated by varying the contributions from the cells associated with soft jets (with
$7<\pT<20$~\gev)  and Cell Out components of
\met\ within their calibration uncertainty.  This procedure was chosen following studies of the dependence of energy resolution on the number of additional interactions.

\paragraph{PDF uncertainties}
The uncertainty from using the selected PDF for MC event production was evaluated by re-weighting  the \ttbar\ ALPGEN+HERWIG MC sample generated with the CTEQ6L1 PDF
to  the nominal and eigenvector sets of the 
MSTW2008lo68cl PDF~\cite{Martin:2009iq}.  The CTEQ6L1 PDF does not provide associated eigenvector sets that can 
can be used for this purpose.  Therefore, the systematic uncertainty was determined from the differences obtained using the MSTW2008lo68cl PDF eigenvector
sets, as well as the difference between the results based on the
best-fit PDF sets of MSTW2008lo68cl and CTEQ6L1.  The total PDF uncertainty was then evaluated by summing each of these orthogonal components in quadrature.

\paragraph{Generator model dependencies}

Systematic uncertainties associated with generator modelling were evaluated from the bias observed after corrections for all detector effects, where the nominal POWHEG+PYTHIA correction factors were used to correct the reconstructed spectra of the different MC samples to particle-level distributions.

The uncertainty due to fragmentation modelling was estimated by comparing ALPGEN+PYTHIA and ALPGEN+HERWIG \ttbar\ samples.  The difference between the biases on the fully corrected spectra was taken as the uncertainty on the final spectra.
The ISR/FSR-modelling uncertainty was evaluated using the ALPGEN+PYTHIA \ttbar\ sample and the corresponding ISR/FSR MC samples \alphas-up and \alphas-down.  The maximum difference between the bias for the fully corrected spectra of ALPGEN+PYTHIA and the bias for the ISR/FSR samples was taken as the uncertainty.

The difference between fixed-order matrix-element calculations and
associated matching schemes (``MC generator'') was estimated by comparing the POWHEG+PYTHIA
and ALPGEN+PYTHIA \ttbar\ samples.  This combination was chosen in preference to a combination with MC@NLO+HERWIG, since
MC@NLO+HERWIG was found not to describe the reconstructed jet multiplicity observed in data for events with $\ge6$~jets. 

\paragraph{$\boldsymbol{W}\!+\mathbf{jets}$ background modelling}
The reconstruction,
charge-misidentification rate, backgrounds, MC generator uncertainties and PDF eigenvector sets were all varied to provide uncertainties on the 
 \wjets\ normalisation scale factors derived
from the charge-asymmetry technique.  In total, these uncertainties were found to vary from 7\% in 3-jet events up
to 15\% in $\ge 5$-jet events.  The uncertainty on each of the heavy-flavour fractions was determined by
reconstruction, background and MC generator variations within their uncertainties and an
additional uncertainty of 25\% for scaling from the 2-jet bin to any
higher jet
multiplicity.  The additional 25\% uncertainty was chosen to cover the variations of different MC predictions.
The uncertainty on the modelling of the kinematic distributions of the
\wjets\ MC samples was estimated by varying the factorisation and renormalisation scales
 and the generator cuts
 in {\sc ALPGEN}.\footnote{using the {\sc ALPGEN} parameters \texttt{iqopt3} and \texttt{ptjmin}.}

\paragraph{Multijet background modelling}
The shape uncertainty on the multijet background in the electron channel was
estimated by varying the maximum \met\ requirement for the background selection region between 15 and 25~\GeV.  The shape uncertainty in the muon channel was taken from the difference between the mean and individual shapes of the two different matrix methods.
A 20\%  normalisation uncertainty was derived  for the muon channel from the comparison of the two background selection regions.  For the electron channel an uncertainty of  50\% was chosen to cover the difference between MC predictions and data in the relevant control distributions. 

\paragraph{Other theoretical uncertainties}
The theoretical uncertainty on the single-top-quark cross-section was taken
from the approximate NNLO cross-section uncertainties to be 4\% for the $t$-channel, 4\% for the $s$-channel  and 8\% for the $Wt$-channel.
The theoretical uncertainty on the diboson cross-section was estimated  to be 5\%
by varying PDFs and comparing NLO calculations of MCFM~\cite{Campbell:1999ah} and MC@NLO. For \zjets\ a normalisation uncertainty  of 4\%  was used for samples with no additional jet and  24\%  for each additional jet was added in quadrature to cover the model uncertainties of this prediction. 

\paragraph{Luminosity}
The integrated luminosity was measured from interaction rates in symmetric forward and backward facing detectors that were calibrated using van der Meer scans~\cite{Aad:2013ucp}.  The systematic
uncertainty on this measurement was estimated to be 1.8\%.
The integrated luminosity of the data and its uncertainty were used to normalise all
MC simulation signal and background samples, with the exception of the
\wjets\ and multijet-background estimates that were extracted from fits to the data.

%% file: results.tex
\section{Reconstructed yields and distributions}\label{sec:results-recolevel}

\begin{figure}[htbp]
\centering
\subfigure[\label{fig:njet-reco-el_25} \ejets, $\pT > 25$~\GeV]{\includegraphics[width=0.49\textwidth]{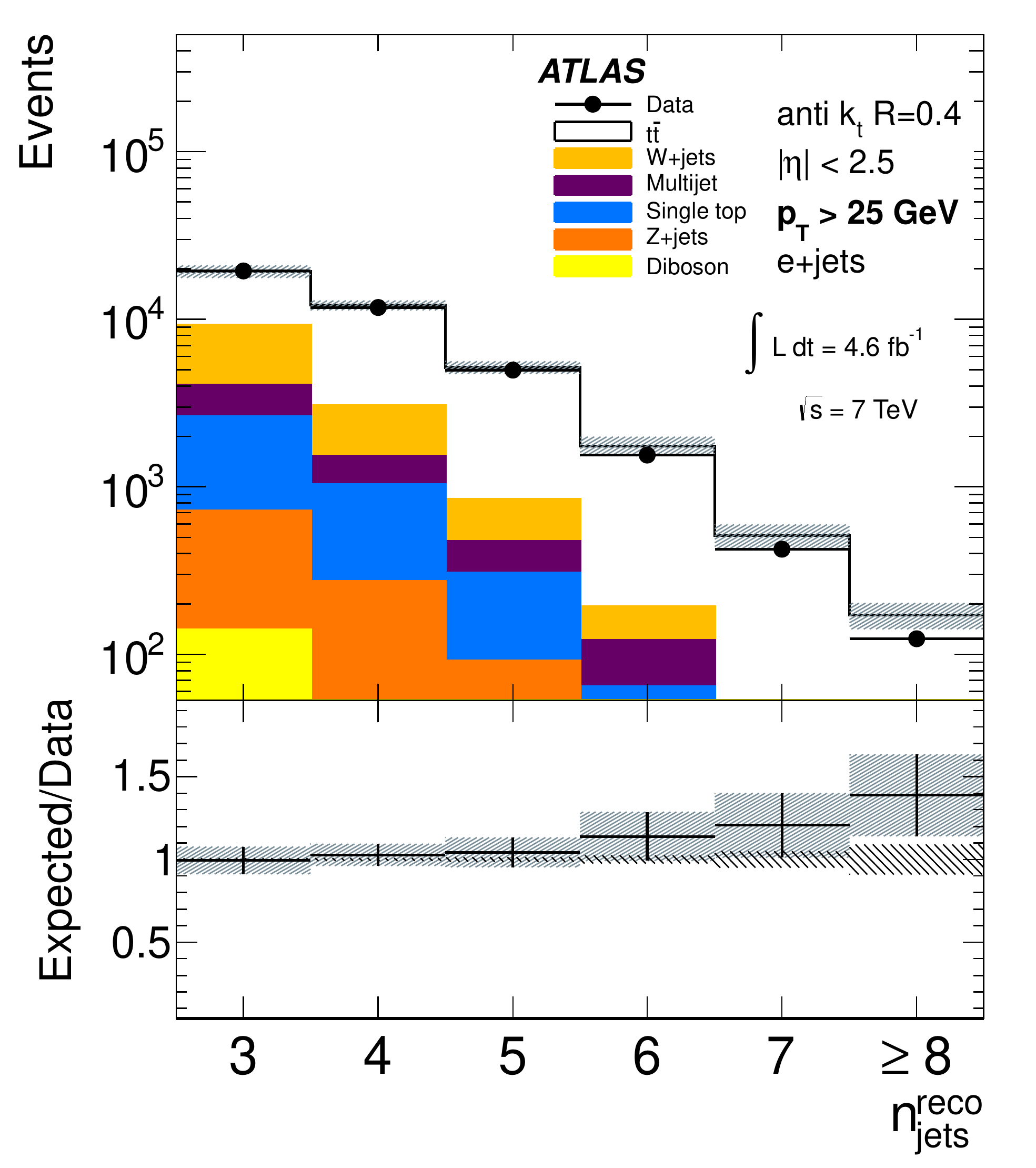}}
\subfigure[\label{fig:njet-reco-mu_25} \mujets, $\pT > 25$~\GeV]{\includegraphics[width=0.49\textwidth]{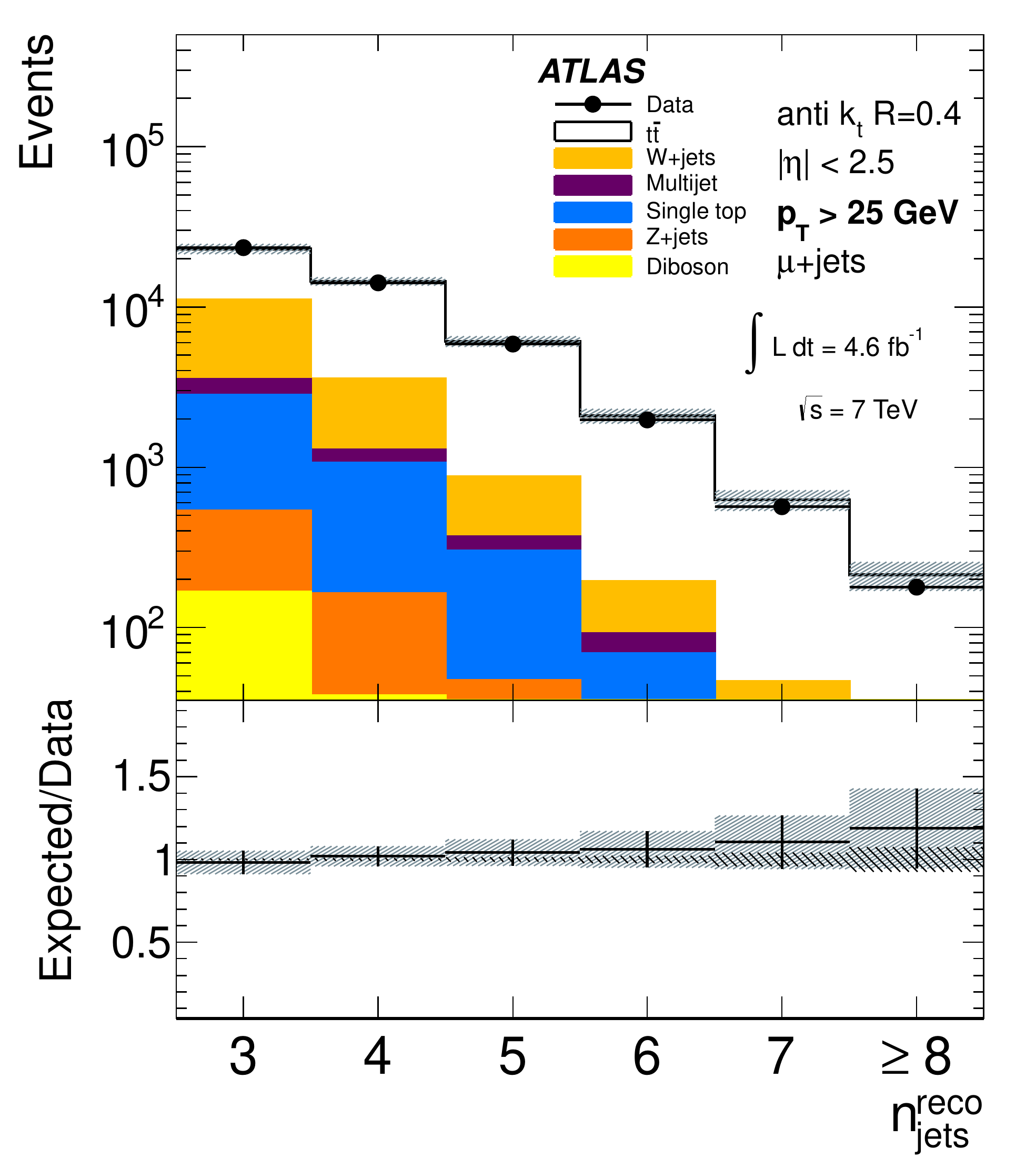}}
\caption{The reconstructed jet multiplicities for the jet \pT\ threshold of 25~\GeV, in the (a) electron (\ejets) and (b) muon (\mujets) channel.
  The data are compared to the sum of the \ttbar\ {\sc POWHEG}+{\sc PYTHIA} MC signal prediction and  
  the background models.  The shaded bands show the total systematic and statistical
  uncertainties on the combined signal and background estimate.  The errors bar on the black points and the hatched area in the ratio, show the statistical uncertainty on the data measurements.}
\label{figure:njets-reco}
\end{figure}

\begin{figure}[htbp]
\centering
\subfigure[\label{fig:jetpt-reco-el_0} \ejets, $\pT > 25$~\GeV, leading jet]{\includegraphics[width=0.49\textwidth]{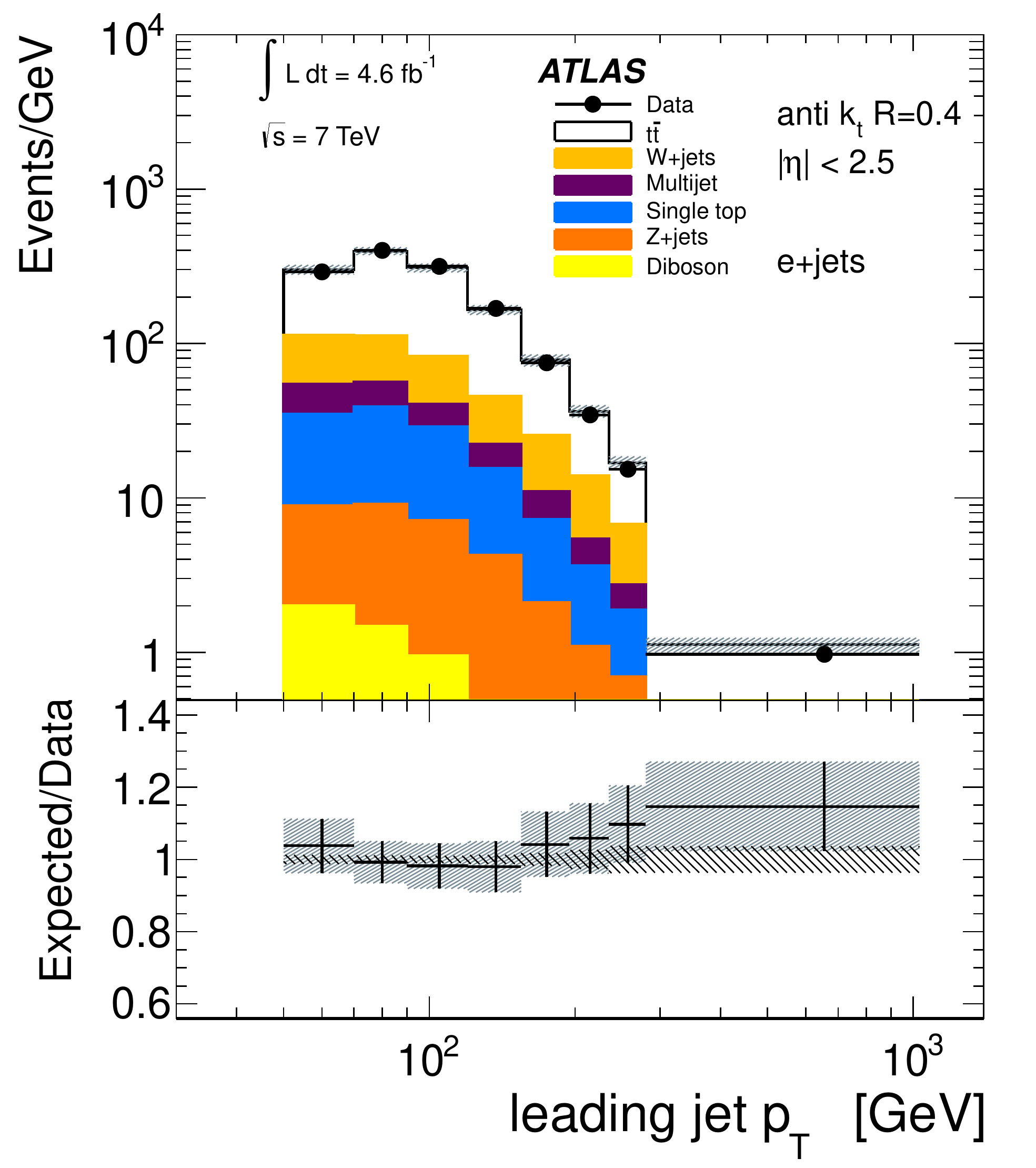}}
\subfigure[\label{fig:jetpt-reco-el_5} \ejets, $\pT > 25$~\GeV, 5th jet]{\includegraphics[width=0.49\textwidth]{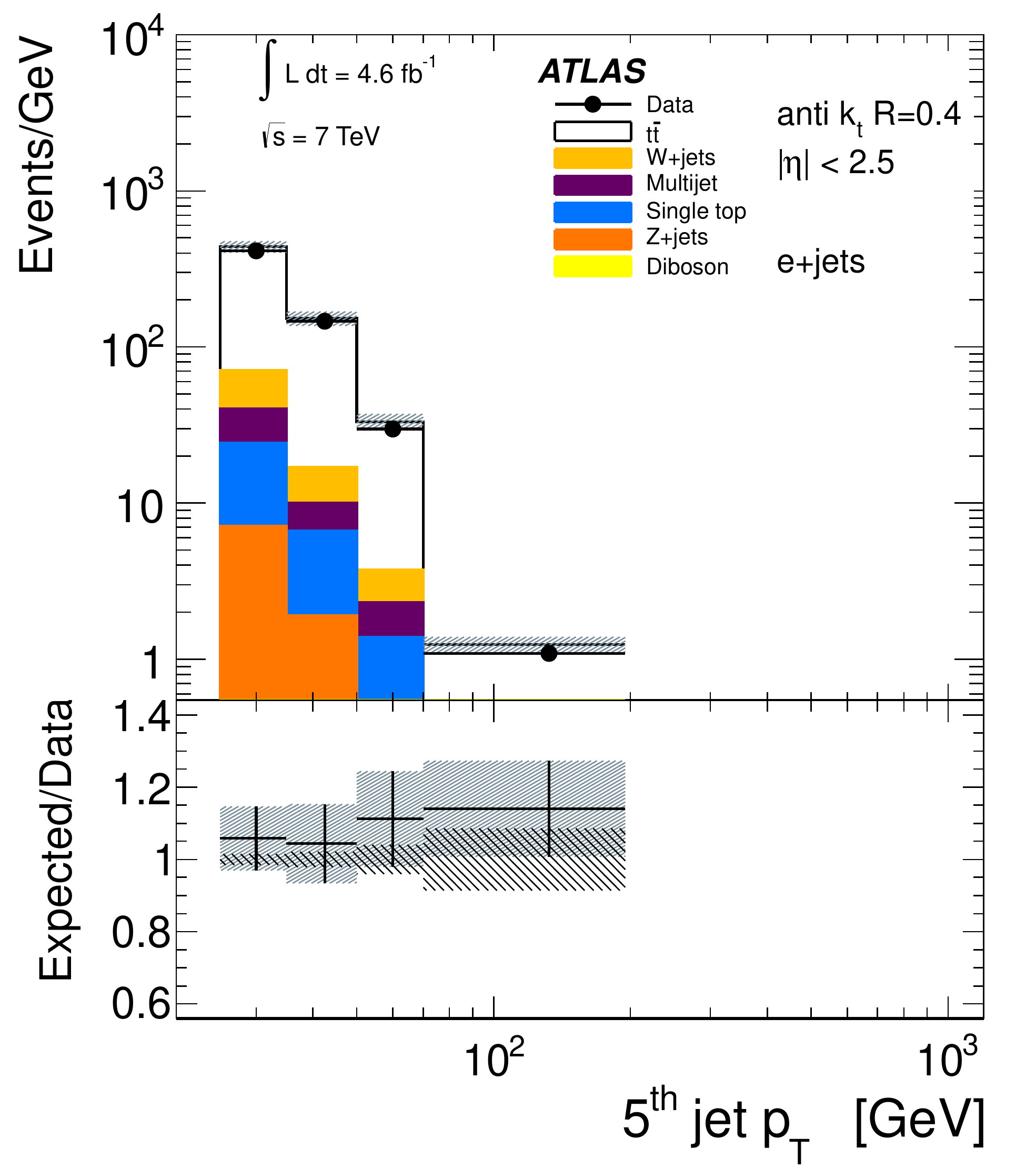}}
\subfigure[\label{fig:jetpt-reco-mu_0} \mujets, $\pT > 25$~\GeV, 5th jet]{\includegraphics[width=0.49\textwidth]{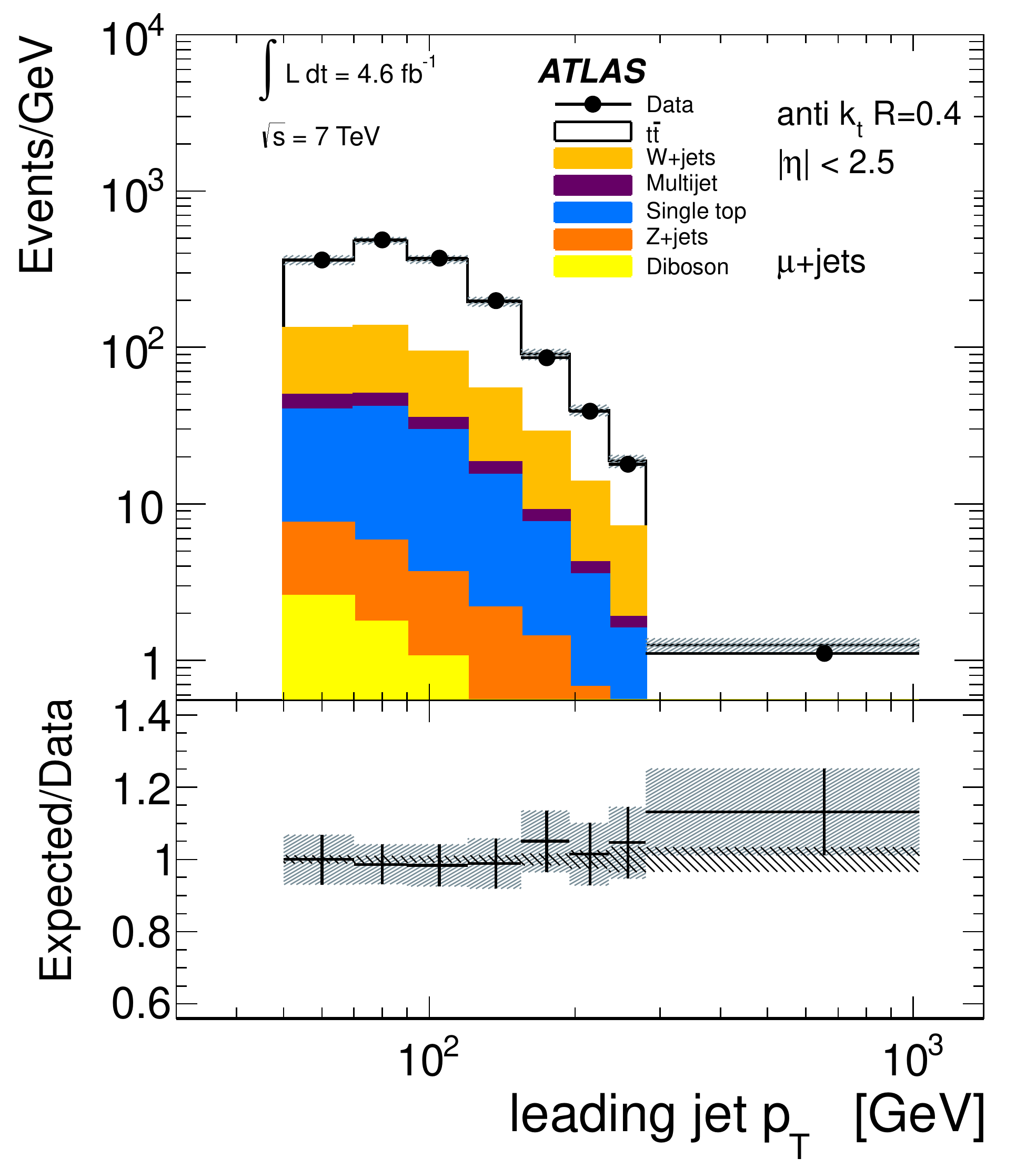}}
\subfigure[\label{fig:jetpt-reco-mu_5}  \mujets, $\pT > 25$~\GeV, 5th jet]{\includegraphics[width=0.49\textwidth]{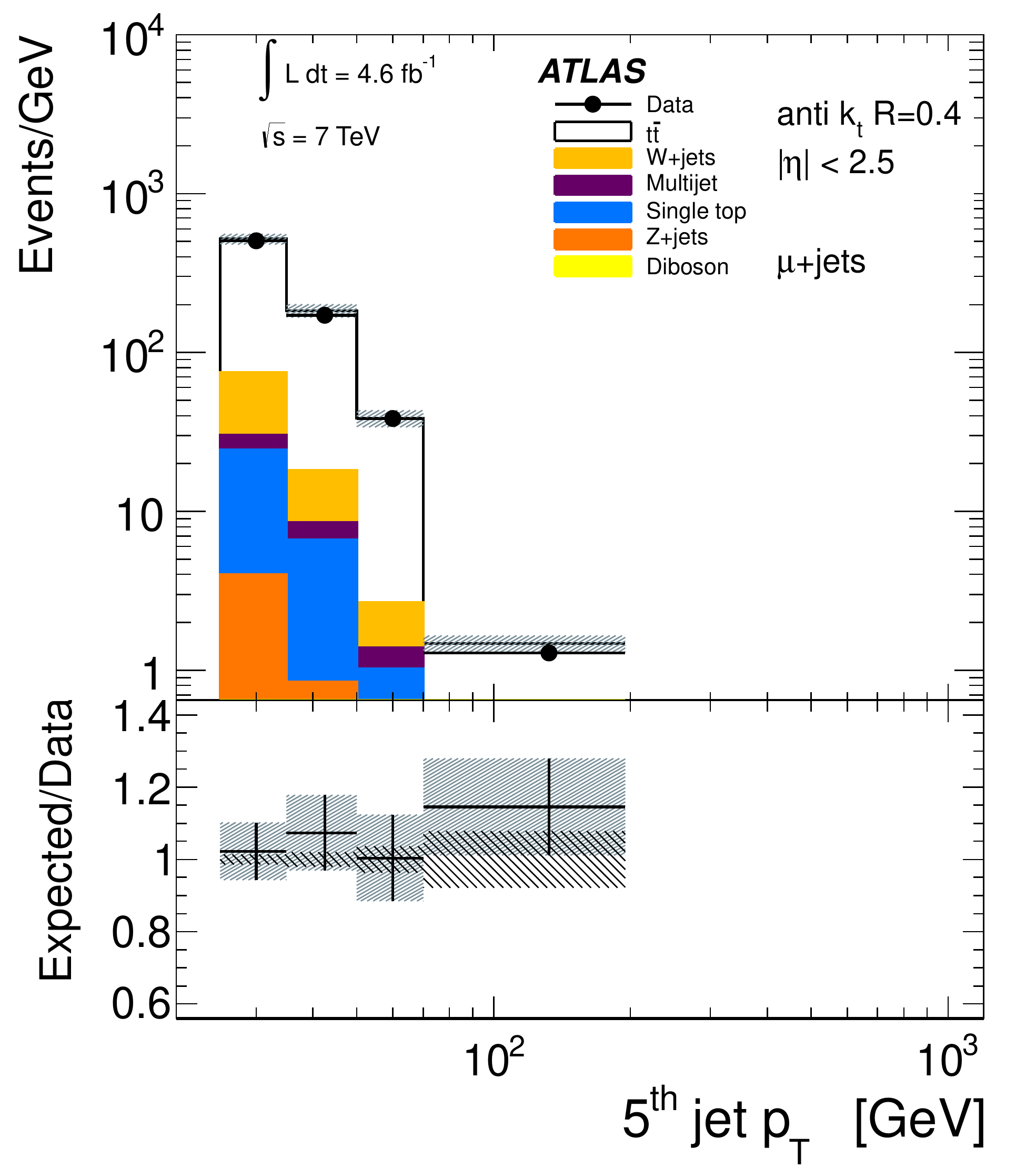}}
\caption{The reconstructed jet \pT\ for the electron (\ejets) channel (a) leading and (b) fifth jet and muon channel (\mujets) (c) leading and (d) fifth jet.
   The data are compared to the sum of the \ttbar\ {\sc POWHEG}+{\sc PYTHIA} MC signal prediction and  
  the background models.  The shaded bands show the total systematic and statistical
  uncertainties on the combined signal and background estimate.  The error bars on the black points and the hatched area in the ratio, show the statistical uncertainty on the data measurements.}
\label{figure:jetspt-reco}
\end{figure}

The predicted and observed reconstructed jet multiplicity yields for the jet \pT\ threshold of 25~\GeV\ are presented in
figure~\ref{figure:njets-reco}.  The uncertainty bands shown correspond to the combination of the uncertainty sources listed in section~\ref{sec:syst-reco}.  The jet multiplicity distributions with jet \pt\ thresholds of 40, 60 and 80~\GeV\ are shown in appendix~\ref{sec:appendixrecoresults}.  The comparison of predicted and observed jet \pT\ spectra for the leading and 5$^\mathrm{th}$ is shown in figure~\ref{figure:jetspt-reco} for events with three or more selected jets.  The bin sizes of the jet \pt\ spectra correspond to approximately one
 standard deviation of the jet energy resolution at low jet \pt.  At high jet \pt, the highest-\pT\ bin is larger to limit the effect of statistical fluctuations.  In a similar manner, the inclusive bin of the jet multiplicity spectra limits the effects of statistical fluctuations.  The predictions from the {\sc POWHEG}+{\sc PYTHIA} \ttbar\ simulation and background estimates agree with the observed jet multiplicity and jet \pt\ spectra within the total uncertainty on the prediction and the statistical uncertainties on the observed data.  The jet \pt\ spectra of the $2^\mathrm{nd}, 3^\mathrm{rd}$ and $4^\mathrm{th}$ leading jet are shown in appendix~\ref{sec:appendixrecoresults}.

\clearpage

%% file: corrections.tex
\section{Corrections for detector effects and channel combinations}

Each reconstructed spectrum was corrected  to the corresponding spectrum at particle level, within the selected kinematic range, by accounting for
detector efficiencies and resolution effects.
To minimise  the corrections of the measured data to particle level, the particles and particle jets  were defined  in a similar manner as the observable experimental objects and 
in  a kinematic phase-space close to the experimental selection, as described in section~\ref{sec:measurement-def}.
The details of the correction procedure are described in section~\ref{sec:correction}.  The propagation of measurement uncertainties through the correction procedure and additional uncertainties from the correction terms are discussed in section~\ref{sec:propagation}.  Finally, 
the combination of the results of the electron and muon channels is described in section~\ref{sec:combination}.

\input{measurement-definition}

\subsection{Correction procedure}\label{sec:correction}

The reconstructed jet multiplicity and momentum spectra were corrected to particle-level spectra, within the selected kinematic range defined in table~\ref{tab:partsel1lep} and \ref{tab:partsel1lepjet}.  The kinematic range of the measurement was chosen to be the same for particle-level and reconstruction-level objects.  However, due to limited efficiencies and detector resolutions, differences between reconstructed and particle-level distributions exist and were corrected for.  
Jet related resolutions and efficiencies that potentially lead to migration effects and bin-to-bin correlations 
 were  taken into account within an iterative Bayesian unfolding~\cite{D'Agostini:1994zf}. 

The reconstructed jet multiplicity measurements were corrected according to
\begin{equation}\label{eqn:corrections}
\Nparti =\fpartrecoi \cdot \sum_\mathrm{j} {\mrecopartij \cdot \frecopartj \cdot \facceptj \cdot (\Nrecoj - \Nbgndj)}
\end{equation}
where  $\Nparti$ is the total number of fully corrected events, i indicates the particle-jet multiplicity  
and $\fpartrecoi$ is an efficiency factor to correct for events that
fulfil the jet multiplicity requirement at particle-level but not at reconstruction level.

 $\Nrecoj$ is the total number of reconstructed events in data,  $\Nbgndj$  is the background contribution discussed in section~\ref{sec:background} and j indicates the reconstructed jet multiplicity. The factor $\facceptj$ corrects for 
all non-jet related efficiencies, such as those stemming from $b$-tagging,  trigger and lepton-reconstruction efficiencies.  It is defined as the ratio of the number of reconstructed jets, where the denominator includes the complete reconstruction-level event selection and the numerator is defined with particle-level objects for all terms other than the jet multiplicity.  The reconstructed jet multiplicity of the numerator of $\facceptj$ is defined using the same jet-electron overlap removal algorithm as described in section~\ref{object-selection}, with the exception of the electron object where the particle-level electron from the $W$ boson decay was used instead.

The factor $\frecopartj$ is a correction for events
passing the jet multiplicity requirement at the reconstruction level, 
but not at the particle level. 
 $\mrecopartij$ is a response matrix applied
iteratively as part of Bayesian unfolding. The correction factor
$\frecopartj$ and the matrix $\mrecopartij$ are defined for
the reconstructed jet multiplicity after the correction for all non-jet
acceptance effects.  They were calculated using the reconstructed jet multiplicity,  within the particle-level acceptance as defined in table~\ref{tab:partsel1lep}.

The corrected spectra were found to converge after four iterations of the Bayesian unfolding algorithm. The resulting jet multiplicity for all events that passed
particle-level lepton and $b$-tagging requirements was used for one
axis of $\mrecopartij$, and the $\facceptj$ numerator.  The $\fpartrecoi$ factor was
derived from the \ttbar\ MC sample, in a similar fashion as $\frecopartj$.

\begin{figure}[htbp]
\centering
\subfigure[\label{fig:closure_el_25GeV}]{\includegraphics[width=0.49\textwidth]{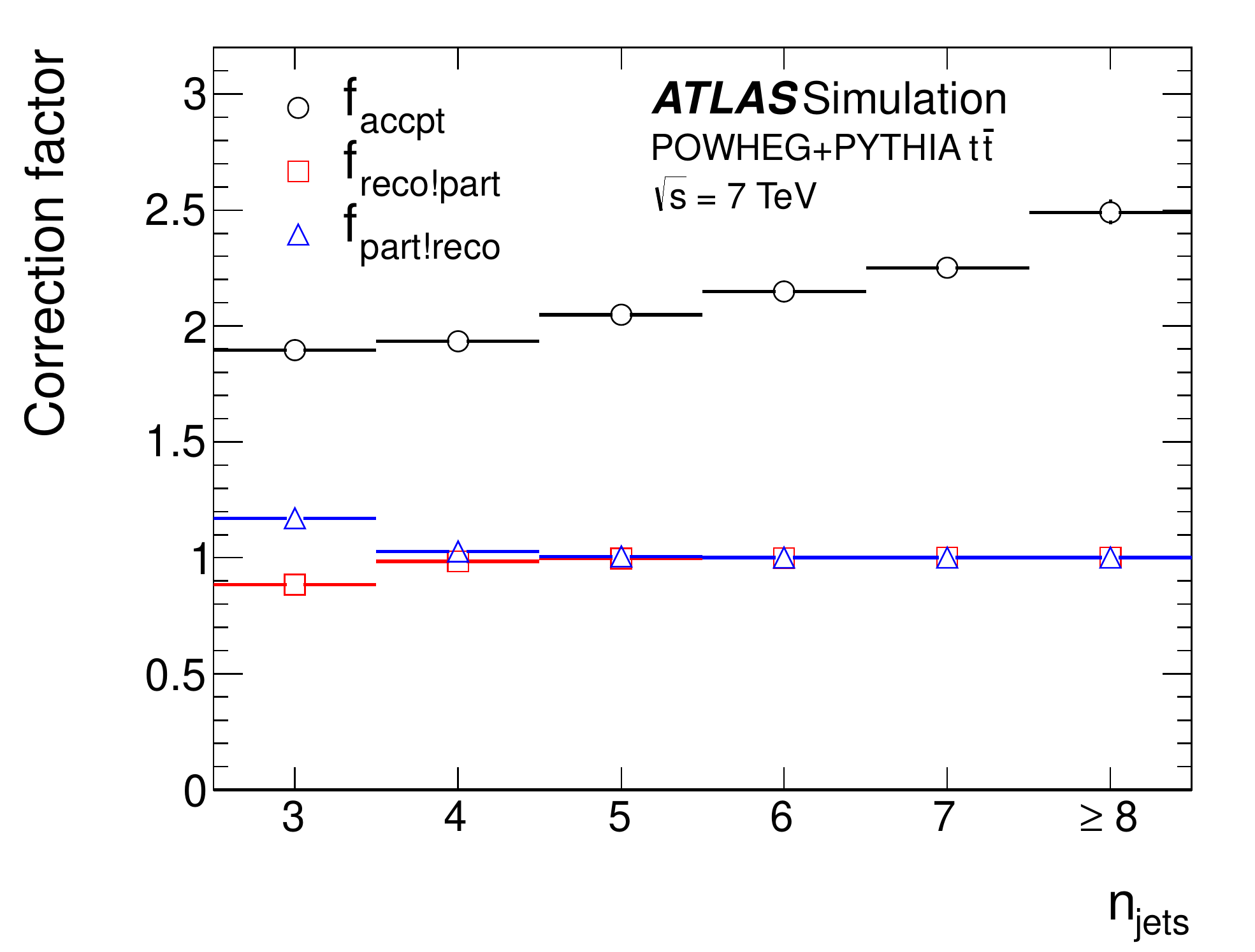}}
\subfigure[\label{fig:closure_mu_25GeV}] {\includegraphics[width=0.49\textwidth]{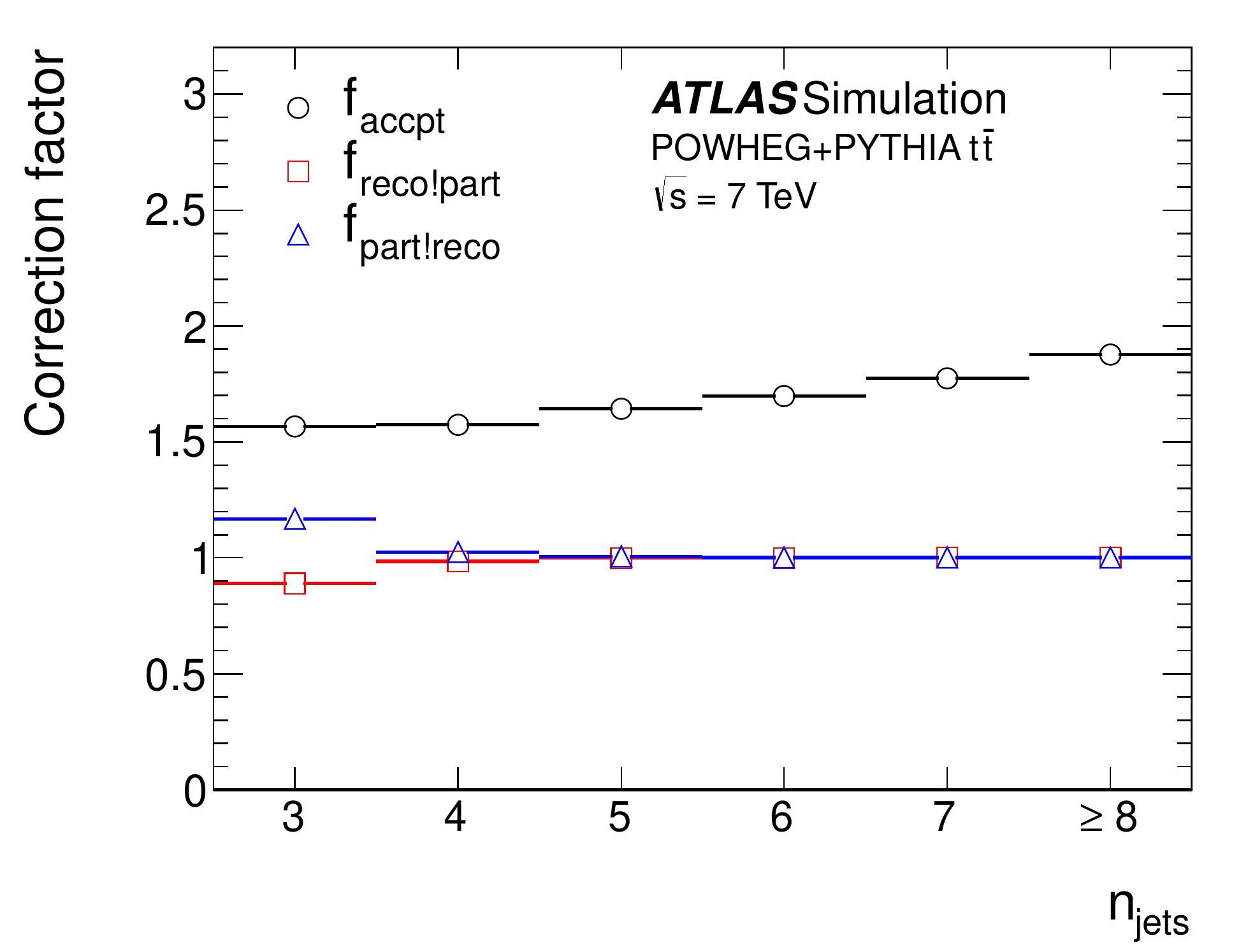}}
\caption{Global correction factors for the acceptance ($f_\mathrm{accpt}$) and particle-level and reconstruction-level inefficiencies ($f_\mathrm{part!reco},f_\mathrm{reco!part}$) to correct the jet multiplicity distribution   with $\pT>25$~\GeV\ to particle level (a) in the electron and (b) in the muon channel as described in the text and in eq.~(\ref{eqn:corrections}).    
The symbol $n_\mathrm{jet}$ refers to the number of particle-level jets for 
$f_\mathrm{accpt}$  and $f_\mathrm{part!reco}$ and to the number of reconstructed jets in case of $f_\mathrm{reco!part}$.  The distributions are shown with statistical uncertainties only, which are too small to be visible.
 \label{fig:correctionfactors}}
\end{figure}

The correction factors are shown as a function of jet multiplicity (for $\pt>25$) in figure~\ref{fig:correctionfactors}.  
In the electron (muon) channel, $\facceptj$ is around 1.9 (1.6) and rises with increasing jet multiplicity by about 40\% (20\%) in the eight-jet inclusive bin. 
Higher values of $\facceptj$ in the electron channel arise from the electron identification efficiency being lower than that of the muon identification.  The electron channel  $\facceptj$ also includes an interpolation across the $\eta$ regions of the calorimeter barrel-endcap transition.  These $\eta$ regions were excluded in the reconstructed electron selection, but not from the definition of the fiducial cross-section.  The factors $\facceptj$ for the \pt\ thresholds of 40--80~\GeV\ are  significantly less dependent on the number of jets, as shown in appendix~\ref{sec:correctionFactorsA}.

All other correction factors are approximately the same  for the electron and muon channel and close to unity for
 jet multiplicities larger than four.  Events with three or four jets are affected by migrations into or out of the fiducial volume, which is visible in the distributions of
$\frecopartj$ and  $\fpartrecoi$.

\begin{figure}[htbp]
\centering
\subfigure[\label{fig:closure_el_25GeV_jetPtN0}]{\includegraphics[width=0.49\textwidth]{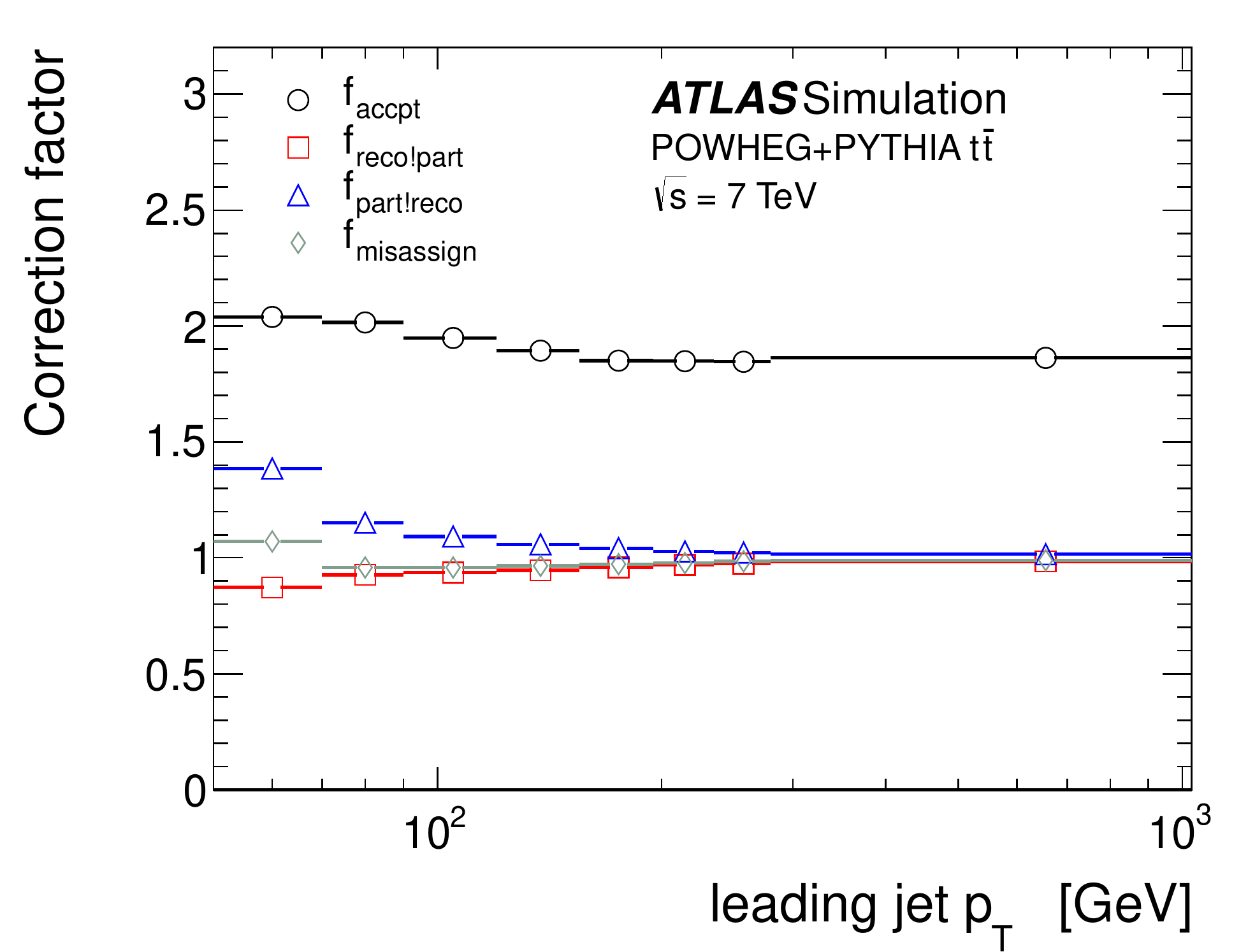}}
\subfigure[\label{fig:closure_el_25GeV_jetPtN1}]{\includegraphics[width=0.49\textwidth]{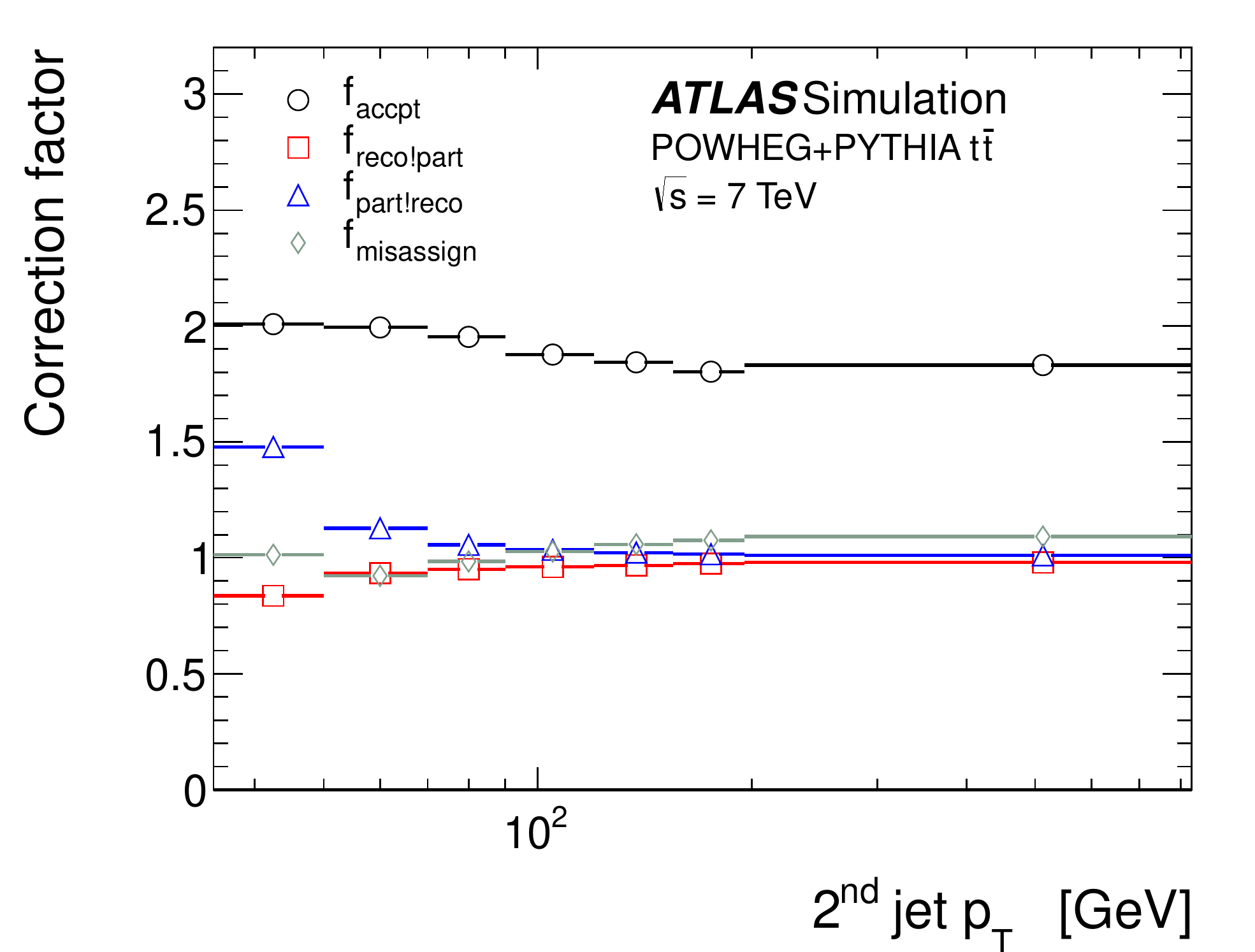}}
\subfigure[\label{fig:closure_el_25GeV_jetPtN2}]{\includegraphics[width=0.49\textwidth]{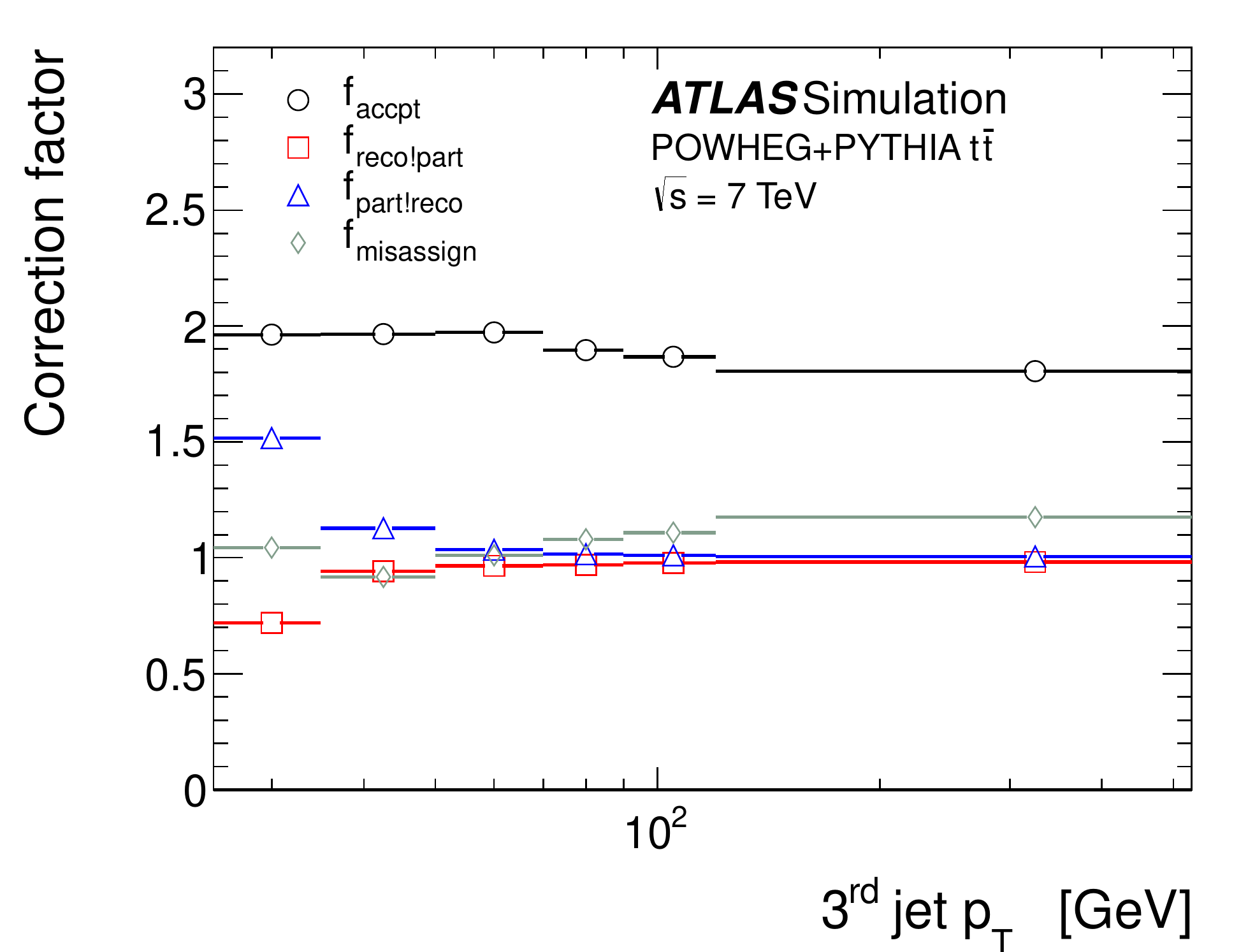}}
\subfigure[\label{fig:closure_el_25GeV_jetPtN3}]{\includegraphics[width=0.49\textwidth]{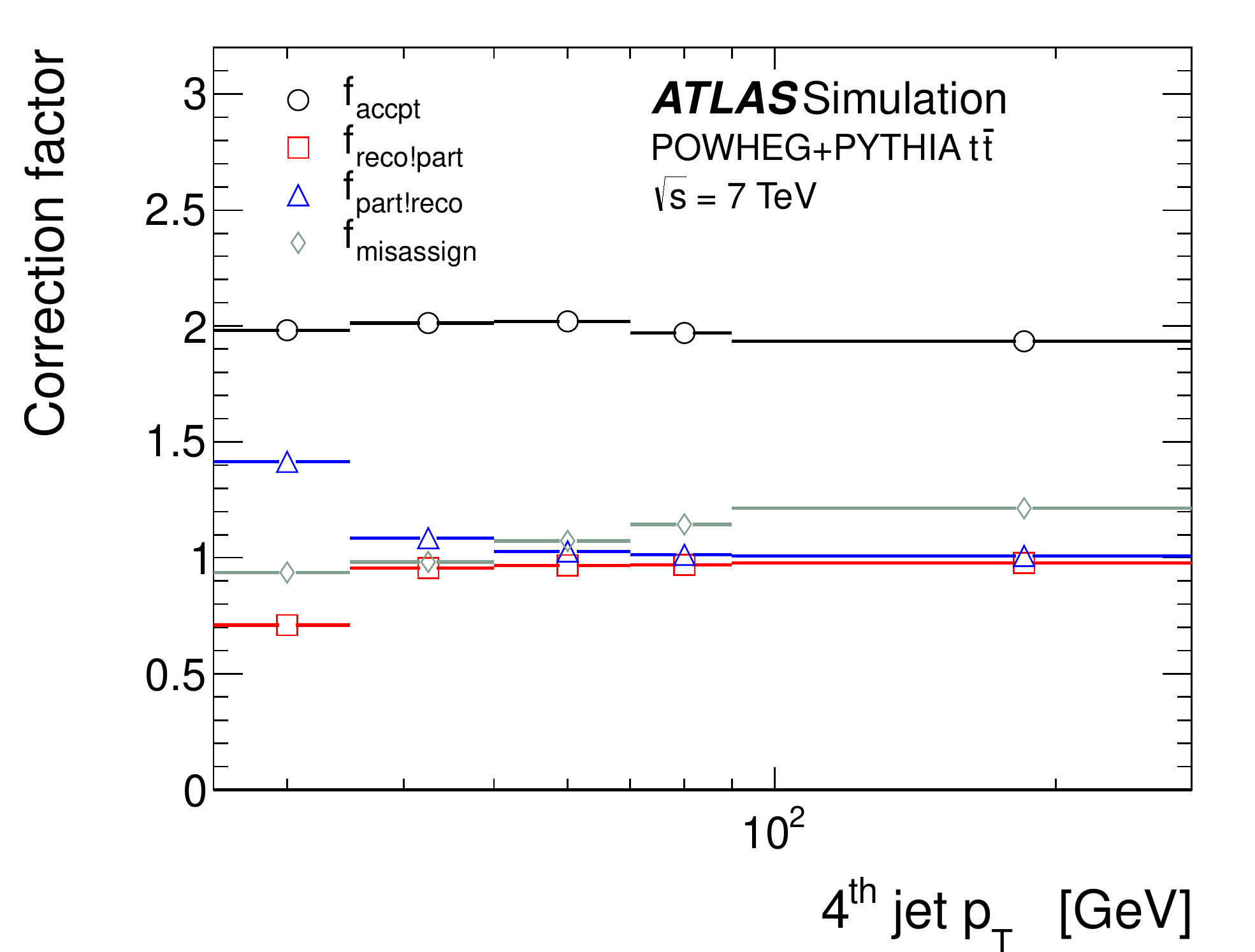}}
\caption{Global correction factors  for the acceptance ($f_\mathrm{accpt}$), particle-level and reconstruction-level inefficiencies ($f_\mathrm{part!reco},f_\mathrm{reco!part}$) and misassignment  in the  \pT\ ordering of the jets ($f_\mathrm{misassign}$), used to correct the jet \pT\ distributions to the particle level as described in the text and in eq.~(\ref{eqn:corrections-jetpt}).  The muon-channel correction factors are shown as an example.  However, the corresponding distributions of the electron channel (not shown) are similar.  The distributions are shown with statistical uncertainties only, which are too small to be visible.
 \label{fig:ptcorrectionfactors}}
\end{figure}

The transverse momentum distribution of each of the \pT-ordered jets was corrected in a similar manner as the jet multiplicity measurements.  Jet \pT\ migrations were separated into migrations between jet \pT-ordering and migrations for the same \pT-ordering.  Reconstructed jets were matched with jets of stable particles within $\Delta R < 0.35$.  Then a bin-by-bin correction (\fmisassignj) was defined as the ratio of the number of events with matching \pT-ordering over all matched jets.  The \pT\ distribution for each jet was then corrected according to
\begin{equation}\label{eqn:corrections-jetpt}
\Nparti = \fpartrecoi \cdot \sum_\mathrm{j} \mrecopartij \cdot \fmisassignj \cdot f^\mathrm{j}_\textrm{reco!part} \cdot f^\mathrm{j}_\textrm{accpt} \cdot (\Nrecoj - \Nbgndj)
\end{equation}
where the correction terms  $\mrecopartij, \fmisassignj, \frecopartj, \facceptj$ and $\Nbgndj$ are functions of the reconstructed jet \pT, $\fpartrecoi$ and $\mrecopartij$ are functions of the particle-jet \pT, and j (i) indicates the bin of reconstructed (particle) jet \pT\ distribution.  Correction factors were derived and applied individually to the \pT\ distributions of the leading, 2$^\mathrm{nd}$, 3$^\mathrm{rd}$ and 4$^\mathrm{th}$ jets.  As demonstrated in figure~\ref{fig:ptcorrectionfactors}, for jet \pT\ above 100~\GeV\ no  correction for missing jets  on particle or reconstruction level   is needed. Softer jets are more likely to fail the reconstruction-level requirements and hence the larger associated correction factor 
of up to 1.5.  However, this is compensated by a factor up to 0.7 for soft reconstructed jets that do not have a matching jet 
at particle level.  The acceptance factor ($\facceptj$) is almost independent of jet \pT; only at low \pT\ can a slight rise be observed. The factor $\fmisassignj$ rises with jet number and with \pT, which follows from the number of jets that can potentially be wrongly assigned and the possible \pT\ difference between the misassigned and the correct matching jet. The $\fmisassignj$ 
correction is very close to unity for the leading jet and within 10\% for the 2$^\mathrm{nd}$ jet.

%% file: measurement-definition.tex
\subsection{Definition of the fiducial cross-section measurement}
\label{sec:measurement-def}

The data were corrected by comparing to leptons and jets from MC generators that were defined  using particles with a mean lifetime greater than $0.3 \times 10^{-10}$~s, directly produced in $pp$ interactions or from subsequent decays of particles with a shorter lifetime.   To select  the leptons from  $W$ boson decay, all leptons ($e$, $\mu$, $\nu_{e}$, $\nu_{\mu}$, $\nu_{\tau}$) for the cross-section definition were required  not to be hadron decay products. Electron and muon four-vectors were calculated after the addition of photon four-vectors within a cone of $\Delta R = 0.1$ around their original directions.  The \met\ was calculated from the four-vector sum of neutrinos from $W$ boson decays.  
Jets were defined using the anti-$k_t$ algorithm with a radius parameter of 0.4.  All particles were considered for jet clustering, except for leptons as defined above (i.e. neutrinos from hadron decays are included in jets)
and any photons associated with the selected electrons or muons.  
Jets initiated by $b$-quarks were identified as such i.e. ``$b$-tagged'' if one or more $b$-hadrons was clustered within the given jet.

The cross-section was defined using events with exactly one electron or muon and at least three jets, each with 
 $\pT > 25$~\GeV\ and $|\eta| < 2.5$. 
One of the jets was required to be $b$-tagged. In addition, $\met > 30$~\GeV\ and 
 $\mtw > 35$~\GeV\ were required.

To reduce the contribution from dilepton \ttbar\ final states, events with additional leptons (electrons or muons) with $\pt > 15$~\GeV\ were excluded.
Following the reconstructed object selection, events with jet-electron pairs or jet-muon pairs with $\Delta R < 0.4$ or jet-jet pairs  with $\Delta R < 0.5$ were excluded. 
  
The differential production cross-section in jet \pt\
was defined using the basic selection with three or more jets with $\pt>25$~\GeV\ and the additional requirement of \pT$ > 50$~\GeV\ and \pT$ > 35$~\GeV\ on the leading and $2^\mathrm{nd}$-leading jet, respectively.  This additional selection was applied to reduce uncertainties that can arise due to a different ordering of the measured  jets with respect to the reference jets used in the correction procedure discussed in section~\ref{sec:correction}. 
The two phase-space definitions are summarised in tables~\ref{tab:partsel1lep} and \ref{tab:partsel1lepjet}.

Additional cross-sections as a function of jet multiplicity were defined by increasing the jet \pT\ thresholds from 25~\GeV\ to 40~\GeV, 60~\GeV\ and 80~\GeV\ in both channels, where the rest of the fiducial-volume definition is as described before. 

\input{TruthEventSelection_1}

%% file: TruthEventSelection_1.tex
\begin{table}[htbp]
\centering
\begin{tabular}{|c|}\hline
\multicolumn{1}{|c|}{$\met > 30$~\GeV\ \& $\mtw > 35$~\GeV} \\
\multicolumn{1}{|c|}{One or more $b$-jets} \\
\multicolumn{1}{|c|}{Three or more jets with $\pt >$25 \GeV\ \& $| \eta | < 2.5$}  \\ \hline
$e\;(\mu)$ with $\pt > 25$~\GeV\ \& $|\eta | < 2.5$  \\ 
No additional $e\;(\mu)$  with $\pt > 15$~\GeV\ \& $|\eta| < 2.5$ \\
No $\mu\;(e)$ with $\pt > 15$~\GeV\ \& $| \eta | < 2.5$  \\ \hline 
\multicolumn{1}{|c|}{No jet-jet pair with $\Delta R < 0.5$} \\
\multicolumn{1}{|c|}{No jet-electron or jet-muon pair with $\Delta R < 0.4$}\\ \hline

\end{tabular}
\caption{Fiducial-volume definition for the electron (muon) channel of the $\ttbar\!+\!\textrm{jets}$ cross-section measurement with the jet \pt\ threshold of 25~\GeV.     These conditions were applied on reconstruction-level and particle-level objects, with the exception of the electron where a veto on the $\eta$-region corresponding to the barrel-endcap transition region was applied on the reconstruction level  (as described in section~\ref{object-selection}), but not included in the fiducial-volume definition. The jet \pt\ threshold in the jet multiplicity distributions was increased to 40, 60 and 80~\GeV, for the corresponding cross-section measurements.}
\label{tab:partsel1lep}
\end{table}

\begin{table}[htbp]
\centering
\begin{tabular}{|C{4cm}|C{4cm}|}\hline
\multicolumn{1}{|c|}{Leading jet with $\pt > 50$~\GeV\ \& $| \eta | < 2.5$} \\
\multicolumn{1}{|c|}{$2^\mathrm{nd}$ leading jet with $\pt > 35$~\GeV\ \& $| \eta | < 2.5$} \\ \hline
\end{tabular}
\caption{Additional fiducial-volume requirements implemented for the \ttbar\ cross-section with respect to the jet \pT.  These requirements were made in addition to those given in table~\ref{tab:partsel1lep} and were applied to the electron and the 
muon channel.}
\label{tab:partsel1lepjet}
\end{table}

%% file: propagation.tex
\subsection{Propagation of uncertainties} \label{sec:propagation}

This section describes how the uncertainties listed in section~\ref{sec:syst-reco} were taken into account in the unfolding and 
 which additional uncertainties appear due to the unfolding procedure.

The response matrix (\mrecopartij) and the correction factors (\fpartrecoi,  \fmisassignj, \frecopartj\ and \facceptj) were
determined using the nominal {\sc POWHEG}+{\sc PYTHIA} \ttbar\ MC
sample.  The  statistical
uncertainty on the size of the MC sample used to derive these  factors was estimated by
smearing  the response matrix according to a Poisson distribution and the correction factors according to a normal distribution.  A Poisson probability density function was chosen for the response matrix, since the matrix contains a number of events in each bin.  The response matrix is also sparsely populated in bins that are far from the diagonal.  Therefore, using a normal distribution is not a valid approximation.  For the correction factor ratios (\fpartrecoi,  \fmisassignj, \frecopartj\ and \facceptj), the statistical uncertainty for the ratio does not correspond to an integer number of events and the number of events in each bin of the ratio is large.  Therefore, a normal probability distribution was used as an approximation for the ratio of the two Poisson distributions.  The statistical uncertainties were propagated by performing 1000 pseudo-experiments, smearing all terms simultaneously.
The difference between the
mean of all 1000 unfolded distributions and the true {\sc POWHEG}+{\sc PYTHIA} \ttbar\ distribution was taken to be the systematic deviation or bias, whereas the standard deviation was taken to be  the statistical uncertainty on the response matrix and the correction factors.

The statistical uncertainty on the reconstructed spectra ($\Nrecoj$) was propagated by performing 1000 pseudo-experiments, following a Poisson distribution corresponding to the number of events in each bin (j), where the number of events in each bin of the reconstructed spectra was independently varied.

The uncertainty on $\Nbgndj$ was determined at the reconstruction level.  The uncertainties related to the \wjets\ and multijet shapes and normalisations were propagated by forming  background subtracted spectra for each of the background-uncertainty terms.  The resulting difference between the nominal and shifted unfolded distributions was taken as the uncertainty.  The statistical significance of this  systematic uncertainty was evaluated by performing 1000 pseudo-experiments, following a normal distribution with a width matching the statistical uncertainty on the shifted input spectrum.  If the root mean square of the pseudo-experiments on the uncertainty prediction was greater than 10\%, then the systematic uncertainty estimate from the neighbouring measurement point was used.  The value of 10\% was established by studying all the systematic uncertainty variations as a function of the statistical uncertainty on the unfolded spectra.  Above a statistical uncertainty of 10\%, discontinuous predictions were observed for some systematic uncertainty variations.  This procedure has a minimal effect on the highest jet-multiplicity bins of a subset of the corrected spectra.

To avoid enlarged uncertainties due to statistical fluctuations of the small background components, all other background uncertainty terms  were combined according to their correlations and then propagated through the corrections by smearing the background subtracted spectra.  The systematic uncertainty on the unfolded spectra from the background was evaluated
by performing 1000 pseudo-experiments, following a normal distribution with a width matching the total uncertainty band.  The
square root of the variance of the unfolded spectra of the pseudo-experiments was taken as the uncertainty on the small background terms.

Systematic uncertainties affecting the \ttbar\ sample used to unfold
the jet multiplicity spectrum were each evaluated as a relative bias, i.e. deviations were determined from differences between the 
bias of the nominal sample and the systematically varied sample.
 For
each variation, a pair of particle and reconstruction-level spectra
was generated.   The bias was evaluated by performing 1000 pseudo-experiments, fluctuating
the reconstructed input-spectrum within its statistical uncertainty.
Each pseudo-experiment was unfolded (using the response matrix 
derived from the nominal {\sc POWHEG}+{\sc PYTHIA} \ttbar\ sample) and the bias was calculated from
the difference between the mean corrected distribution and the true distribution.  The systematic uncertainty estimation was taken from the relative bias, the difference between the bias evaluated with
the nominal {\sc POWHEG}+{\sc PYTHIA}   \ttbar\ sample and the bias evaluated using each reconstructed and true systematic uncertainty variation sample.  This applies to all 
cases except the {\sc ALPGEN}+{\sc PYTHIA} \alphas\ variations, where the relative bias between
the ALPGEN+PYTHIA central and shifted samples was used.  The
uncertainty on the fixed-order matrix-element calculation and
matching scheme (the generator uncertainty)  was estimated from the
relative bias of unfolding {\sc ALPGEN}+{\sc HERWIG} with respect to the {\sc POWHEG}+{\sc PYTHIA} nominal \ttbar\ sample.  The MC@NLO sample was not used for this uncertainty, since it does not describe reconstructed data well at higher jet multiplicities.  Each of the \ttbar\ model uncertainties was propagated individually and
symmetrised before being combined.  

The effect on the measured multiplicity spectra due to the JES uncertainty rises with the jet multiplicity from   3\%  to 40\% for the 25~\GeV\ jet \pT\ threshold. This uncertainty decreases in the higher jet multiplicity bins for the higher jet \pT\ thresholds, to values of around 15\%. For the 25~\GeV\ jet \pT\ threshold, the background uncertainty is 18\%(3\%) for events with low (high) jet multiplicities.  The effect of the ISR/FSR-modelling uncertainty varies from 1--6\%.  The next most significant uncertainties are the matrix-element generator and $b$-tagging uncertainties.  These are of a similar magnitude as the ISR/FSR uncertainty.  
The systematic uncertainty from the MC statistical uncertainties of each of the correction fractions is within the range 1--11\% (25~\GeV\ \pT\ threshold) and becomes significant (40\%) in events with 7(6) jets for the 60 (80)~\GeV\ \pT\ thresholds. 
Statistical uncertainties from the data are not dominant in any region.

The systematic uncertainties on the jet \pT\ spectra are 10--16\% and increase with  \pT\ except for the lowest jet \pT\ bin.  There are many sources of  uncertainties of approximately  2--7\%  depending on jet \pT.  For example, there are uncertainties   from the  $b$-jet related systematic uncertainties, i.e. uncertainty on the $b$-jet energy scale (2--5\%) and the  $b$-tagging efficiency (4--7\%), the uncertainty on the $W$+jets background (2--8\% each for normalisation and  flavour composition), and the uncertainty components of the jet energy calibration related to the detector, the close-by jet correction and the intercalibration  (each 1--3\%).  The statistical error rises with jet \pT\ and with the order of the jet for a given jet \pT\ bin.  The lowest values are 1.5\% and the highest are 14\%, which is only slightly smaller than the systematic uncertainty.

%% file: results-particlelevel.tex
\subsection{Combination of lepton channels} \label{sec:combination}

The  particle-level jet multiplicity and jet \pt\ spectra  were combined by using the Best Linear Unbiased Estimate 
 (BLUE) method~\cite{Valassi2003391,Lyons:1988rp}.  The BLUE method determines the coefficients 
(weights) to be used in a linear combination of the input measurements by minimising the total uncertainty of the combined 
result.  All uncertainties were assumed to be distributed according to a Gaussian probability density function. The algorithm 
takes both statistical and systematic uncertainties and their correlations into account.  The BLUE combination was cross-checked against an average performed using the algorithm discussed in~\cite{Aaron:2009bp}.  The two methods were found to agree within their uncertainties.  The averaging procedure was also used to probe the compatibility of the electron and the muon channel, resulting in a $\chi^2 / dof \approx 1$.

The systematic uncertainties related to the  measurements of the leptons, the multijet-background normalisation and the overall \wjets\ background normalisation were treated as uncorrelated between the two channels, but bin-to-bin correlated within one channel.
The data selected with the two different lepton event selections constitute independent samples, for which the multijet and overall \wjets\ normalisation were determined separately.  The MC statistical uncertainties on the correction factors for the two samples were also assumed to be uncorrelated.  All other systematic uncertainties were treated as fully correlated.

The uncertainty of the combined jet multiplicity measurement at low values is dominated by the uncorrelated background sources that are smaller in the muon channel than in the electron channel, due to the smaller multijet background in the muon channel (see section~\ref{sec:propagation}).  The uncertainty of the combined result is therefore similar to the uncertainty of the muon 
channel result itself.  At high multiplicity, the uncertainty is dominated by correlated sources, such as the uncertainty on the jet energy scale and model uncertainties of fragmentation and colour reconnection.  
The combined cross-section measurement has a 3\% uncertainty improvement with respect to the muon channel result and approximately a 20\% improvement with respect to the electron channel result.

The uncertainty of the cross-section measurements as a function of jet \pT\ are about 20\% smaller in  the muon channel measurement than in the electron-channel measurement, because of the significantly smaller uncertainty on the muon identification and energy scale compared to electrons. Therefore, the data selected in the muon channel have a statistically higher impact on the combined results.
The uncertainty on the combined jet \pT\ measurements is 7--14\% for the leading jet and up to 17\%
for the highest \pt\ region of the other jets.  This corresponds to an uncertainty improvement of 15--30\%,  compared to the uncertainty on the electron channel measurement and 4--7\% compared to the uncertainty on the muon-channel measurement.

A summary of systematic uncertainty components, statistical uncertainty and the total uncertainty after the channel combination is given in appendix~\ref{sec:appendixsystematics}.

\section{Results}
\label{sec:results-particlelevel}

The result of the combinations of the fully-corrected distributions for jet multiplicity and \pT\ were converted into fiducial cross-section measurements using
$\sigma^\mathrm{fid}\left(n_\mathrm{jet}\right) = \frac{N_\mathrm{part}}{\int{ L \mathrm{d}t}}$
and 
$\sigma^\mathrm{fid}\left(\pT\right) = \frac{N\left(p_\mathrm{T,part}\right)}{\int{ L \mathrm{d}t}}$,
where ${\int{ L \mathrm{d}t}}$ is the integrated luminosity, $N_\mathrm{part}$ represents the fully-corrected distributions for the number 
of particle jets, $N(p_\mathrm{T,part})$ is the fully-corrected distribution of the number of jets as a function of \pT\ for each \pT-ordered distribution, and $\sigma^\mathrm{fid}\left(n_\mathrm{jet}\right)$ and $\sigma^\mathrm{fid}\left(\pT\right)$ are the differential fiducial cross-sections.

\begin{figure}[htbp]
\centering
\subfigure[\label{fig:njet-part-lep_25} $\pT > 25$~\GeV]{\includegraphics[width=0.49\textwidth]{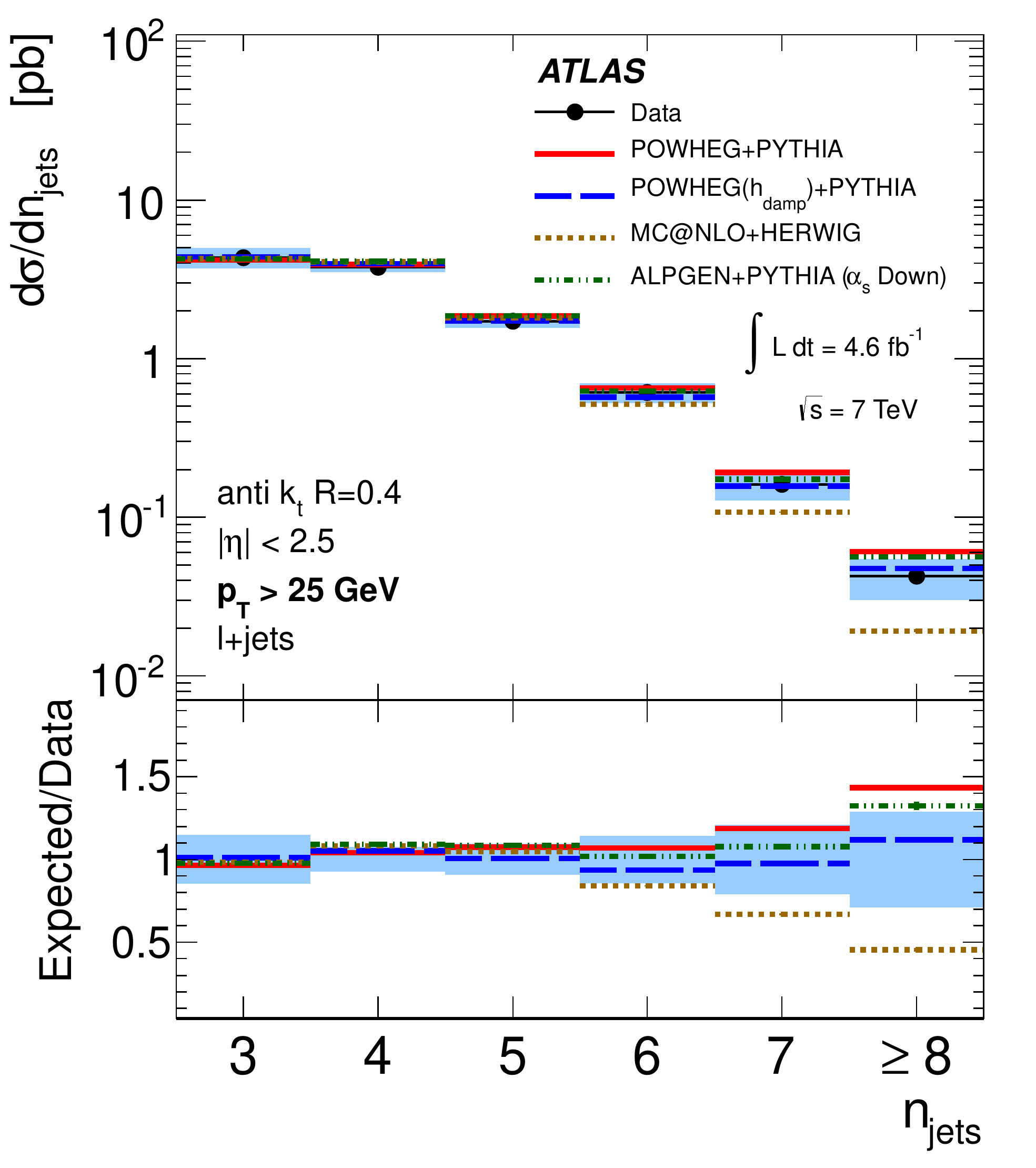}}
\subfigure[\label{fig:njet-part-lep_40} $\pT > 40$~\GeV]{\includegraphics[width=0.49\textwidth]{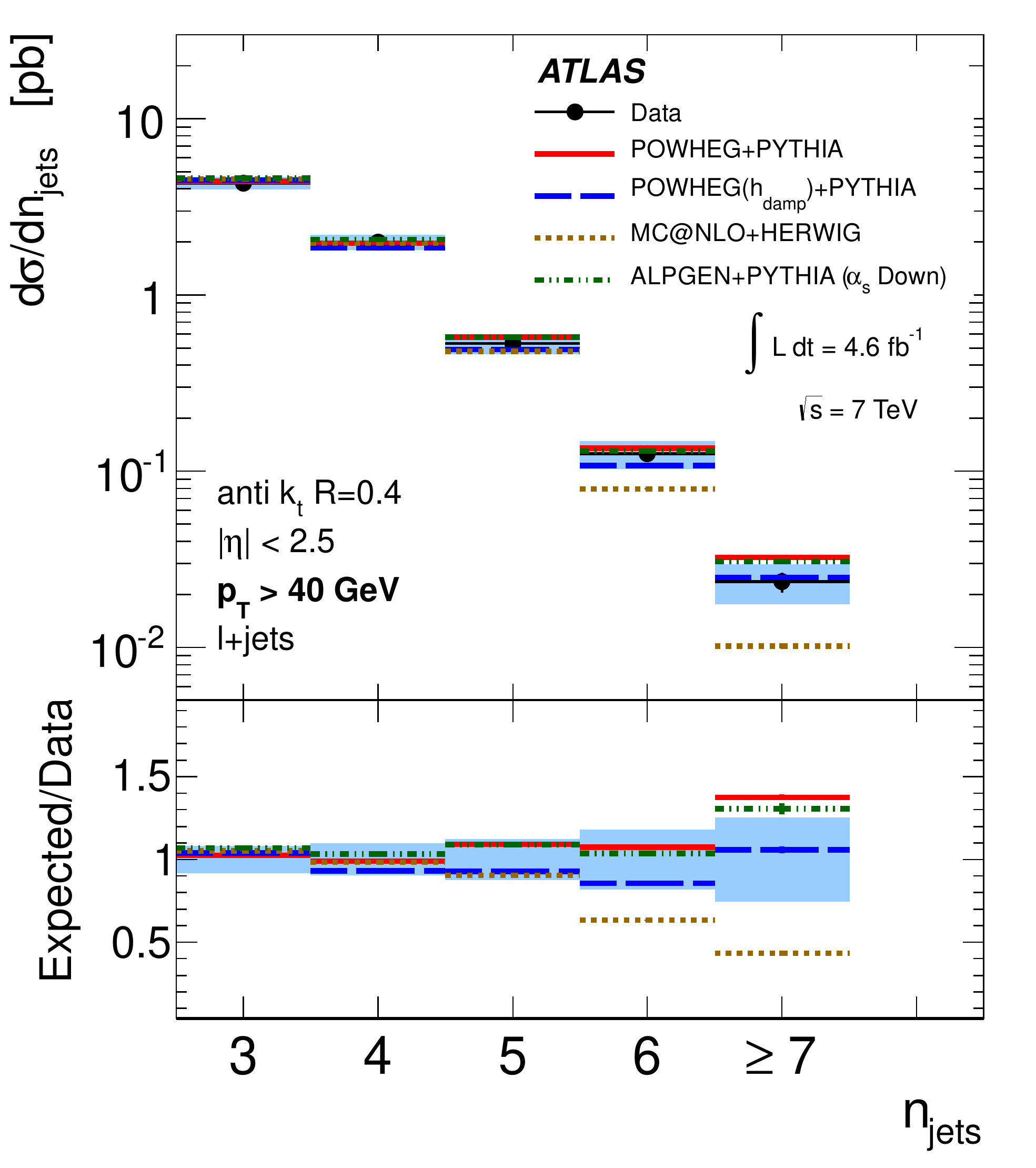}}
\subfigure[\label{fig:njet-part-lep_60} $\pT > 60$~\GeV]{\includegraphics[width=0.49\textwidth]{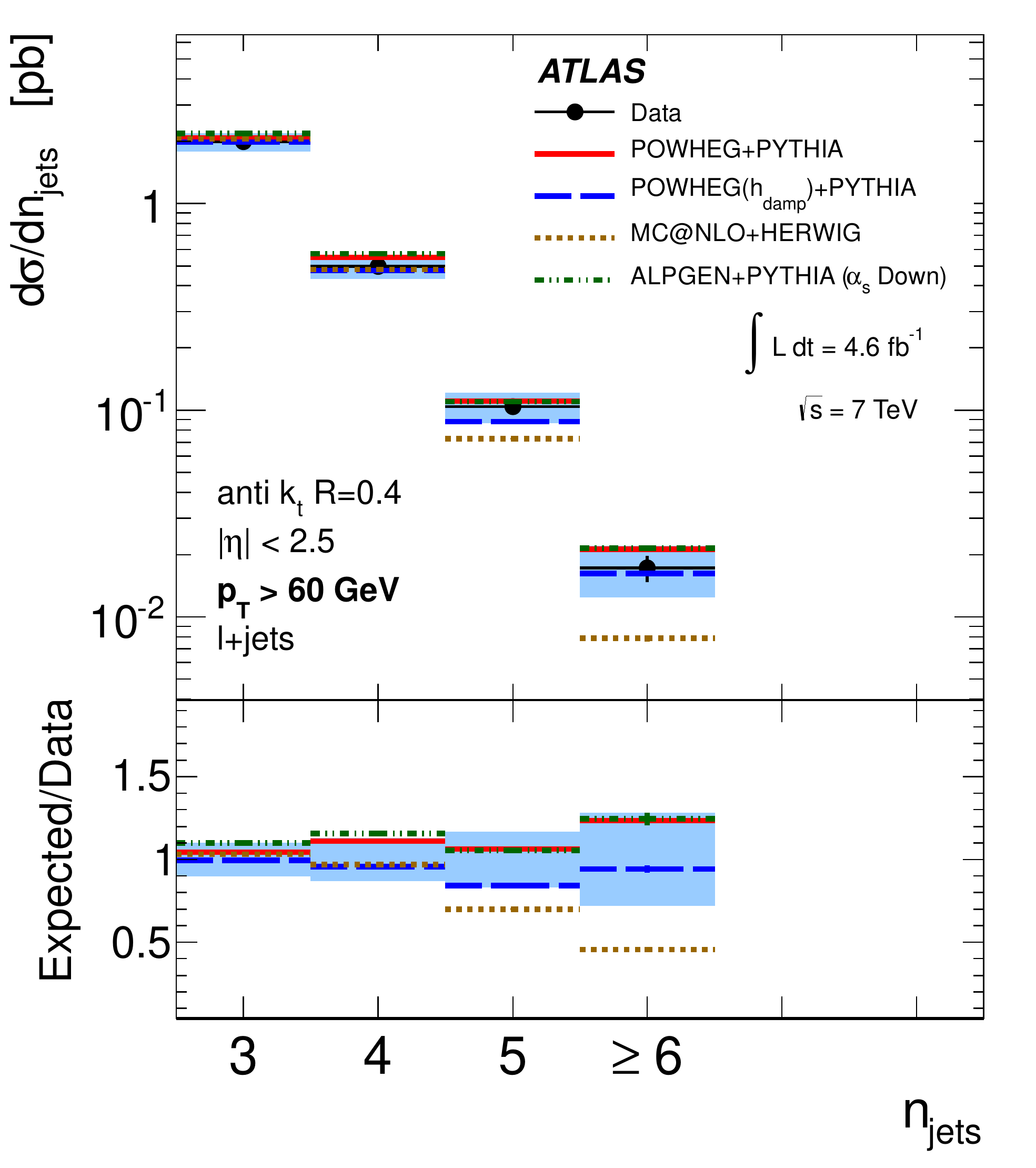}}
\subfigure[\label{fig:njet-part-lep_80} $\pT > 80$~\GeV]{\includegraphics[width=0.49\textwidth]{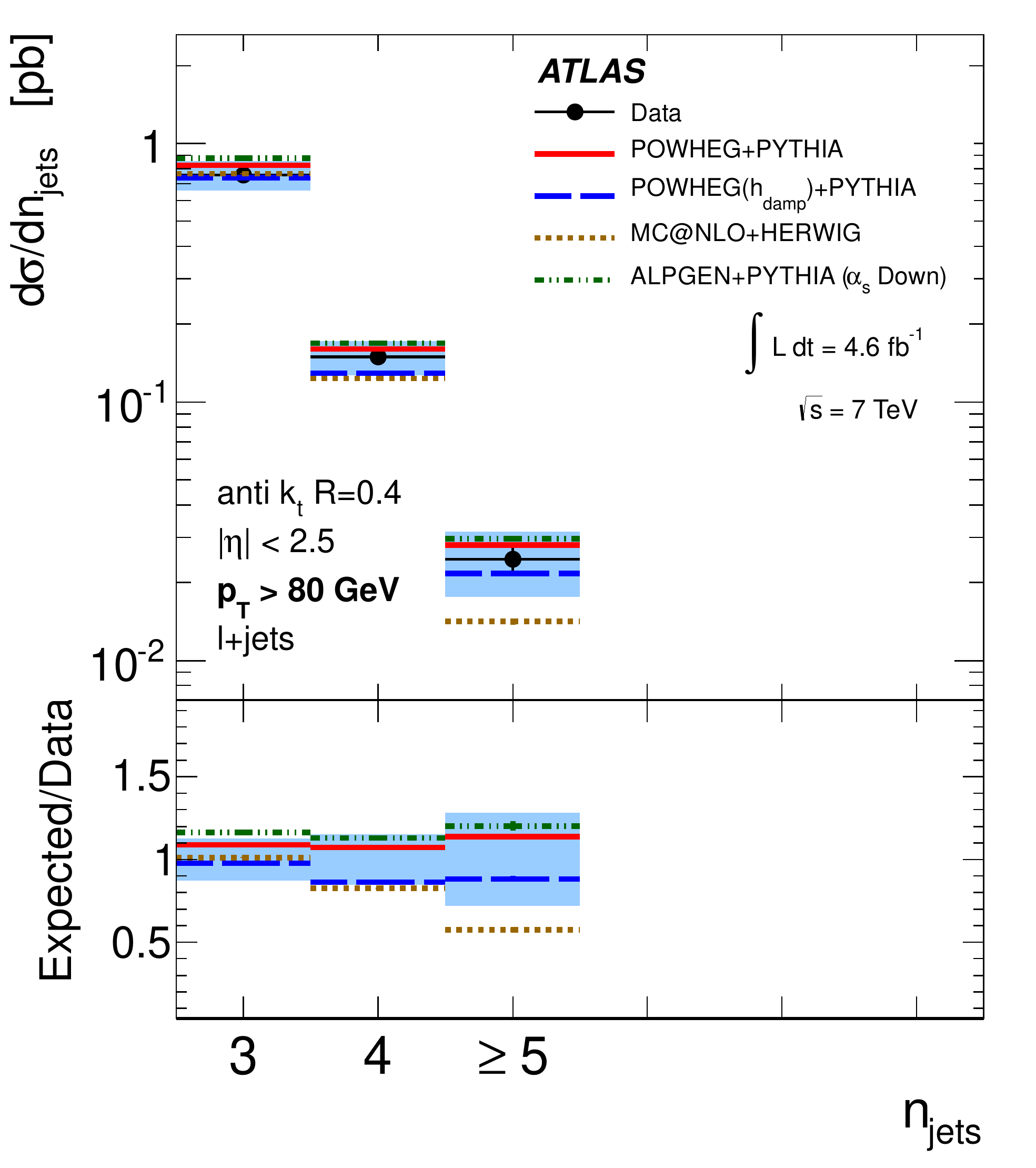}}
\caption{The \ttbar\ cross-section as a function of the jet multiplicity for the average of the electron and muon channels for the jet \pT\ thresholds (a) 25, (b) 40, (c) 60, and (d) 80~\GeV.  The data are shown in comparison to POWHEG+PYTHIA, POWHEG($h_\mathrm{damp}$)+PYTHIA, MC@NLO+HERWIG and  ALPGEN+PYTHIA (\alphas\ down) predictions.  The data points and their corresponding total statistical and systematic uncertainties added  in quadrature is shown as a shaded band.  The MC predictions are shown with their statistical uncertainty.}    
\label{figure:njets-part-lep}
\end{figure}

\begin{figure}[htbp]
\centering
\subfigure[\label{fig:njet-part-lep_25-isr} $\pT > 25$~\GeV]{\includegraphics[width=0.49\textwidth]{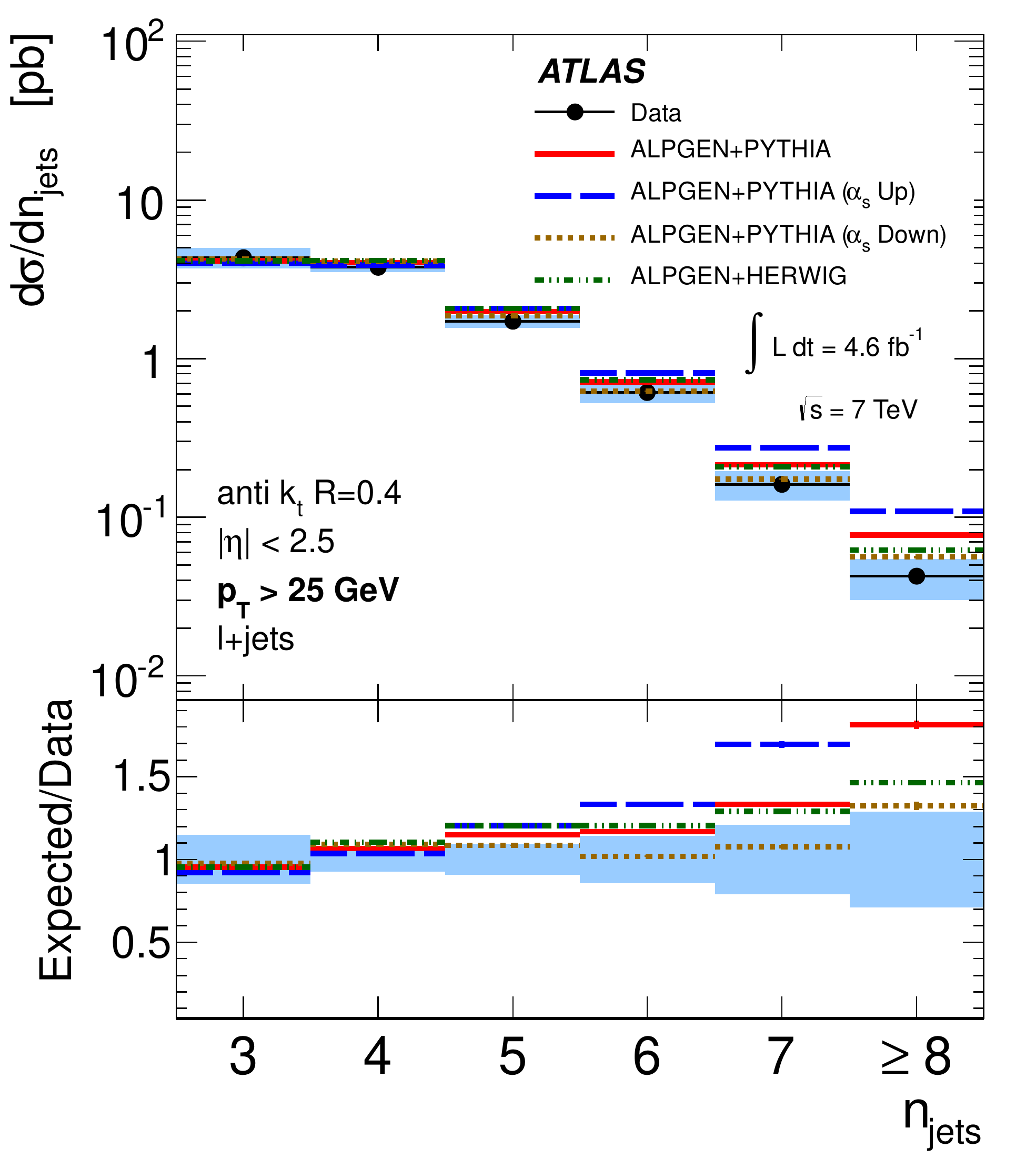}}
\subfigure[\label{fig:njet-part-lep_40-isr} $\pT > 40$~\GeV]{\includegraphics[width=0.49\textwidth]{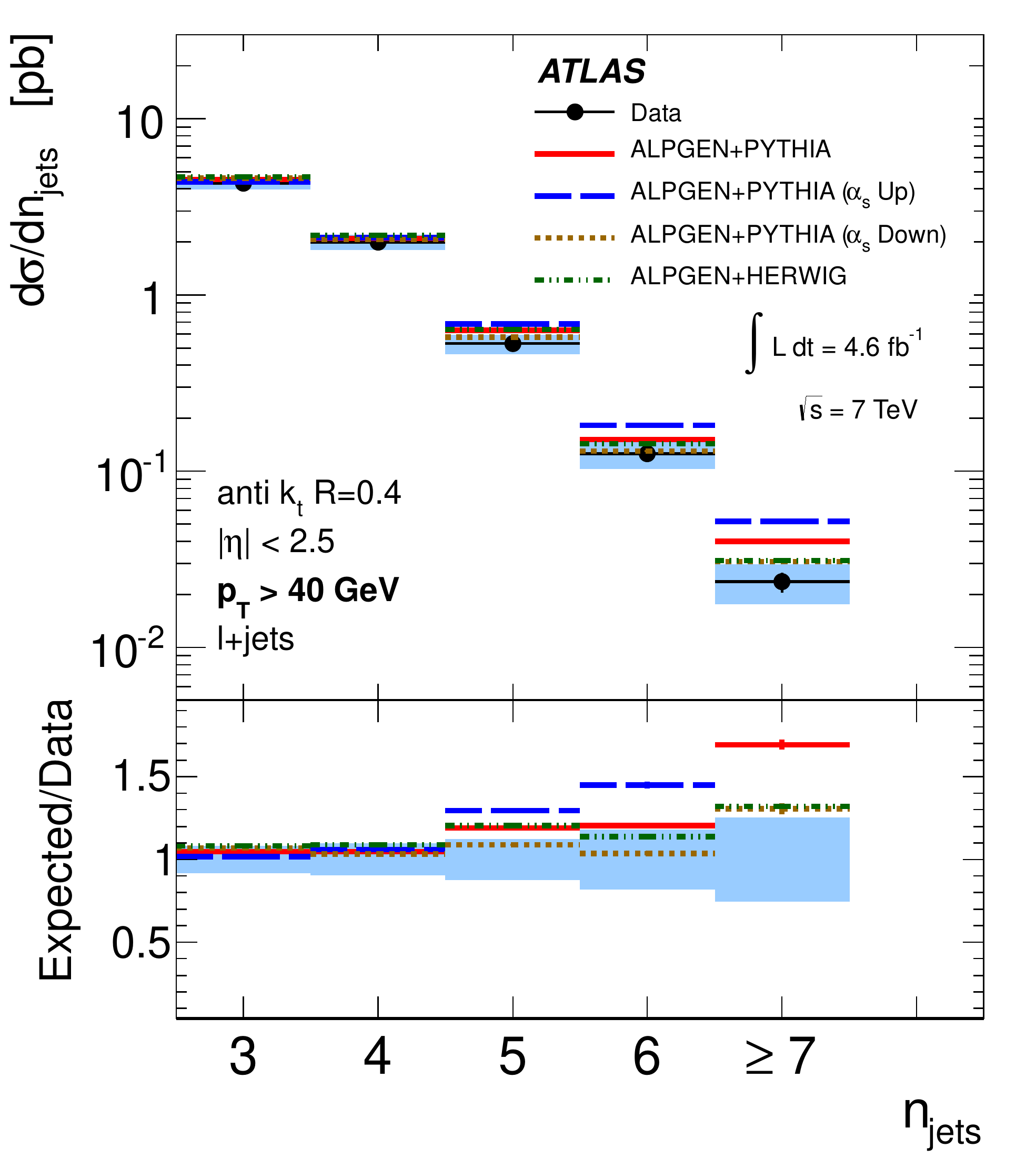}}
\subfigure[\label{fig:njet-part-lep_60-isr} $\pT > 60$~\GeV]{\includegraphics[width=0.49\textwidth]{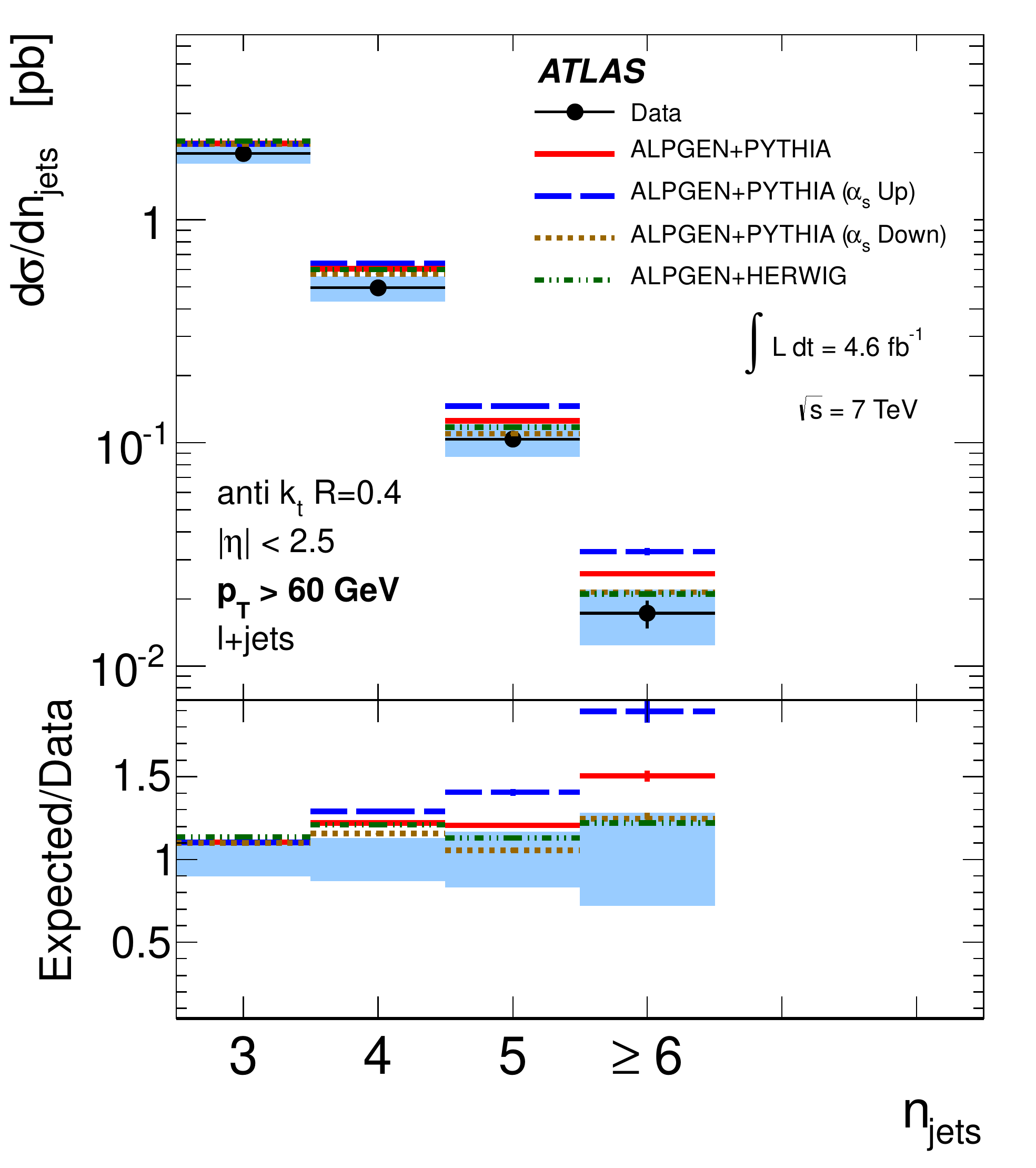}}
\subfigure[\label{fig:njet-part-lep_80-isr} $\pT > 80$~\GeV]{\includegraphics[width=0.49\textwidth]{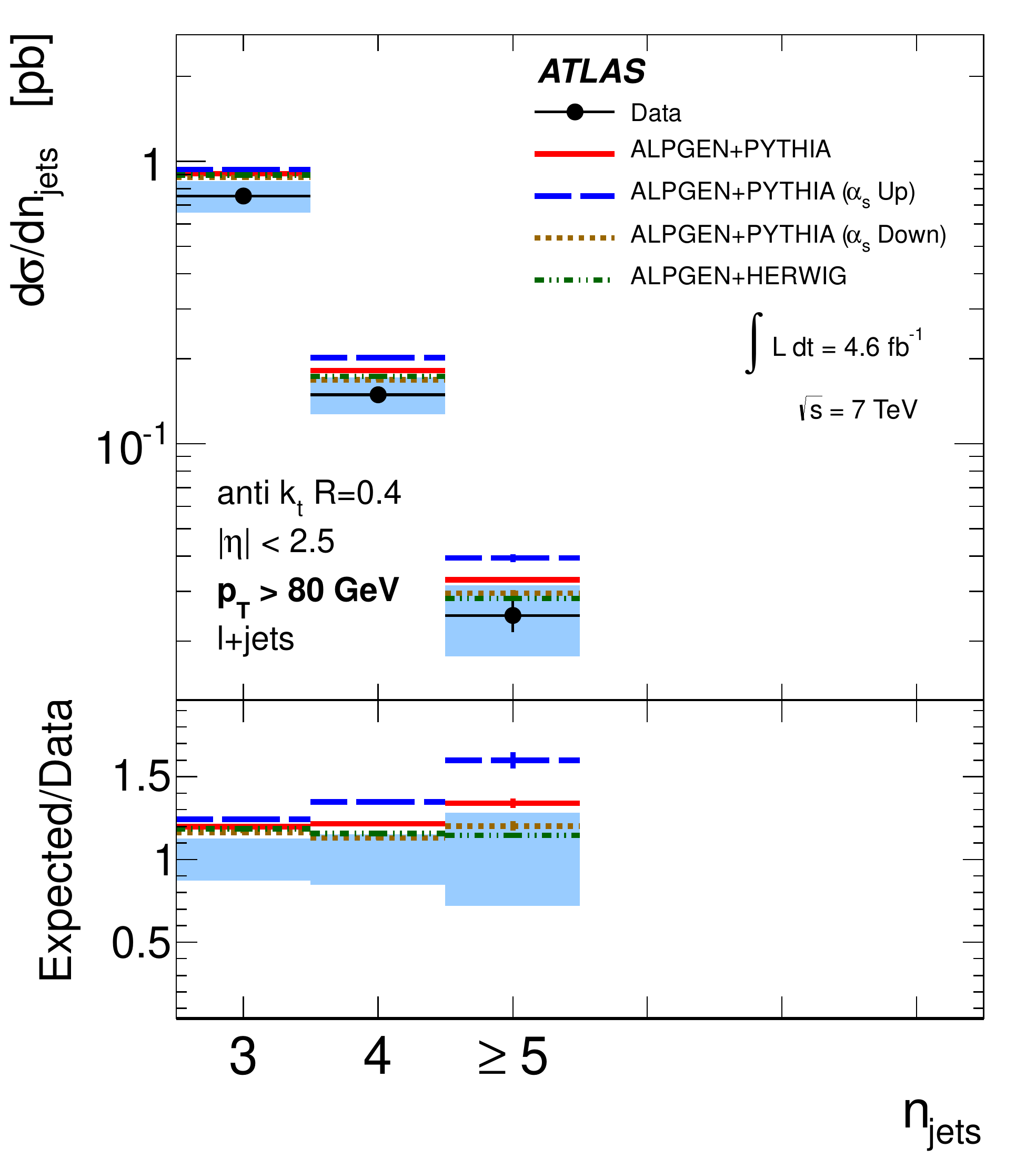}}
\caption{The \ttbar\ cross-section as a function of the jet multiplicity for the average of the electron and muon channels for the jet \pT\ thresholds (a) 25, (b) 40, (c) 60, and (d) 80~\GeV.  The data are shown in comparison to the ALPGEN+PYTHIA,  ALPGEN+PYTHIA ISR/FSR variations and ALPGEN+HERWIG. The data points and their corresponding total statistical and systematic uncertainties added in quadrature is shown as a shaded band.  The MC predictions are shown with their statistical uncertainty.}            
\label{figure:njets-part-lep-isr}
\end{figure}

\begin{figure}[htbp]
\centering
\subfigure[\label{fig:jetpt-part-lep_1} $\pT > 25$~\GeV, leading jet]{\includegraphics[width=0.49\textwidth]{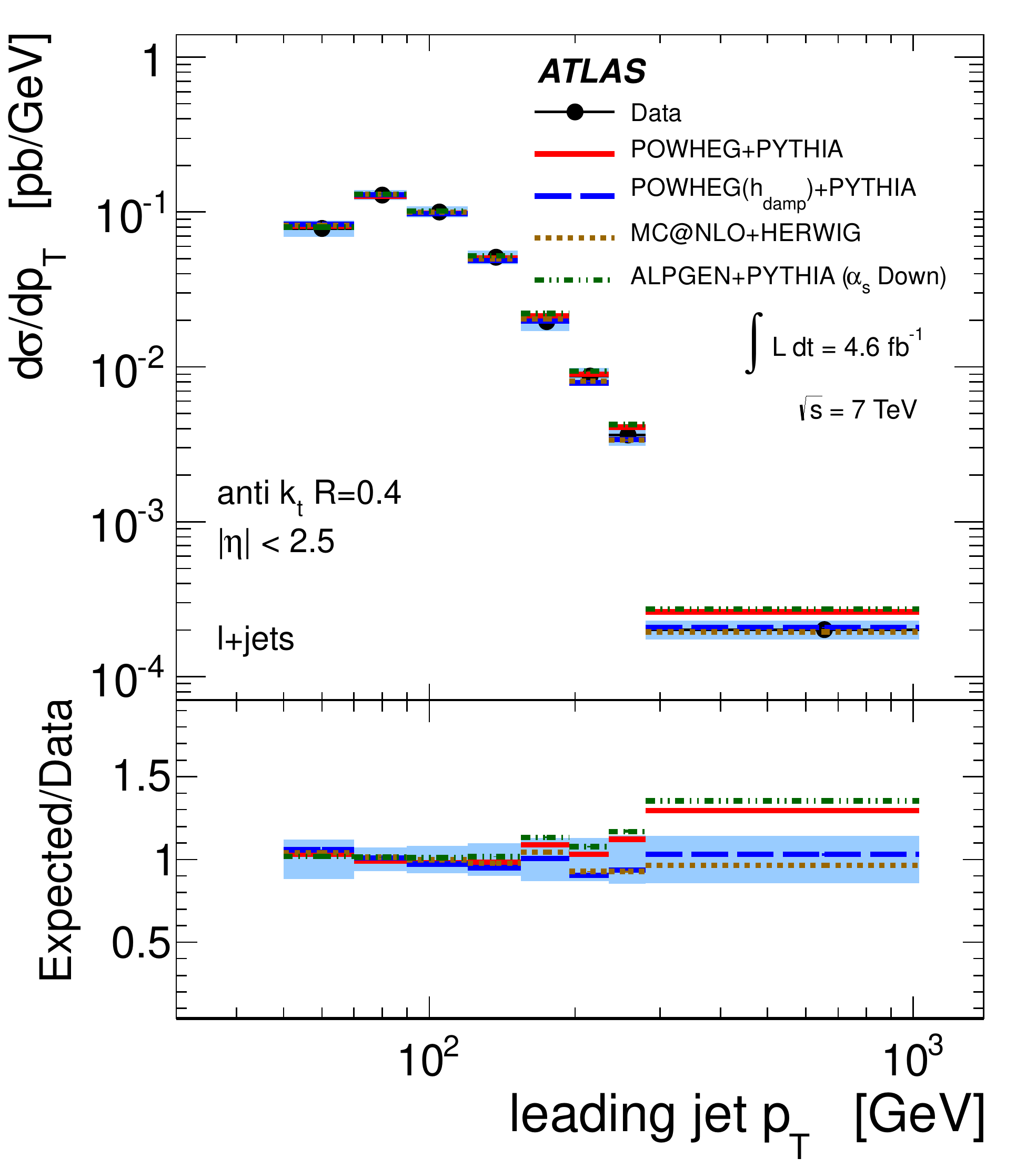}}
\subfigure[\label{fig:jetpt-part-lep_2} $\pT > 25$~\GeV, 2nd jet]{\includegraphics[width=0.49\textwidth]{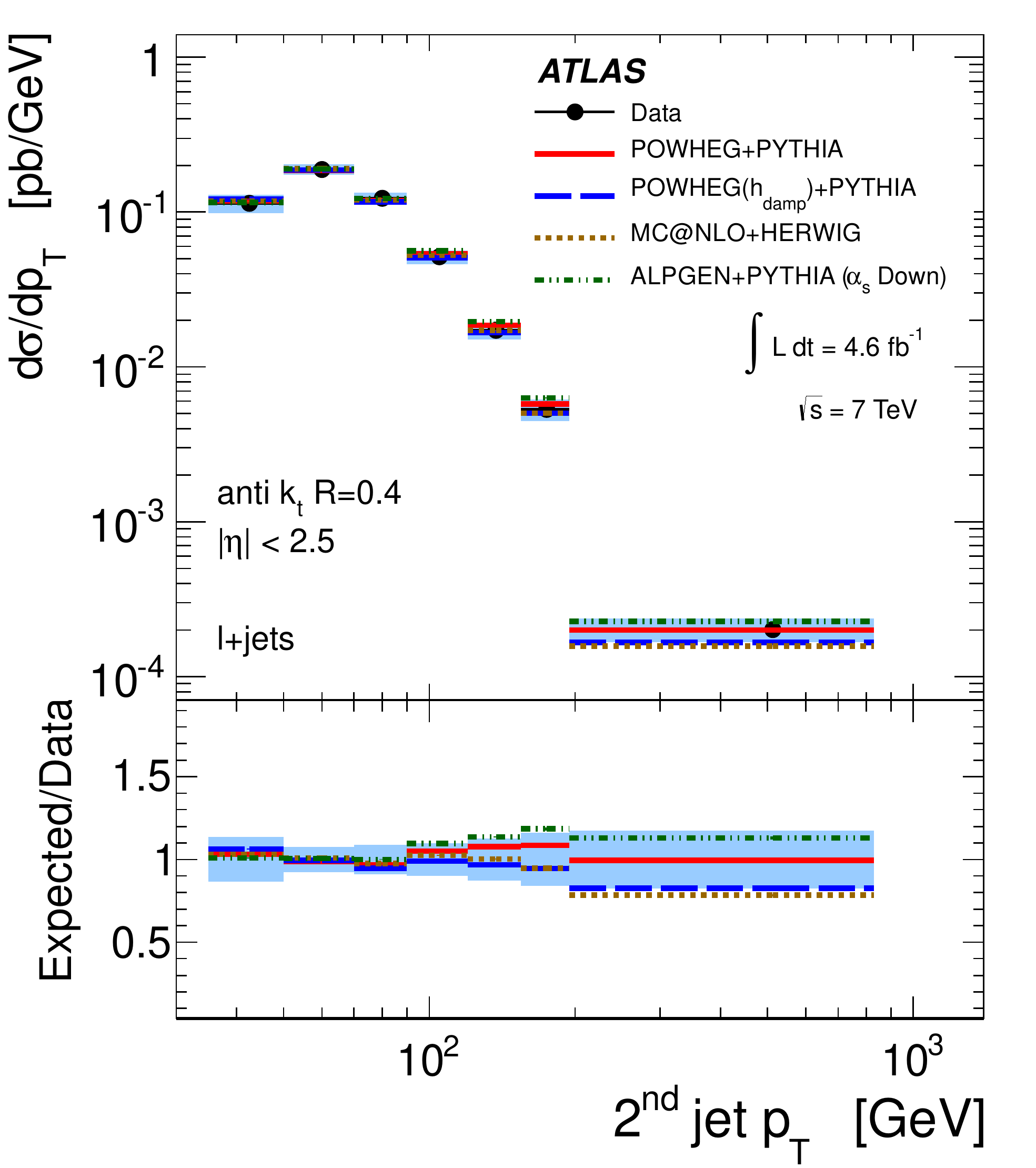}}
\subfigure[\label{fig:jetpt-part-lep_3} $\pT > 25$~\GeV, 3rd jet]{\includegraphics[width=0.49\textwidth]{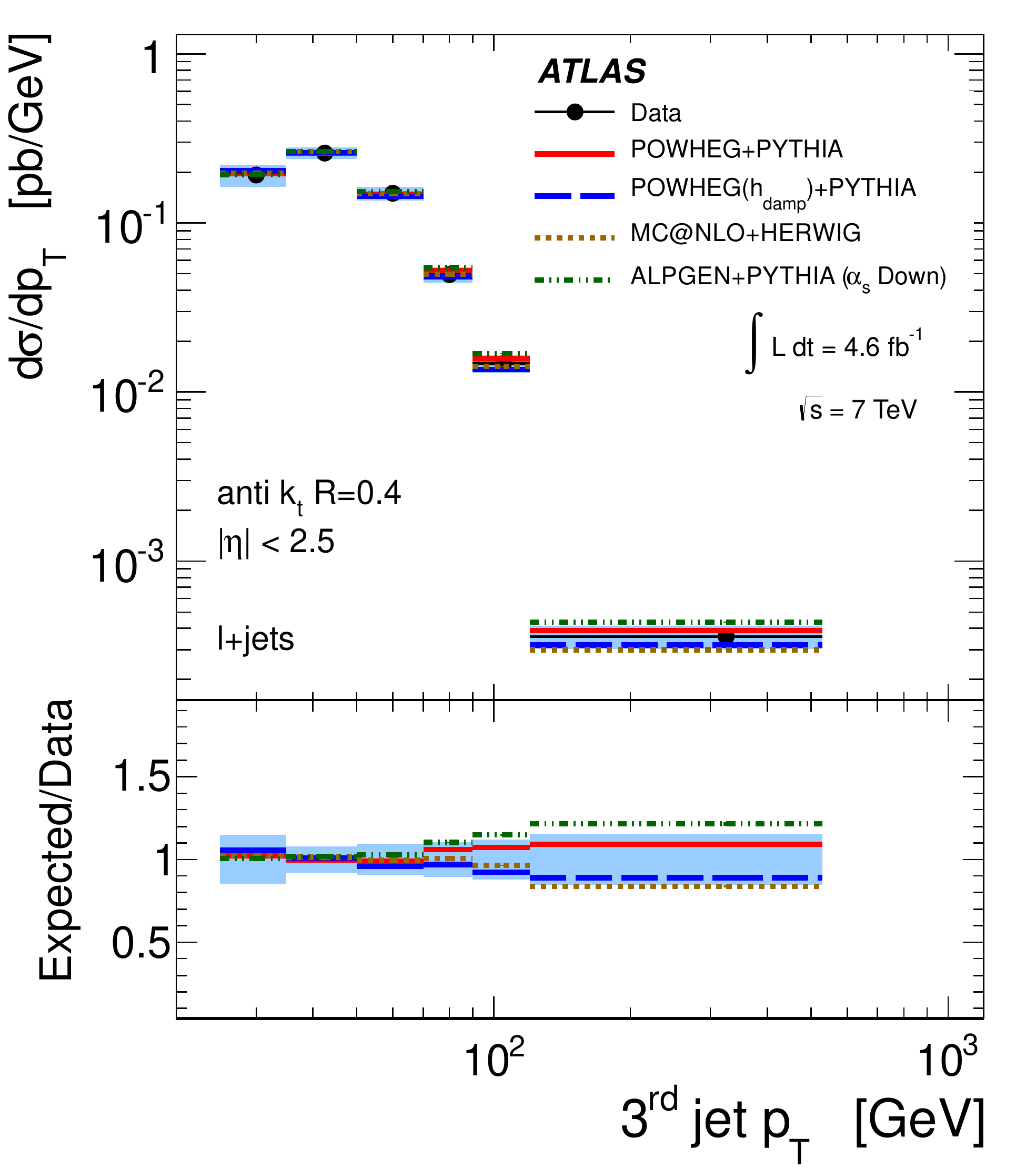}}
\subfigure[\label{fig:jetpt-part-lep_4} $\pT > 25$~\GeV, 4th jet]{\includegraphics[width=0.49\textwidth]{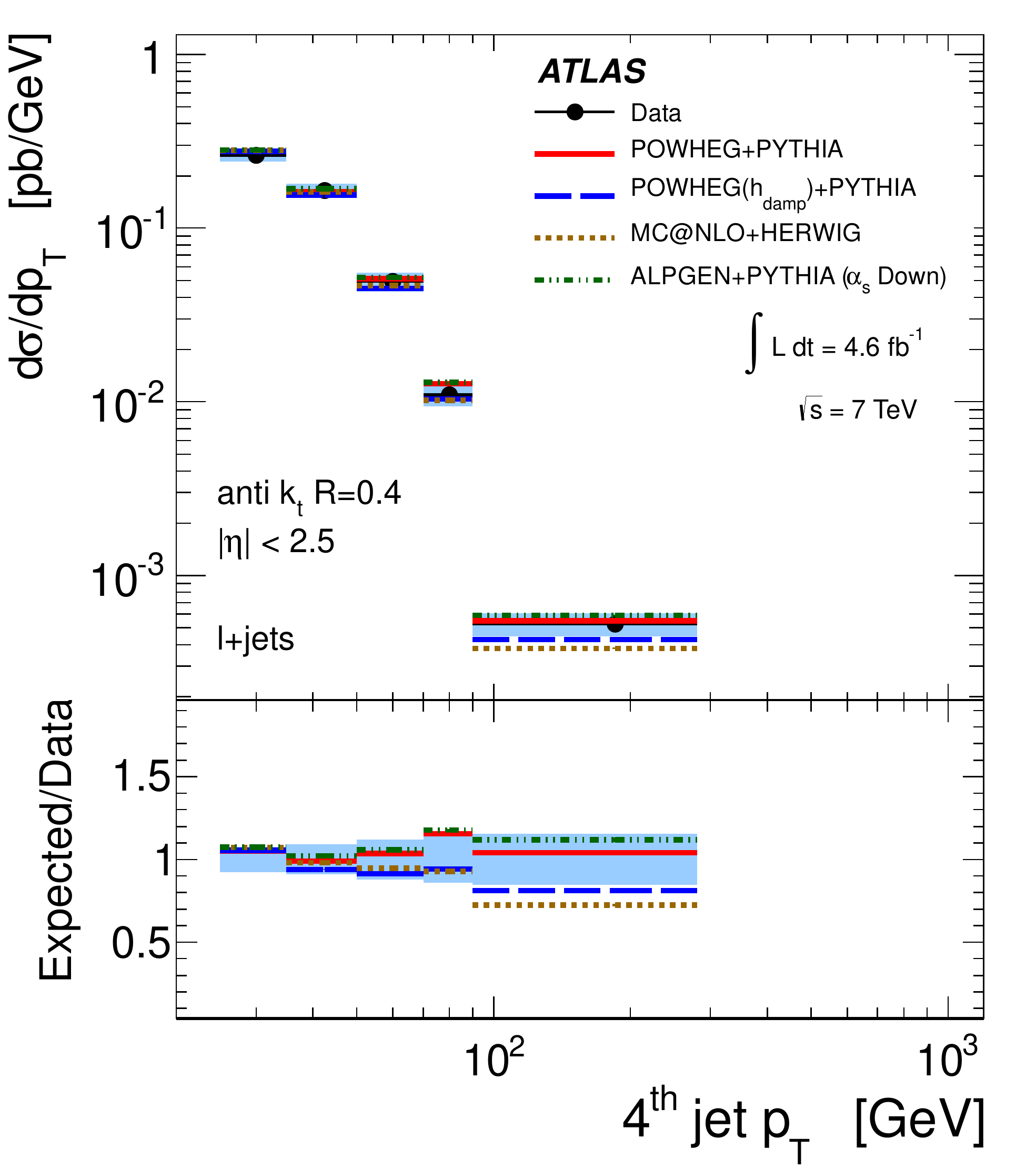}}
\caption{The \ttbar\ cross-section as a function of the jet \pT\ for the average of the electron and muon channels for the (a) leading, (b) 2nd, (c) 3rd, and (d) 4th jet.  The data are shown in comparison to POWHEG+PYTHIA, POWHEG($h_\mathrm{damp})$+PYTHIA, MC@NLO+HERWIG and  ALPGEN+PYTHIA (\alphas\ down)  predictions.   The data points and their corresponding total statistical and systematic uncertainties added in quadrature is shown as a shaded band.  The MC predictions are shown with their statistical uncertainty.}    
\label{figure:jetpt-part-lep}
\end{figure}

The fully corrected fiducial \ttbar\ production cross-section is shown as a function of jet multiplicity for the jet \pT\ thresholds of 25, 40, 60,
and 80~\GeV\ in figures~\ref{figure:njets-part-lep} and \ref{figure:njets-part-lep-isr} and as a function of the jet \pT\ in figures~\ref{figure:jetpt-part-lep}--\ref{figure:jetpt-part-lep-5}. Tabulated results with systematic uncertainties are given in appendix~\ref{sec:appendixsystematics}.  In these figures, the data are compared to predictions from  {\sc POWHEG+PYTHIA},
 POWHEG($h_\mathrm{damp}$)+PYTHIA with varied amount of hard radiation,  {\sc ALPGEN}+HERWIG and  ALPGEN+PYTHIA with \alphas\ variations, MC@NLO
 +HERWIG and the POWHEG+PYTHIA MC models.

The MC@NLO+HERWIG model is seen to be disfavoured by the jet-multiplicity spectra, since it predicts too few events with six or more $\pT > 25$~\GeV\ jets.  This disagreement is visible for the higher jet \pT\ thresholds for events with five or more jets, although with less significance due to the larger uncertainty in these measurements.
 The ALPGEN+PYTHIA \alphas-down variation is seen to best describe the data.  The ALPGEN+HERWIG curve produces slightly more jets than the observed  jet multiplicity.

The ALPGEN+PYTHIA \alphas-up variation and the central tune are found to be disfavoured by the  jet-multiplicity measurements.  The ALPGEN+PYTHIA \alphas-up variation deviates from data with five or more jets with $\pT > 25$~\GeV\ in the final state,  whereas the ALPGEN+PYTHIA central sample deviates in the case of events with six or more jets with $\pT > 25$~\GeV.  Similar disagreements are seen at higher jet \pT\ thresholds. 
The MC@NLO+HERWIG predictions underestimate the cross-section for  six jets in the \ttbar\ final state.  The underestimate of the higher
jet multiplicity bins for MC@NLO compared to {\sc ALPGEN} is also observed in~\cite{mangano2007}, where the 
difference is explained by a significantly smaller 
contribution of the $t\bar{t}+q(g)$ hard matrix-element calculation  to the multijet final-states and a higher fraction of 
 additional jets from the parton shower~\cite{mangano2007}.  

In contrast to MC@NLO, the prediction from {\sc POWHEG}+{\sc PYTHIA} is in reasonable agreement with the data for
all jet \pt\ thresholds and jet multiplicities.  {\sc POWHEG}($h_\mathrm{damp}$)
+{\sc PYTHIA} provides the best description of the leading-jet \pT\ and 
the higher jet multiplicities.
However, due to the damping of the hardest emissions, {\sc POWHEG}($h_\mathrm{damp}$)+{\sc PYTHIA} predicts a softer $5^\mathrm{th}$ jet \pT\ spectrum and a correspondingly slightly lower jet multiplicity spectrum for the  80~\GeV\ threshold.

As shown in figures~\ref{figure:jetpt-part-lep} to~\ref{figure:jetpt-part-lep-5}, all models predict a similar cross-section as a function of  jet \pT\ below approximately 100~GeV for the four leading jets. 
However, the ISR/FSR model variations differ significantly for higher  jet \pT\ and for the full \pT\ spectrum of the 5$^\mathrm{th}$ leading jet.
The conclusions drawn from the 5$^\mathrm{th}$ jet comparisons of data versus predictions are similar to the ones from the jet multiplicity measurements: the MC@NLO+HERWIG MC program generates a \pT\ spectrum that is softer than the observed data. The detailed study of POWHEG+PYTHIA in~\cite{Corke:2010zj} shows that the probability of the emission at high \pt\ 
largely depends on the modelling of the ISR evolution and its upper limit of the virtuality on the ISR parton.  The setting used in this analysis
 yields slightly higher predictions than the observed data, which could 
potentially be improved by tuning the free parameters of the ISR model. 
The ALPGEN+PYTHIA \alphas\ variations demonstrate the sensitivity of the predictions to the value of \alphas\ used in the calculation of the hard matrix element and the parton shower.   All ISR variations are higher than the data, where \alphas-down provides the best description.

\newpage

\begin{figure}[htbp]
\centering
\subfigure[\label{fig:jetpt-part-lep_1-isr} $\pT > 25$~\GeV, leading jet]{\includegraphics[width=0.49\textwidth]{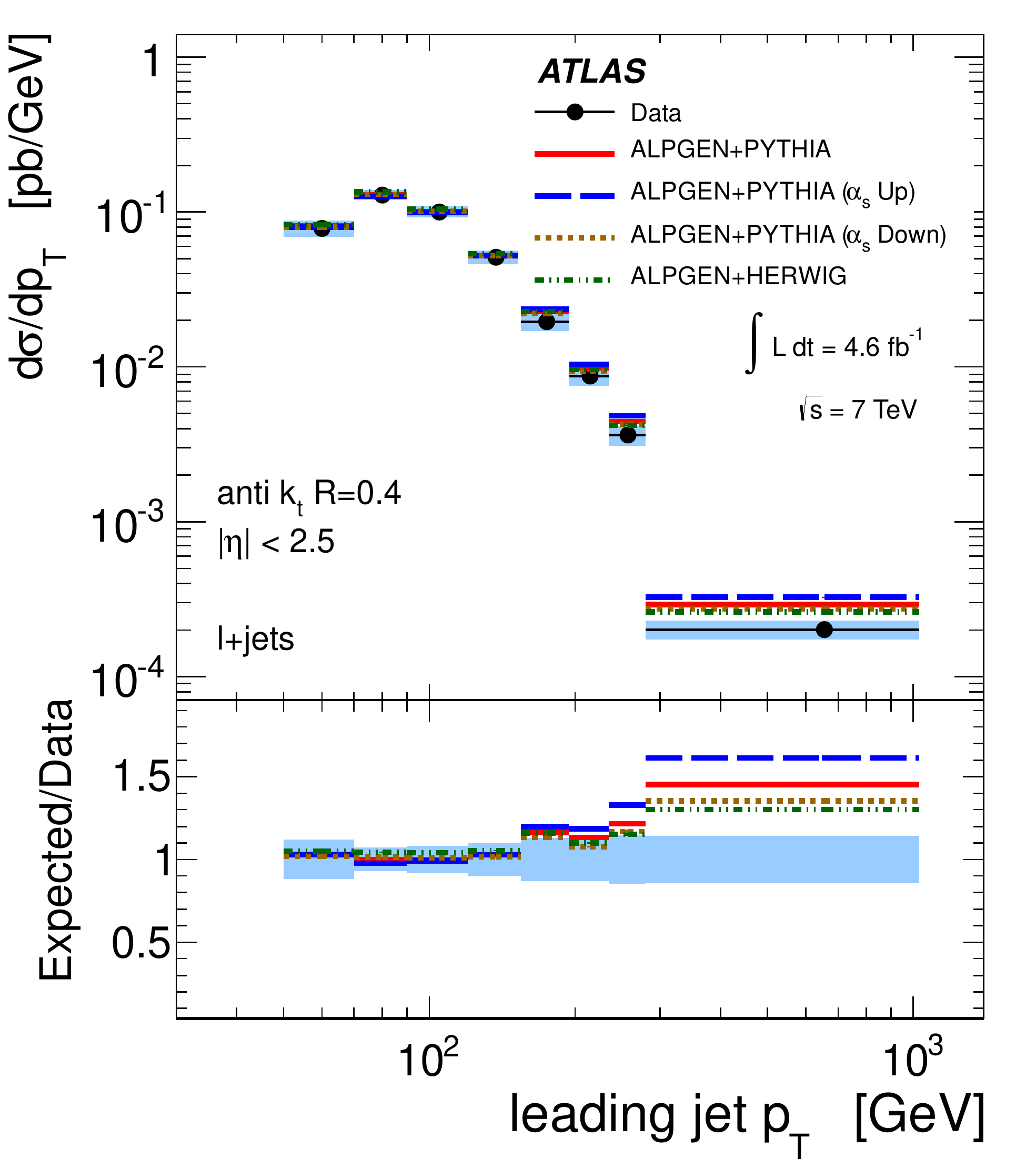}}
\subfigure[\label{fig:jetpt-part-lep_2-isr} $\pT > 25$~\GeV, 2nd jet]{\includegraphics[width=0.49\textwidth]{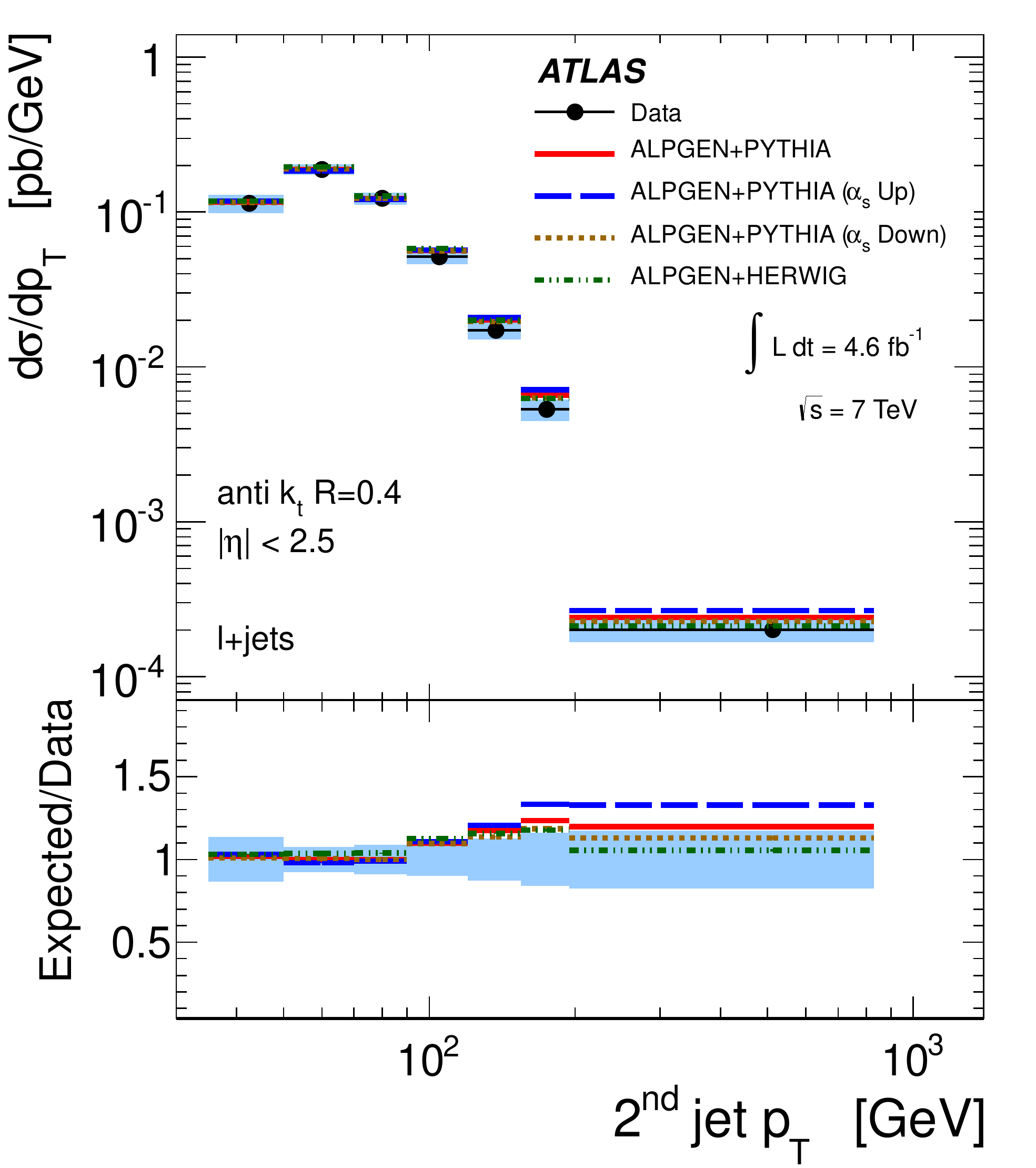}}
\subfigure[\label{fig:jetpt-part-lep_3-isr} $\pT > 25$~\GeV, 3rd jet]{\includegraphics[width=0.49\textwidth]{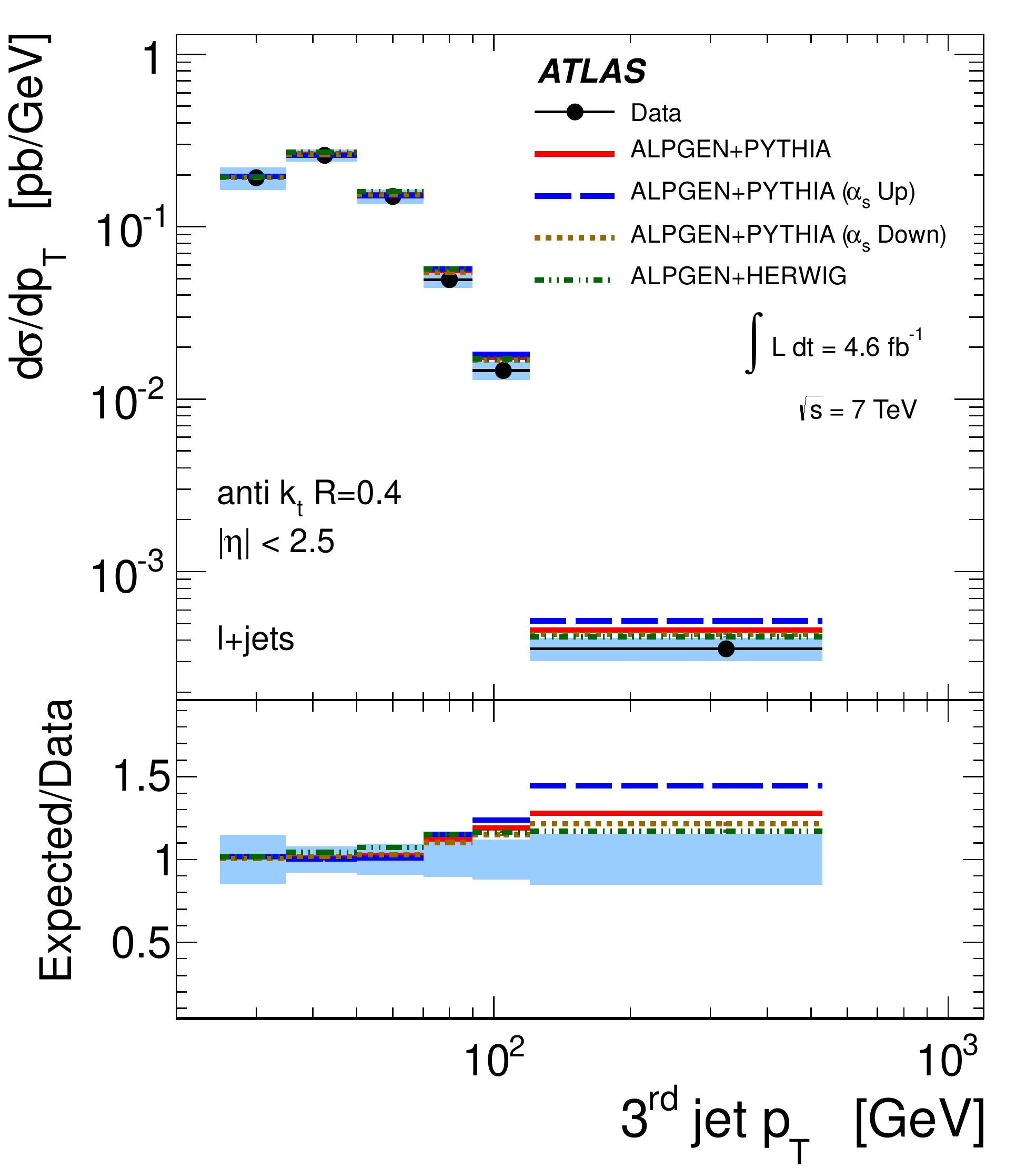}}
\subfigure[\label{fig:jetpt-part-lep_4-isr} $\pT > 25$~\GeV, 4th jet]{\includegraphics[width=0.49\textwidth]{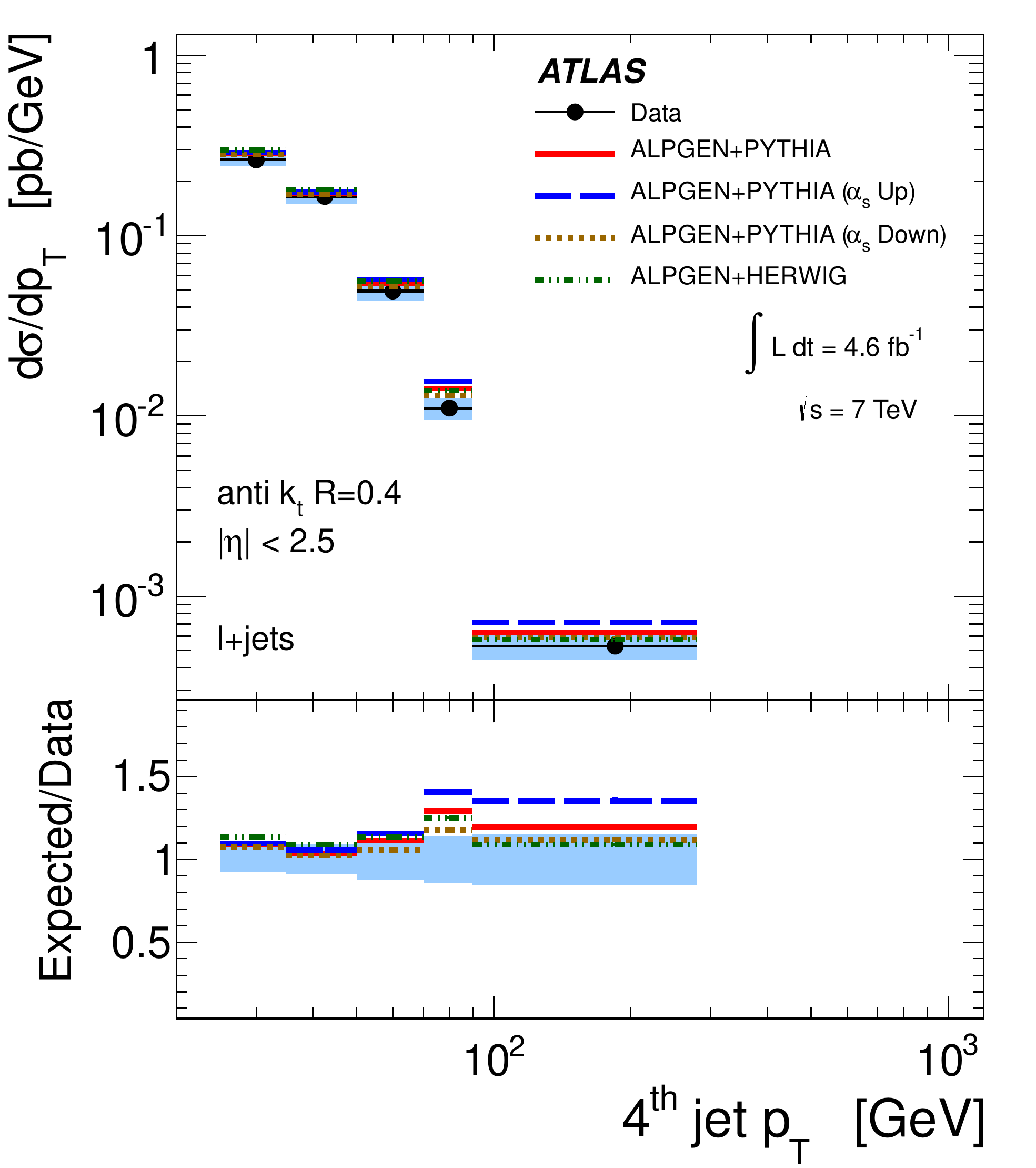}}
\caption{The \ttbar\ cross-section as a function of the jet \pT\ for the average of the electron and muon channels for the (a) leading, (b) 2nd, (c) 3rd, and (d) 4th jet.  The data are shown in comparison to the ALPGEN+PYTHIA, ALPGEN+PYTHIA ISR/FSR variations and ALPGEN+HERWIG. The data points and their corresponding total statistical and systematic uncertainties added in quadrature is shown as a shaded band.  The MC predictions are shown with their statistical uncertainty.}            
\label{figure:jetpt-part-lep-isr}
\end{figure}

\begin{figure}[htbp]
\centering
\subfigure[\label{fig:jetpt-part-lep_5}]{\includegraphics[width=0.49\textwidth]{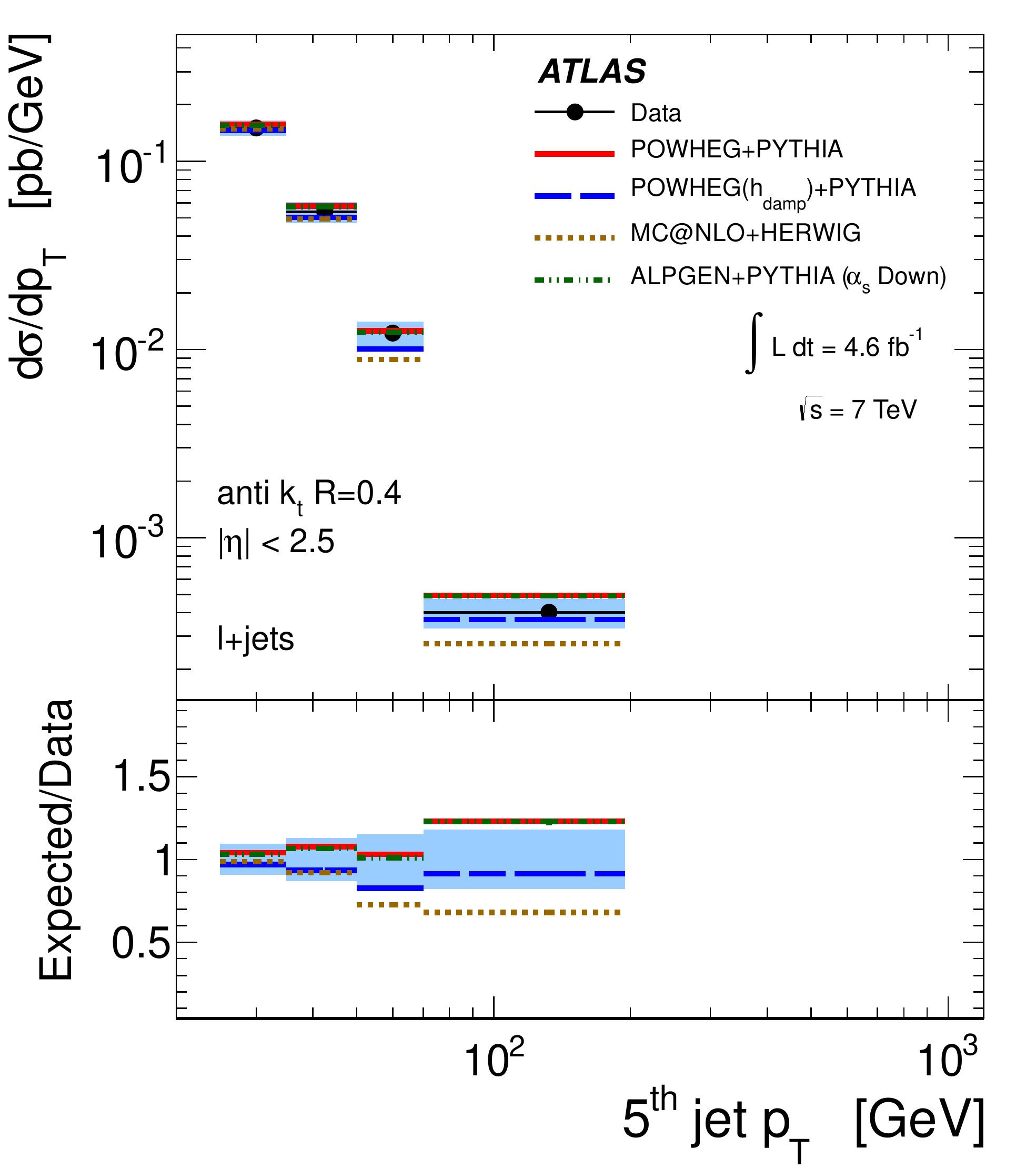}}
\subfigure[\label{fig:jetpt-part-lep_5-isr}]{\includegraphics[width=0.49\textwidth]{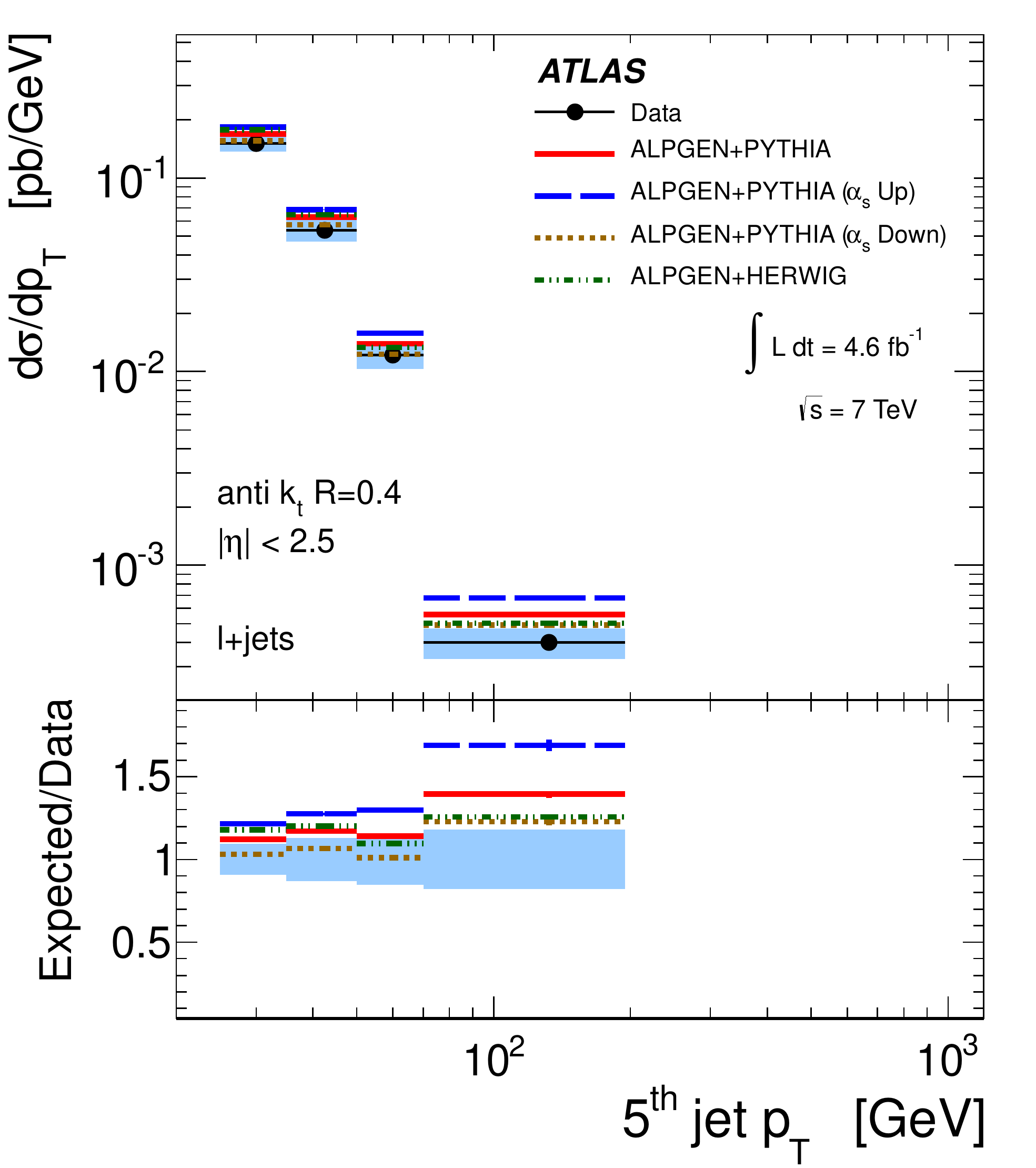}}
\caption{The \ttbar\ cross-section as a function of the jet \pT\ for the average of the electron and muon channels for the 5th jet.  The data are shown in comparison to  (a) POWHEG+PYTHIA, POWHEG($h_\mathrm{damp})$+PYTHIA, MC@NLO+HERWIG and  ALPGEN+PYTHIA (\alphas\ down) predictions and in comparison to (b) the ALPGEN+PYTHIA, ALPGEN+PYTHIA ISR/FSR variations and ALPGEN+HERWIG.  The data points and their corresponding total statistical and systematic uncertainties added in quadrature is shown as a shaded band.  The MC predictions are shown with their statistical uncertainty.}            
\label{figure:jetpt-part-lep-5}
\end{figure}

%% file: conclusions.tex
\section{Conclusions}

The fiducial \ttbar\ production cross-section in $pp$ collisions at 7~\TeV\ is presented as a function of the jet multiplicity for up to eight jets with jet \pT\ thresholds of 25, 40,
60, and 80~\GeV\ using 4.6~\ifb\ of data.  The precision is between approximately 10\% and 30\%, with the largest uncertainty at highest jet multiplicity.
The fiducial \ttbar\ production cross-section is shown as a function of the jet \pT\ 
 separately for each jet up to the fifth jet.  The measured jet \pT\ spectra have 
 a precision between approximately
10\% and 16\%.  The measurement precision is limited in most kinematic regions by systematic uncertainties, from
background modelling (at lower jet multiplicities) to jet energy scale 
(at higher jet multiplicities).

The conclusions drawn from the comparisons
of data versus theory predictions are similar at high 
jet multiplicity, high leading jet \pt\ and in the full spectrum of the $5^\mathrm{th}$ jet.
The presented measurements  have discriminating power for MC model predictions. 
At high jet multiplicities, which are dominated by parton-shower emissions, MC@NLO is disfavoured by the data.  A similar finding applies to the additional jet \pT\ distributions, which are too soft at high \pT.  In contrast, predictions from {\sc POWHEG} showered with {\sc PYTHIA} are
consistent with the data within the total uncertainties of the
measurements. This agreement can be further improved by limiting the hard radiations in {\sc POWHEG} using free model parameters.

The comparison to different \alphas\  settings using  the {\sc  ALPGEN}+{\sc PYTHIA} sample indicates that 
 the data prefer a softer parton-shower, i.e. a smaller value of \alphas.  The prediction of  {\sc  ALPGEN}+{\sc HERWIG} that uses  
a similar \alphas\ in the matrix-element calculation as the lower \alphas\ {\sc  ALPGEN}+{\sc PYTHIA} configuration also yields a
 similar good agreement with the data.  For the lowest jet \pT\ threshold the multiplicity distribution of the lower \alphas\ {\sc  ALPGEN}+{\sc PYTHIA} configuration is closest to the data.  However, at  high leading jet \pt\ the  model predictions  that describe the higher jet multiplicities well are at the upper limit of the uncertainty band or above the data. Only {\sc POWHEG} with reduced hard radiation is able to describe both observables consistently with high accuracy.

The present measurements provide important tests of higher-order QCD effects in \ttbar\ production at the LHC.   They provide 
important inputs for MC developments, in particular for the recent developments of  NLO QCD calculations of $\ttbar\!+\!\textrm{jets}$ matched to parton-shower algorithms as discussed in~\cite{Lonnblad:2012ix}.  An improved understanding in this area is highly important for searches for new particles or new interactions.

%% file: acknowledgements.tex
\acknowledgments

We thank CERN for the very successful operation of the LHC, as well as the
support staff from our institutions without whom ATLAS could not be
operated efficiently.

We acknowledge the support of ANPCyT, Argentina; YerPhI, Armenia; ARC,
Australia; BMWF and FWF, Austria; ANAS, Azerbaijan; SSTC, Belarus; CNPq and FAPESP,
Brazil; NSERC, NRC and CFI, Canada; CERN; CONICYT, Chile; CAS, MOST and NSFC,
China; COLCIENCIAS, Colombia; MSMT CR, MPO CR and VSC CR, Czech Republic;
DNRF, DNSRC and Lundbeck Foundation, Denmark; EPLANET, ERC and NSRF, European Union;
IN2P3-CNRS, CEA-DSM/IRFU, France; GNSF, Georgia; BMBF, DFG, HGF, MPG and AvH
Foundation, Germany; GSRT and NSRF, Greece; ISF, MINERVA, GIF, I-CORE and Benoziyo Center,
Israel; INFN, Italy; MEXT and JSPS, Japan; CNRST, Morocco; FOM and NWO,
Netherlands; BRF and RCN, Norway; MNiSW and NCN, Poland; GRICES and FCT, Portugal; MNE/IFA, Romania; MES of Russia and ROSATOM, Russian Federation; JINR; MSTD,
Serbia; MSSR, Slovakia; ARRS and MIZ\v{S}, Slovenia; DST/NRF, South Africa;
MINECO, Spain; SRC and Wallenberg Foundation, Sweden; SER, SNSF and Cantons of
Bern and Geneva, Switzerland; NSC, Taiwan; TAEK, Turkey; STFC, the Royal
Society and Leverhulme Trust, United Kingdom; DOE and NSF, United States of
America.

The crucial computing support from all WLCG partners is acknowledged
gratefully, in particular from CERN and the ATLAS Tier-1 facilities at
TRIUMF (Canada), NDGF (Denmark, Norway, Sweden), CC-IN2P3 (France),
KIT/GridKA (Germany), INFN-CNAF (Italy), NL-T1 (Netherlands), PIC (Spain),
ASGC (Taiwan), RAL (UK) and BNL (USA) and in the Tier-2 facilities
worldwide.

%% file: appendixRecoLevelResults.tex
\section{Reconstruction-level results}\label{sec:appendixrecoresults}

\begin{figure}[htbp]
\centering
\subfigure[\label{fig:njet-reco-el_40}  \ejets, $\pT > 40$~\GeV]{\includegraphics[width=0.48\textwidth]{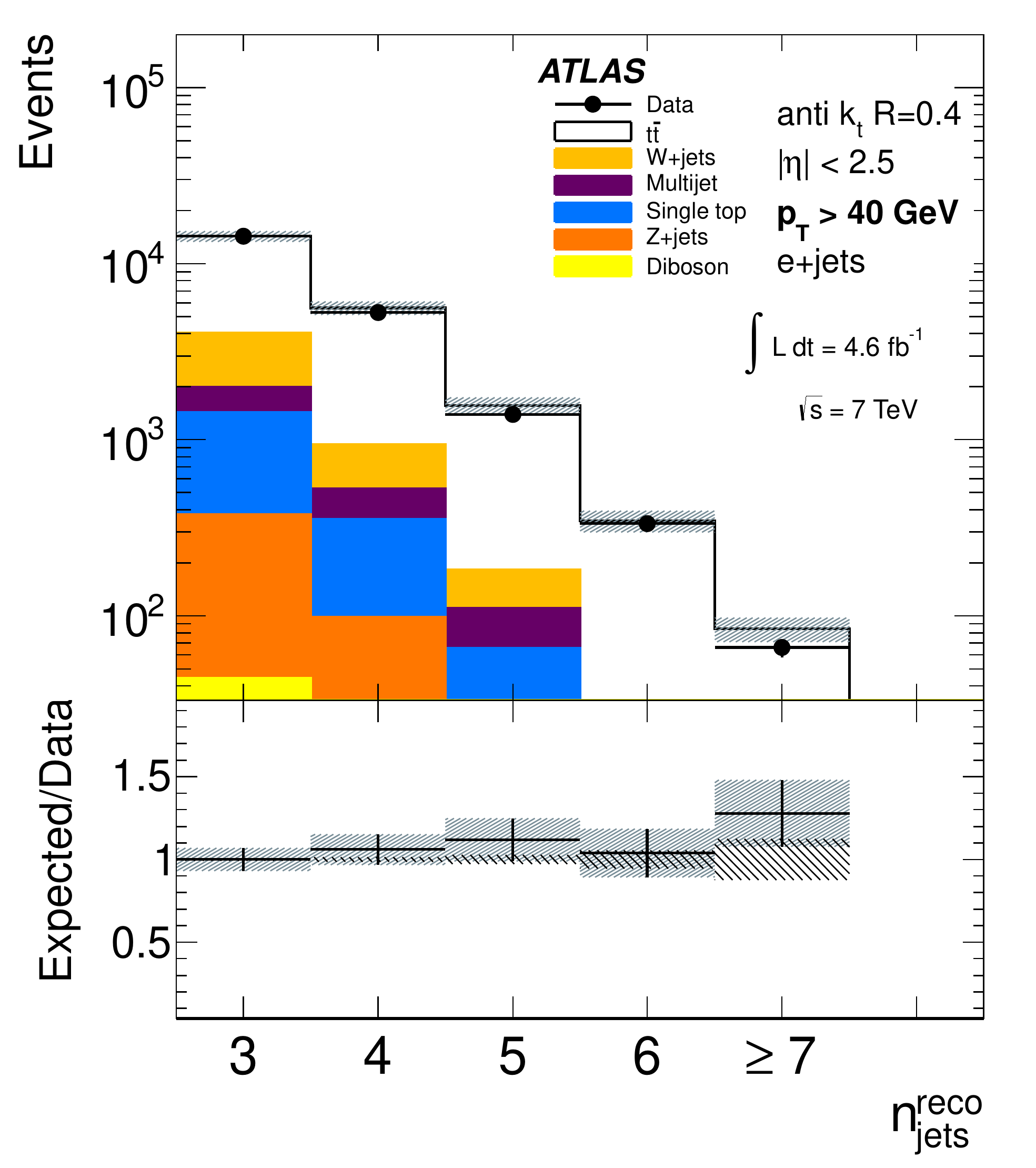}}
\subfigure[\label{fig:njet-reco-el_60}  \ejets, $\pT > 60$~\GeV]{\includegraphics[width=0.48\textwidth]{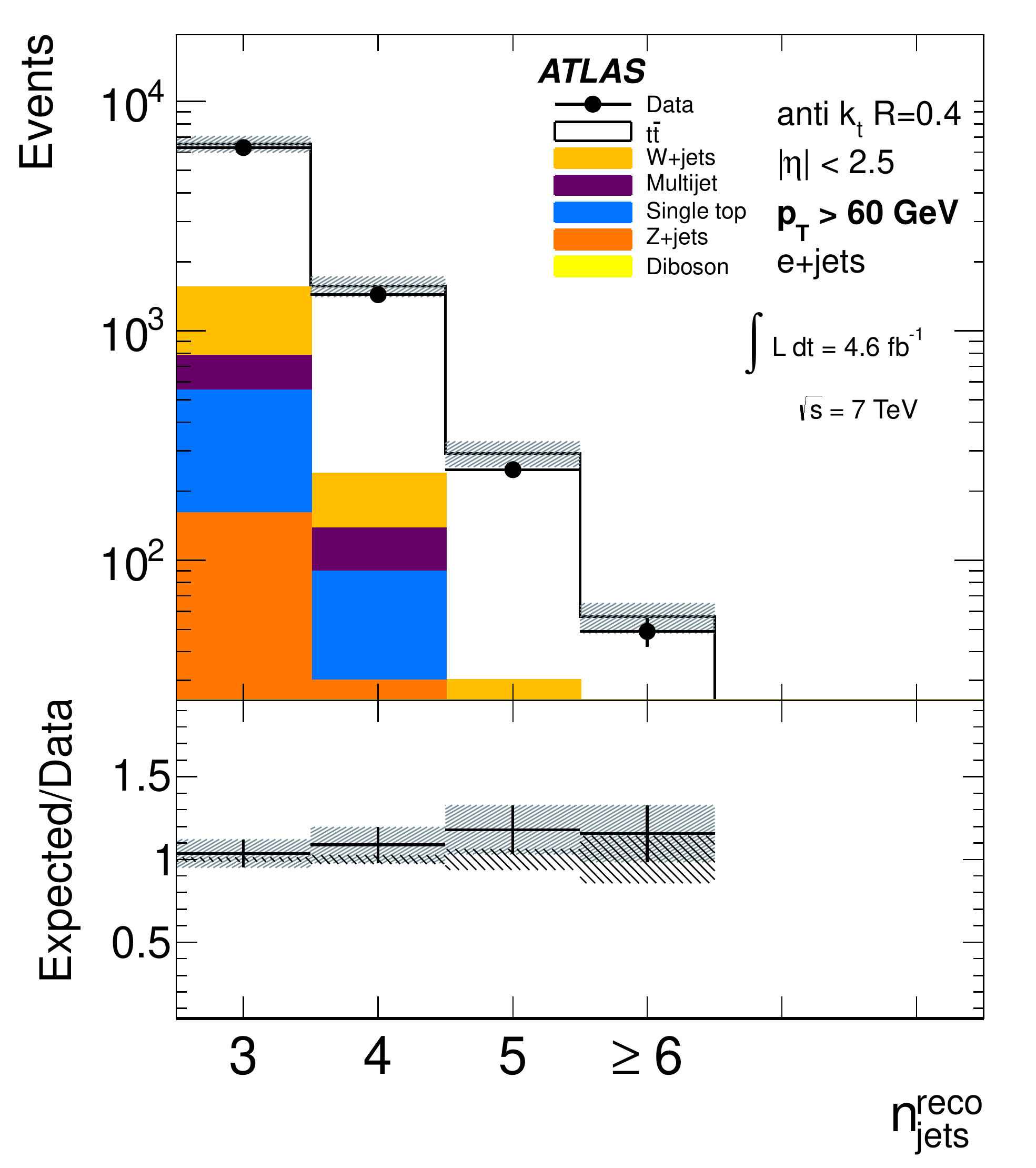}}
\subfigure[\label{fig:njet-reco-el_80}  \ejets, $\pT > 80$~\GeV]{\includegraphics[width=0.48\textwidth]{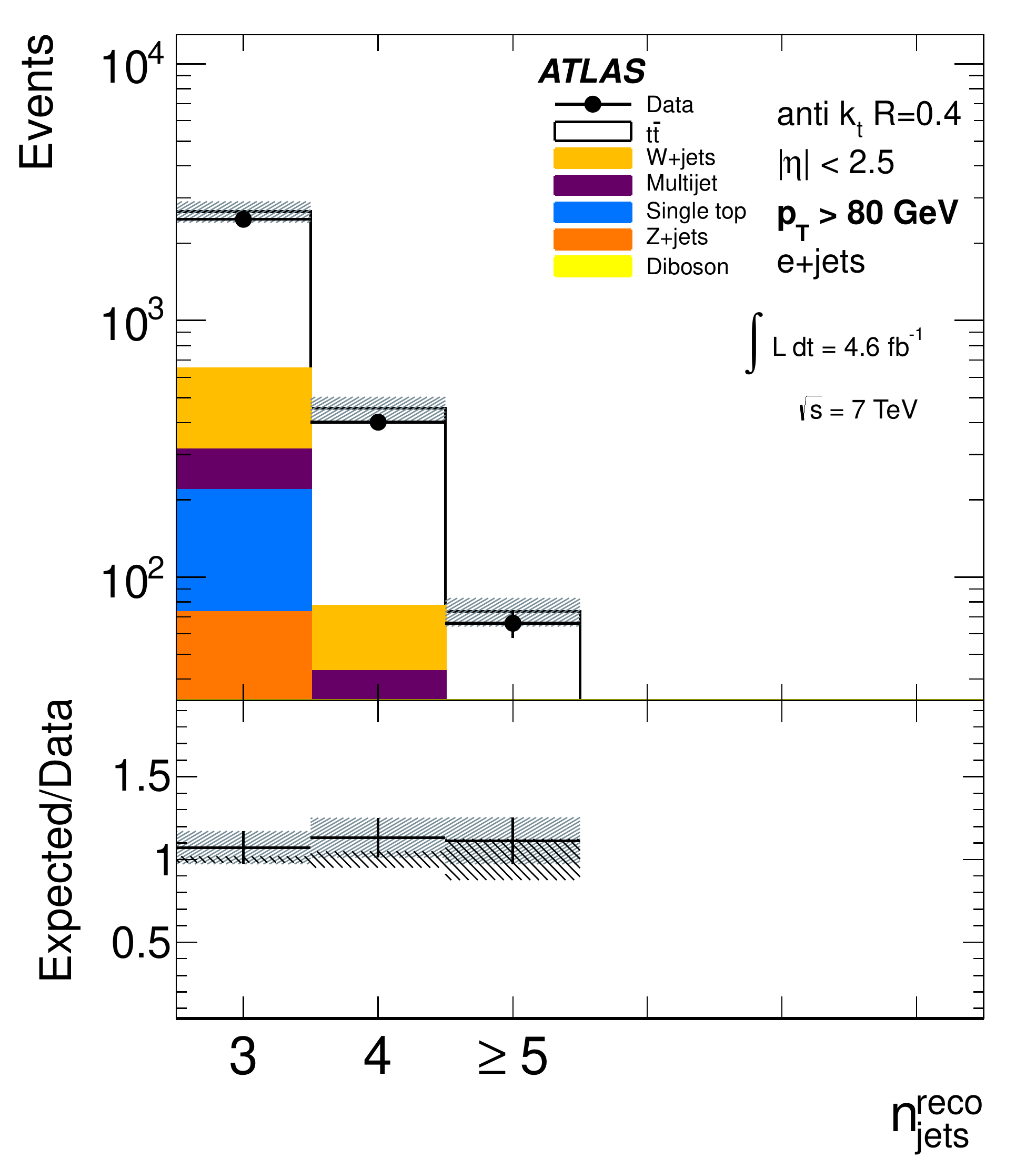}}
\caption{The reconstructed jet multiplicities in the electron (\ejets) channel
  for the jet \pT\ thresholds
 (a) 40, (b) 60, and (c) 80~\GeV.
   The data are compared to the sum of the \ttbar\ {\sc POWHEG}+{\sc PYTHIA} MC signal prediction and  
  the background models.  The shaded bands show the total systematic and statistical
  uncertainties on the combined signal and background estimate. The errors bar on the black points and the hatched area in the ratio, show the statistical uncertainty on the data measurements.}
\label{figure:njets-reco-el}
\end{figure}

\begin{figure}[htbp]
\centering
\subfigure[\label{fig:njet-reco-mu_40} \mujets, $\pT > 40$~\GeV]{\includegraphics[width=0.48\textwidth]{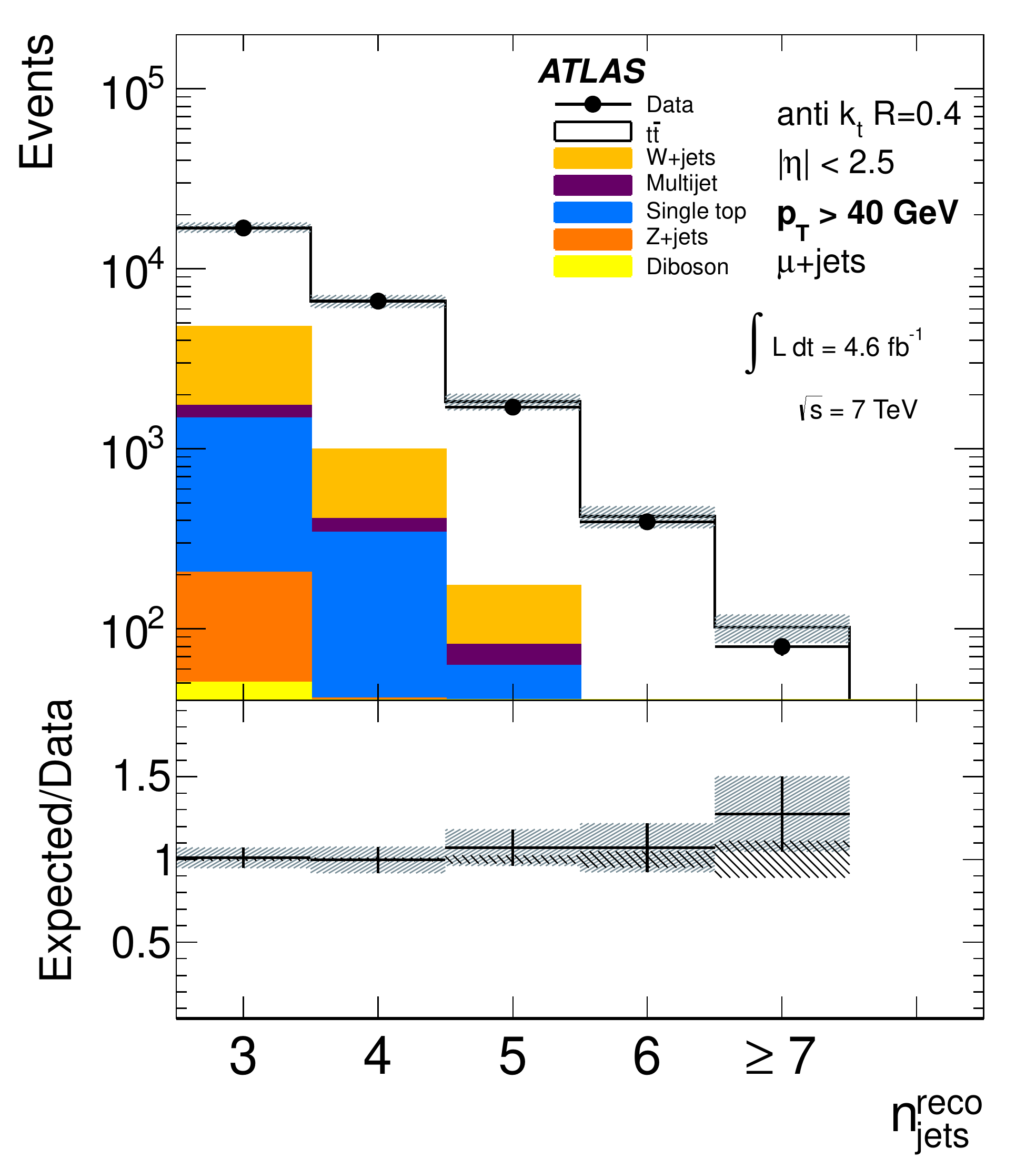}}
\subfigure[\label{fig:njet-reco-mu_60} \mujets, $\pT > 60$~\GeV]{\includegraphics[width=0.48\textwidth]{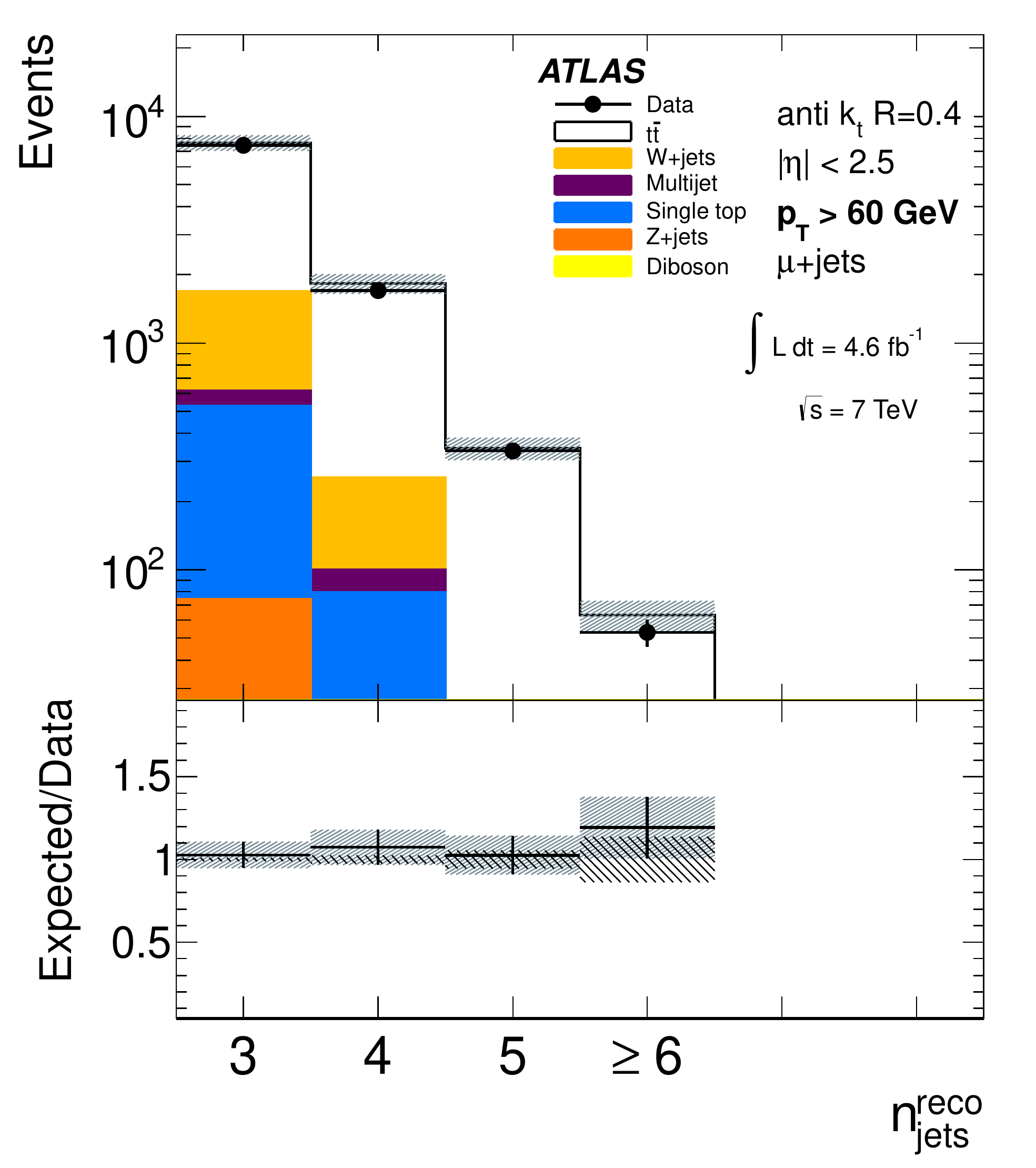}}
\subfigure[\label{fig:njet-reco-mu_80} \mujets, $\pT > 80$~\GeV]{\includegraphics[width=0.48\textwidth]{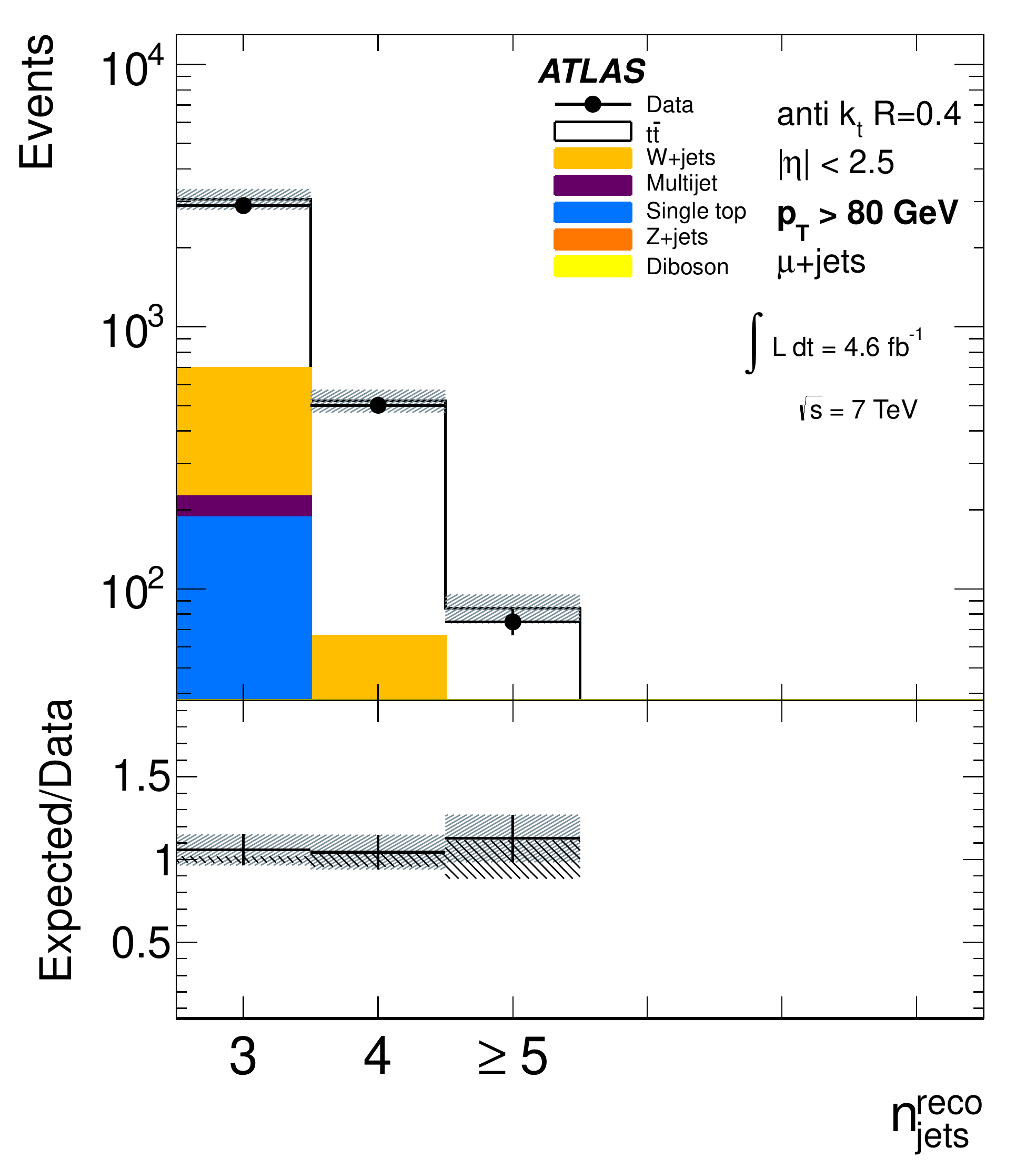}}
\caption{The reconstructed jet multiplicities in the muon ($\mujets$) channel for
  the jet \pT\ thresholds 
(a) 40, (b) 60, and (c) 80 GeV. 
  The data are compared to the sum of the \ttbar\ {\sc POWHEG}+{\sc PYTHIA} MC signal prediction and  
  the background models.  The shaded bands show the total systematic and statistical
  uncertainties on the combined signal and background estimate. The errors bar on the black points and the hatched area in the ratio, show the statistical uncertainty on the data measurements. }
\label{figure:njets-reco-mu}
\end{figure}
\begin{figure}[htbp]
\centering
\subfigure[\label{fig:jetpt-reco-el_2} \ejets, $\pT > 25$~\GeV, 2$^\mathrm{nd}$ jet]{\includegraphics[width=0.48\textwidth]{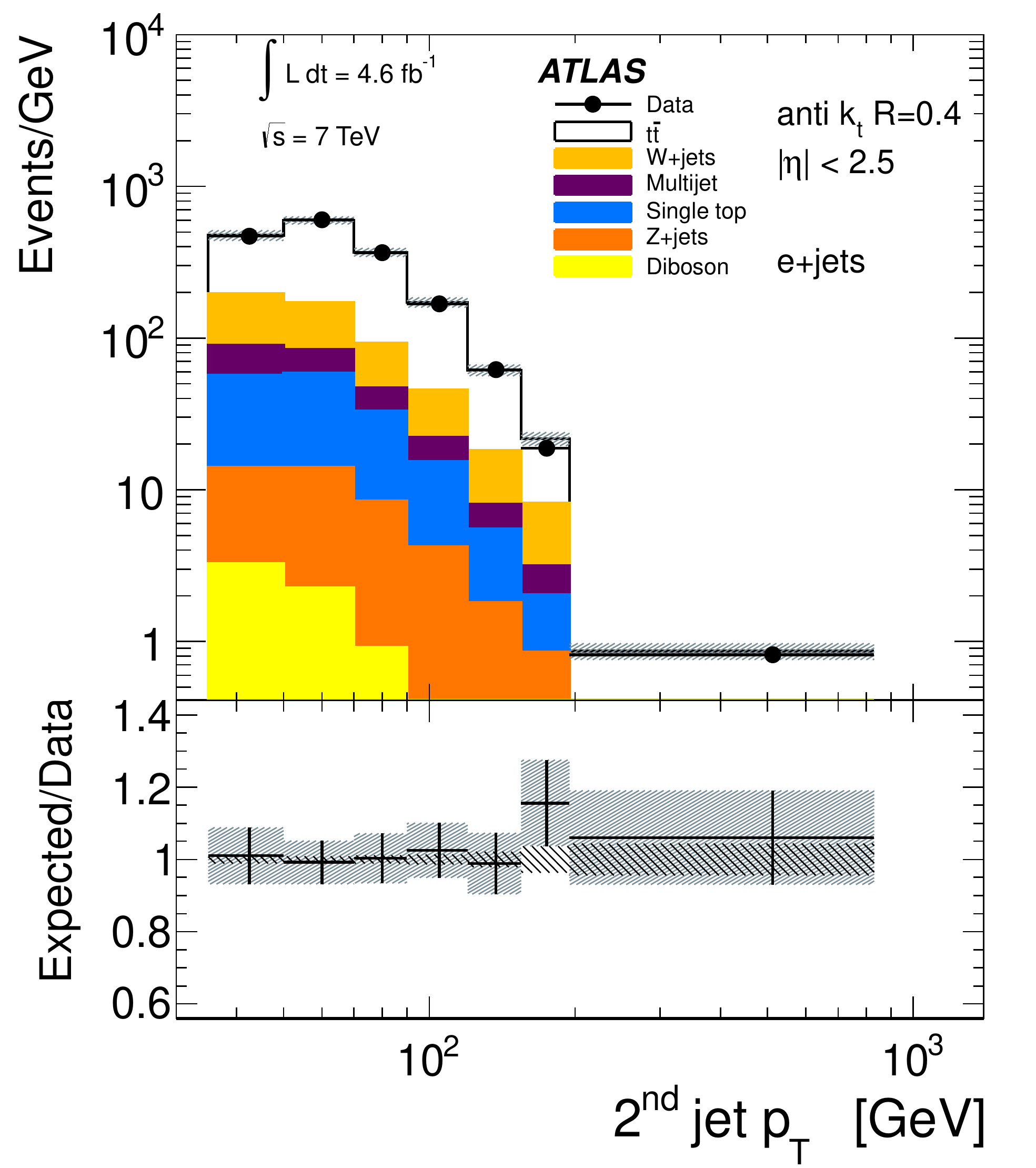}}
\subfigure[\label{fig:jetpt-reco-el_3} \ejets, $\pT > 25$~\GeV, 3$^\mathrm{rd}$ jet]{\includegraphics[width=0.48\textwidth]{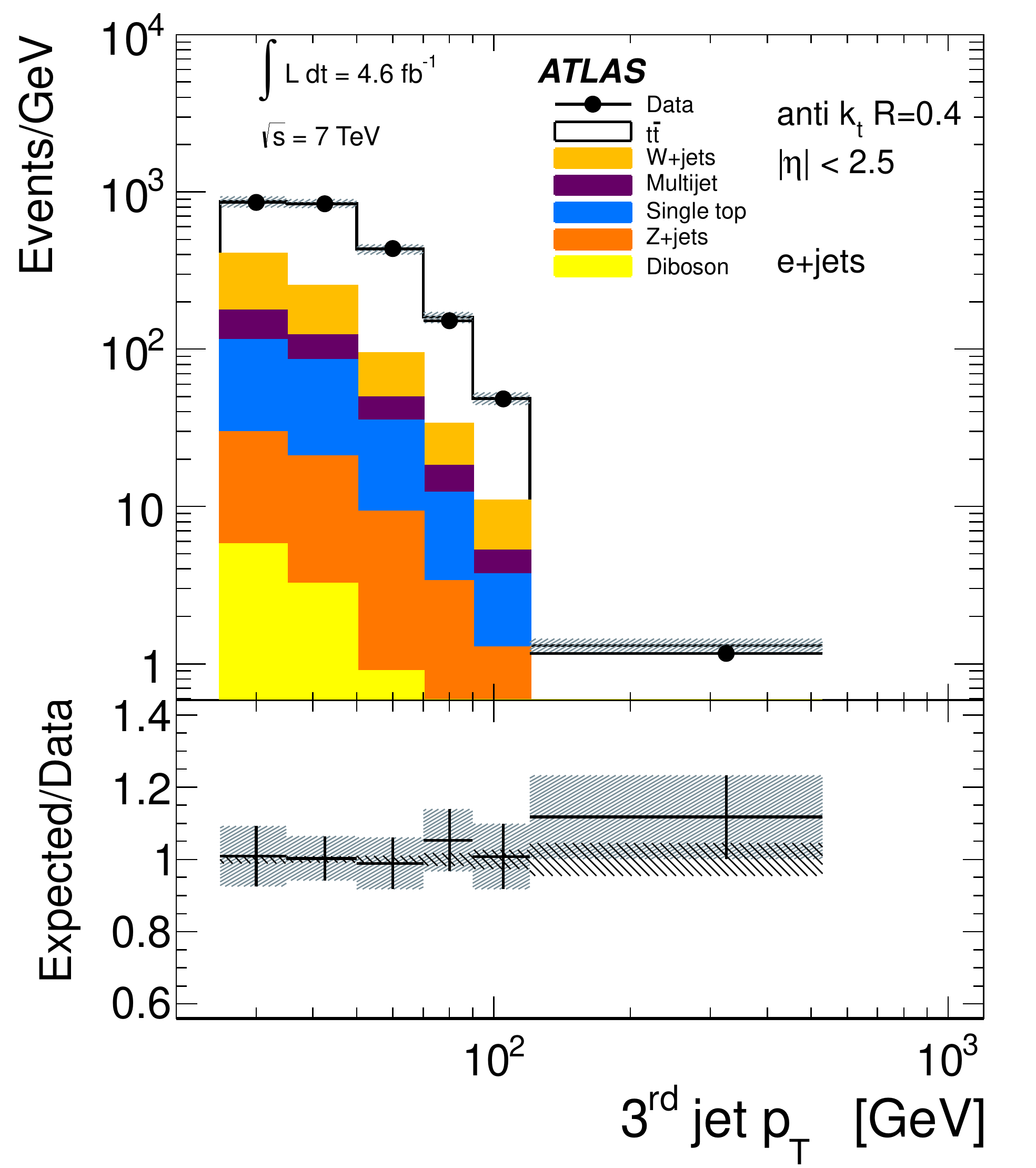}}
\subfigure[\label{fig:jetpt-reco-el_4} \ejets, $\pT > 25$~\GeV, 4$^\mathrm{th}$ jet]{\includegraphics[width=0.48\textwidth]{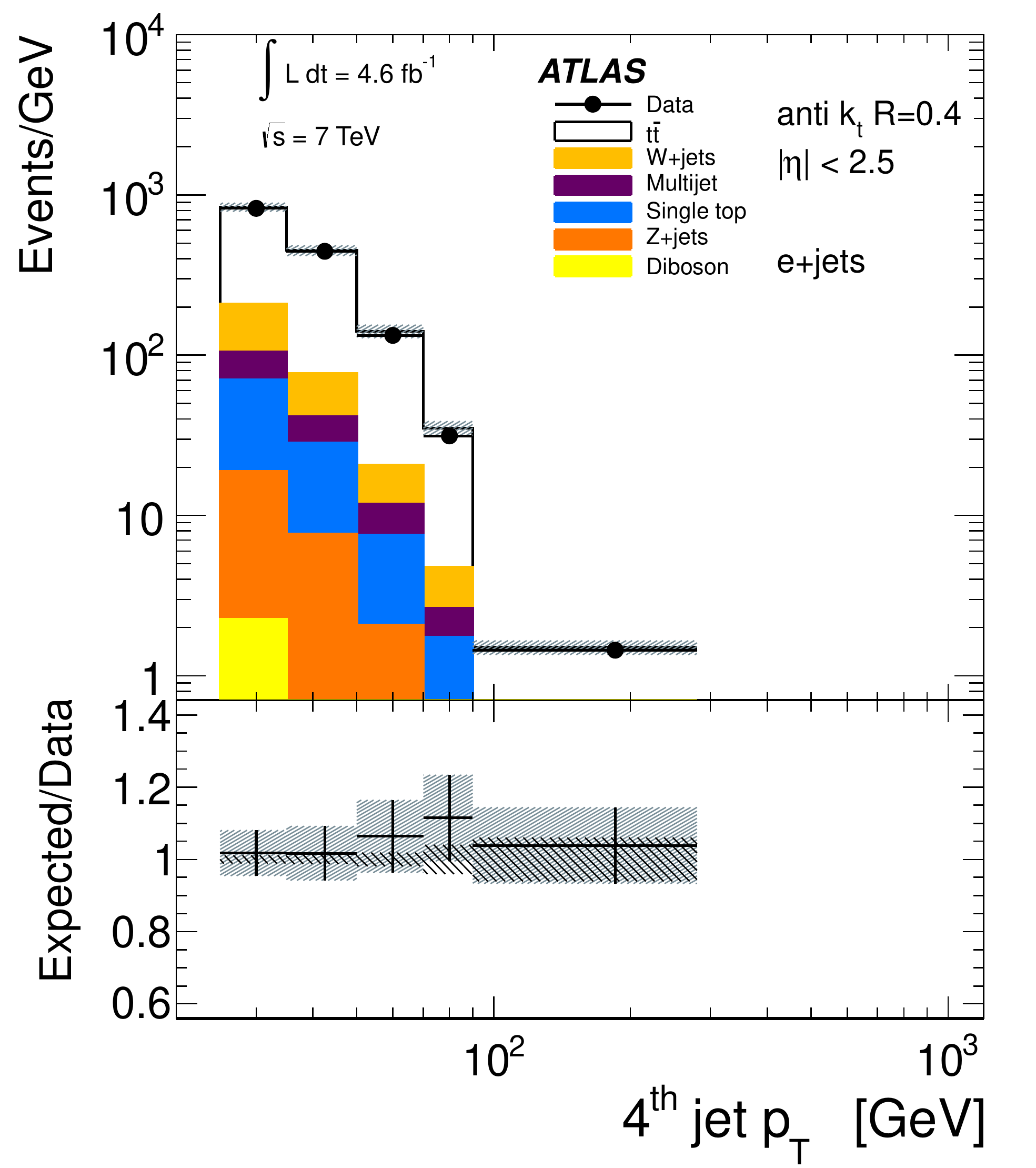}}
\caption{The reconstructed jet \pT\ for the 2$^\mathrm{nd}$ (a), 3$^\mathrm{th}$ (b) and 4$^\mathrm{th}$ (c) jets in the electron (\ejets)  channel.
  The data are compared to the sum of the \ttbar\ {\sc POWHEG}+{\sc PYTHIA} MC signal prediction and  
  the background models.  The shaded bands show the total systematic and statistical
  uncertainties on the combined signal and background estimate. The errors bar on the black points and the hatched area in the ratio, show the statistical uncertainty on the data measurements.}
\label{figure:jetspt-reco-el}
\end{figure}

\begin{figure}[htbp]
\centering
\subfigure[\label{fig:jetpt-reco-mu_2} \mujets, $\pT > 25$~\GeV, 2$^\mathrm{nd}$ jet]{\includegraphics[width=0.48\textwidth]{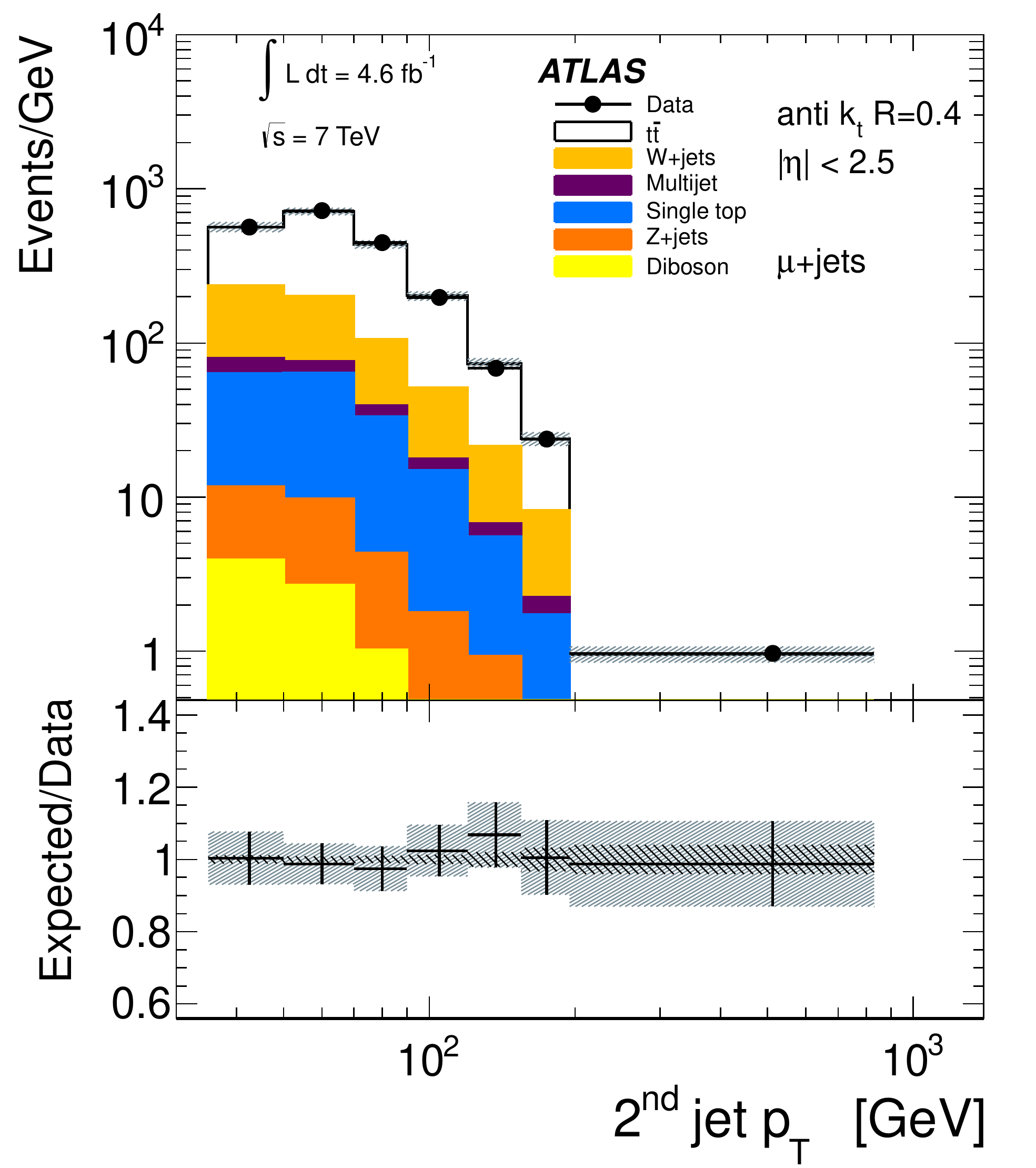}}
\subfigure[\label{fig:jetpt-reco-mu_3} \mujets, $\pT > 25$~\GeV, 3$^\mathrm{rd}$ jet]{\includegraphics[width=0.48\textwidth]{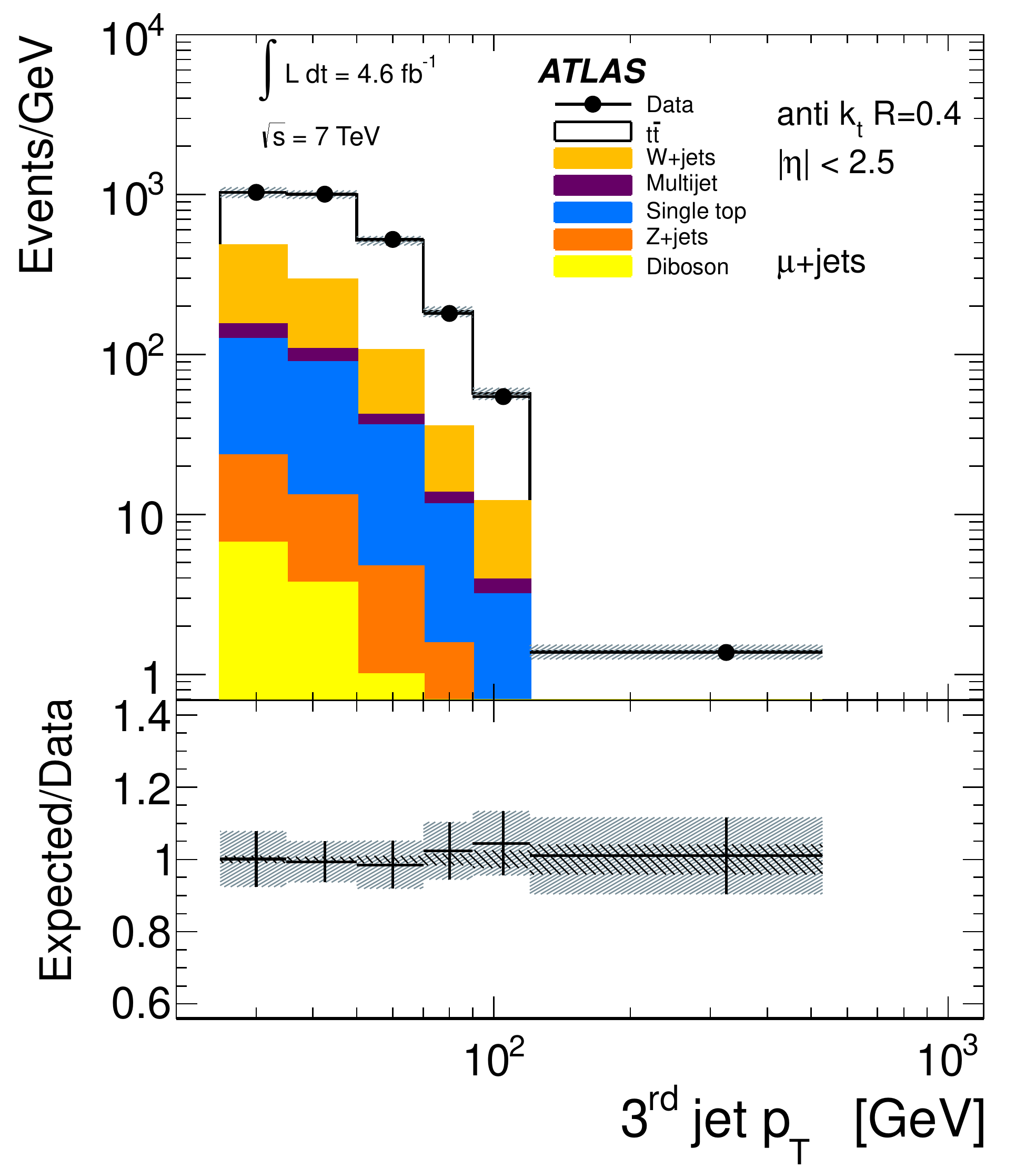}}
\subfigure[\label{fig:jetpt-reco-mu_4} \mujets, $\pT > 25$~\GeV, 4$^\mathrm{th}$ jet]{\includegraphics[width=0.48\textwidth]{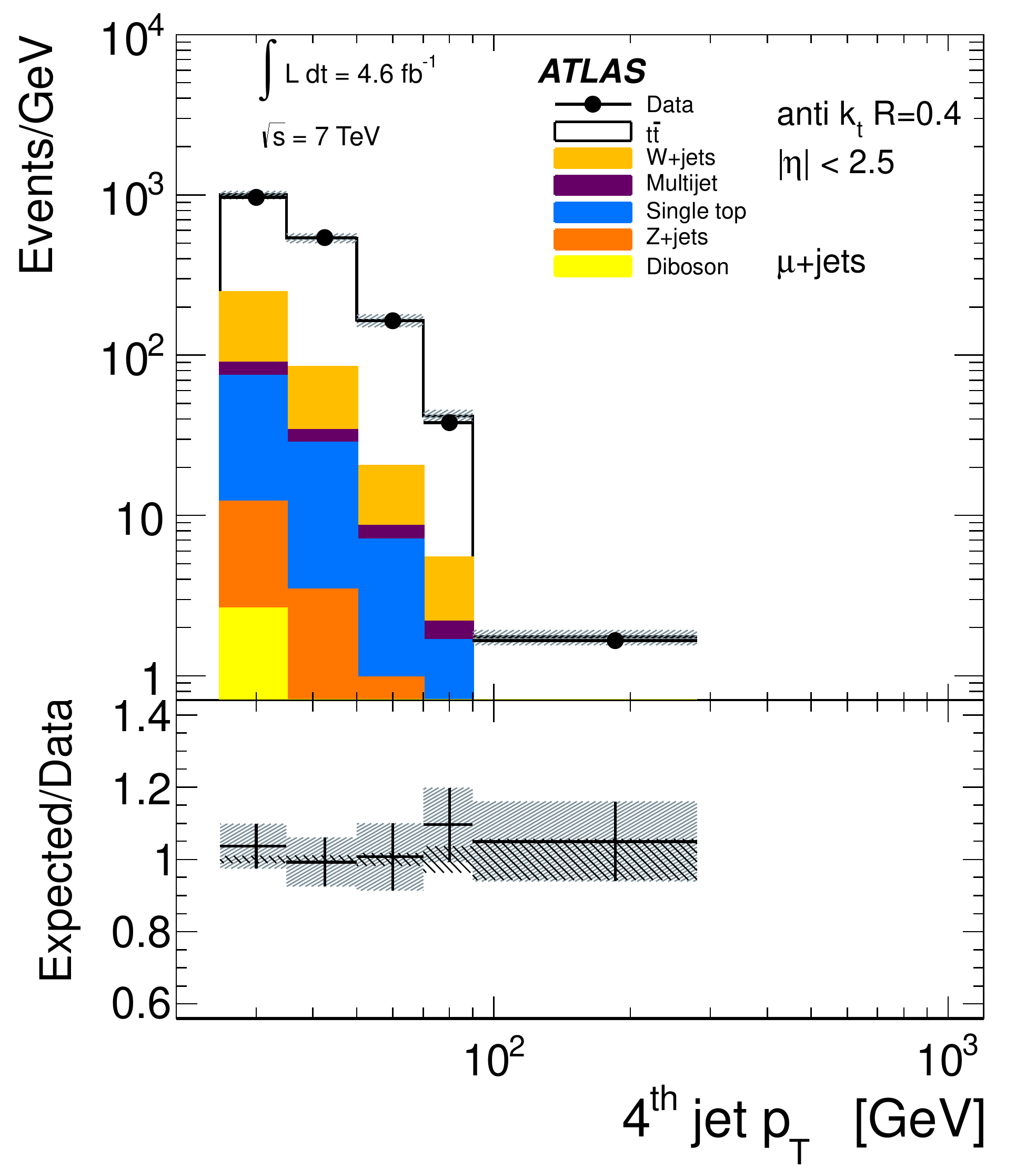}}
\caption{The reconstructed jet \pT\ for the 2$^\mathrm{nd}$ (a), 3$^\mathrm{th}$ (b) and 4$^\mathrm{th}$ (c) jets in the muon (\mujets)  channel.
  The data are compared to the sum of the \ttbar\ {\sc POWHEG}+{\sc PYTHIA} MC signal prediction and  
  the background models.  The shaded bands show the total systematic and statistical
  uncertainties on the combined signal and background estimate. The errors bar on the black points and the hatched area in the ratio, show the statistical uncertainty on the data measurements.}
\label{figure:jetspt-reco-mu}
\end{figure}

\clearpage

%% file: appendixCorrectionFactors.tex
\section{Global correction factors }
\label{sec:correctionFactorsA}

\begin{figure}[htbp]
\centering
\subfigure[\label{fig:closure_el_40GeV}]{\includegraphics[width=0.49\textwidth]{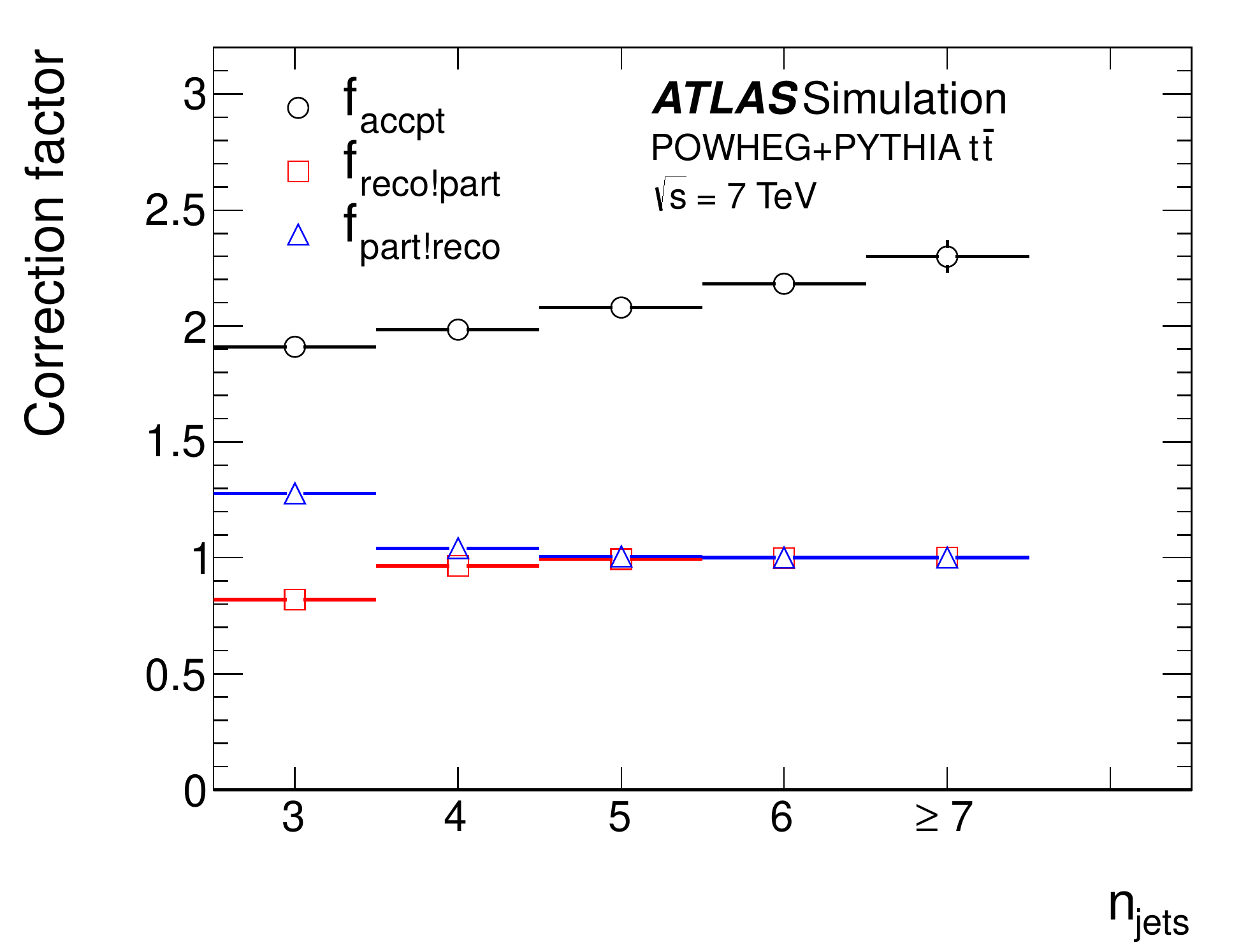}}
\subfigure[\label{fig:closure_mu_40GeV}]{\includegraphics[width=0.49\textwidth]{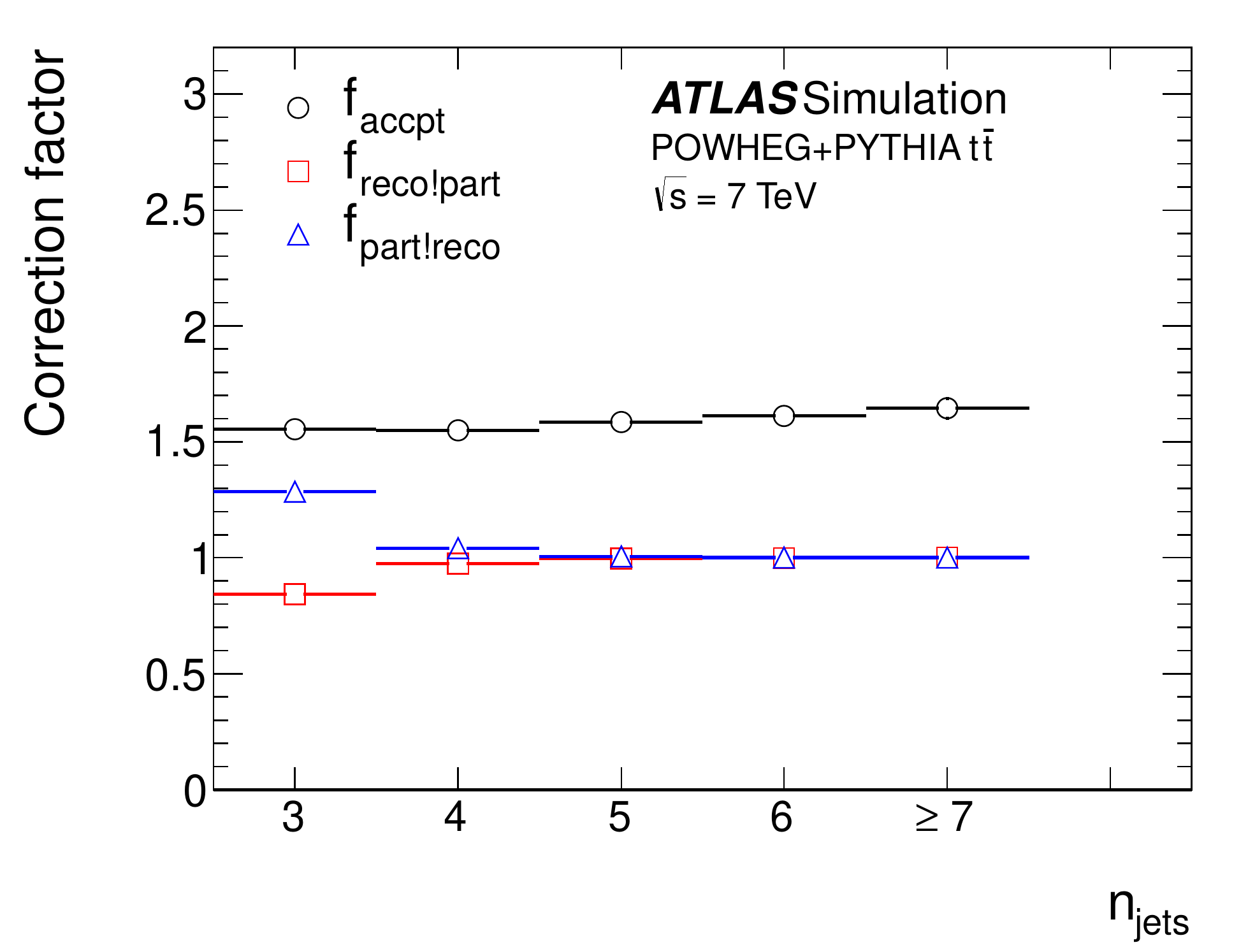}}
\caption{Global correction factors used in the unfolding of jets with $\pT>40$~\GeV\ to particle level in the electron (a) and muon (b) channel as described in the text and in eq.~(\ref{eqn:corrections}).   
The axis label $n_\mathrm{jets}$ refers to the number of particle-level jets for 
$f_\mathrm{accpt}$  and $f_\mathrm{part!reco}$ and to the number of reconstructed jets in the case of $f_\mathrm{reco!part}$.
 \label{fig:correctionfactorsA1}}
\end{figure}

\begin{figure}[htbp]
\centering
\subfigure[\label{fig:closure_el_60GeV}]{\includegraphics[width=0.49\textwidth]{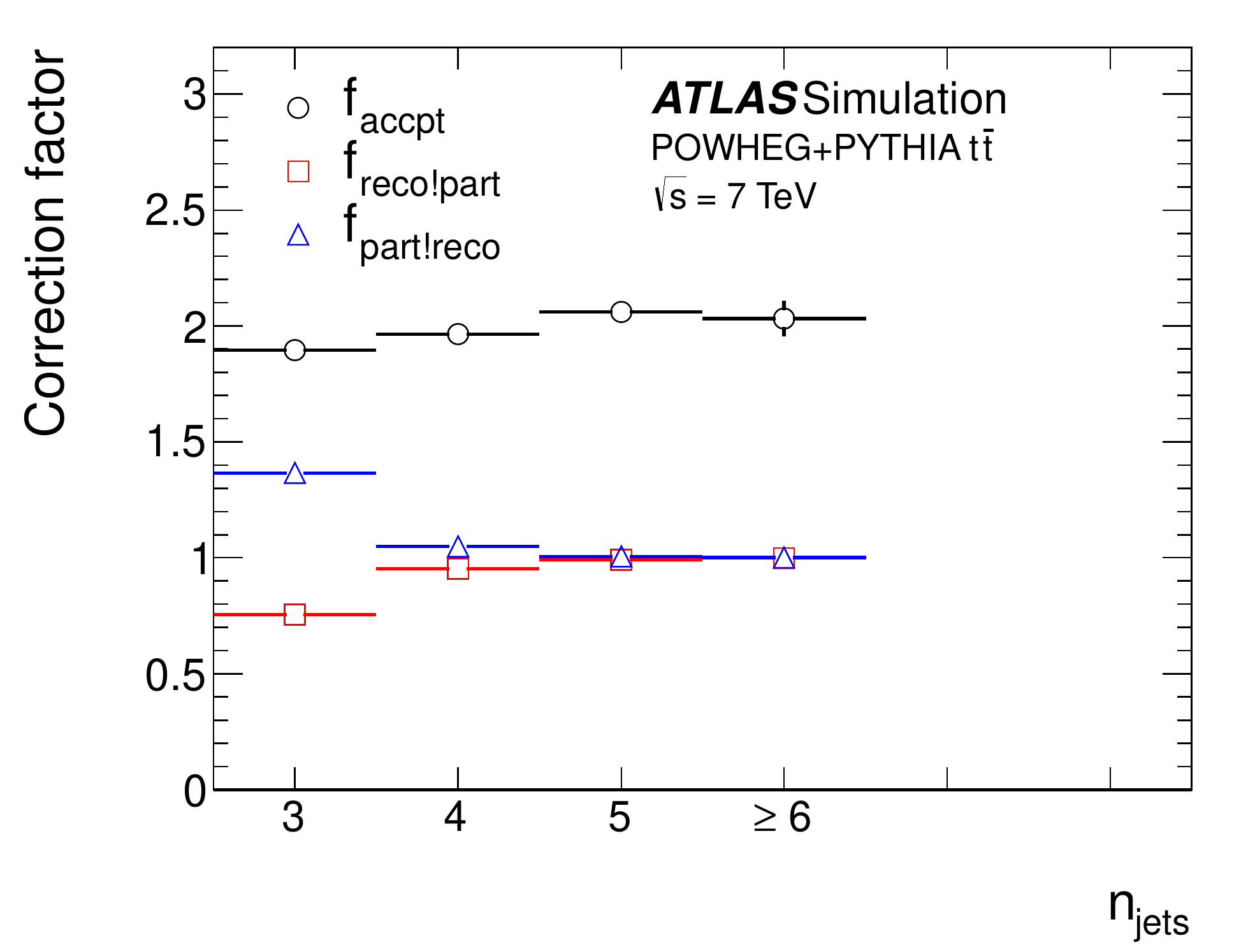}}
\subfigure[\label{fig:closure_mu_60GeV}]{\includegraphics[width=0.49\textwidth]{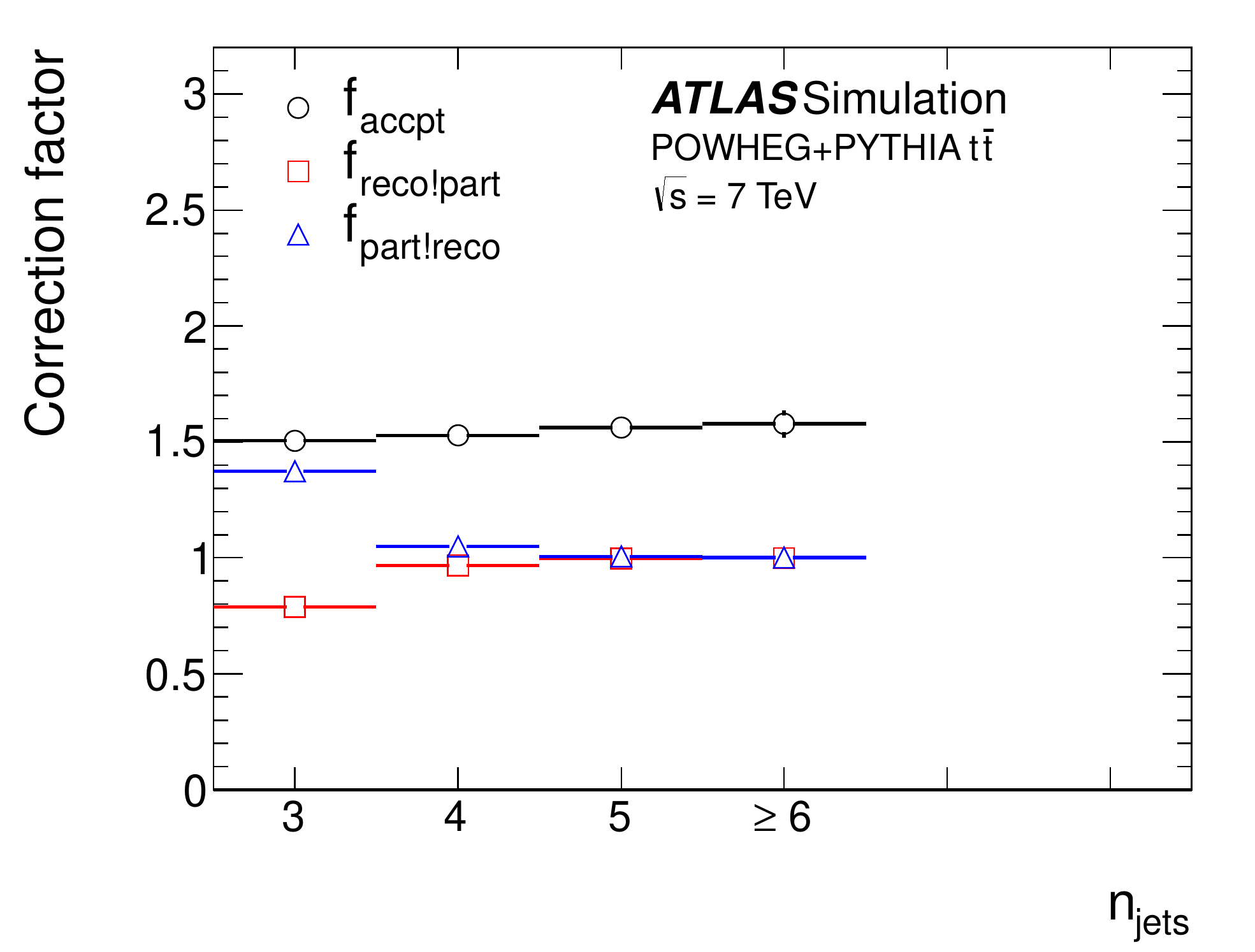}}
\caption{Global correction factors used in the unfolding of jets with $\pT>60$~\GeV\ to particle level in the electron (a) and muon (b) channel as described in the text and in eq.~(\ref{eqn:corrections}).   
The axis label $n_\mathrm{jets}$ refers to the number of particle-level jets for 
$f_\mathrm{accpt}$  and $f_\mathrm{part!reco}$ and to the number of reconstructed jets in the case of $f_\mathrm{reco!part}$.
 \label{fig:correctionfactorsA2}}
\end{figure}

\begin{figure}[htbp]
\centering
\subfigure[\label{fig:closure_el_80GeV}]{\includegraphics[width=0.49\textwidth]{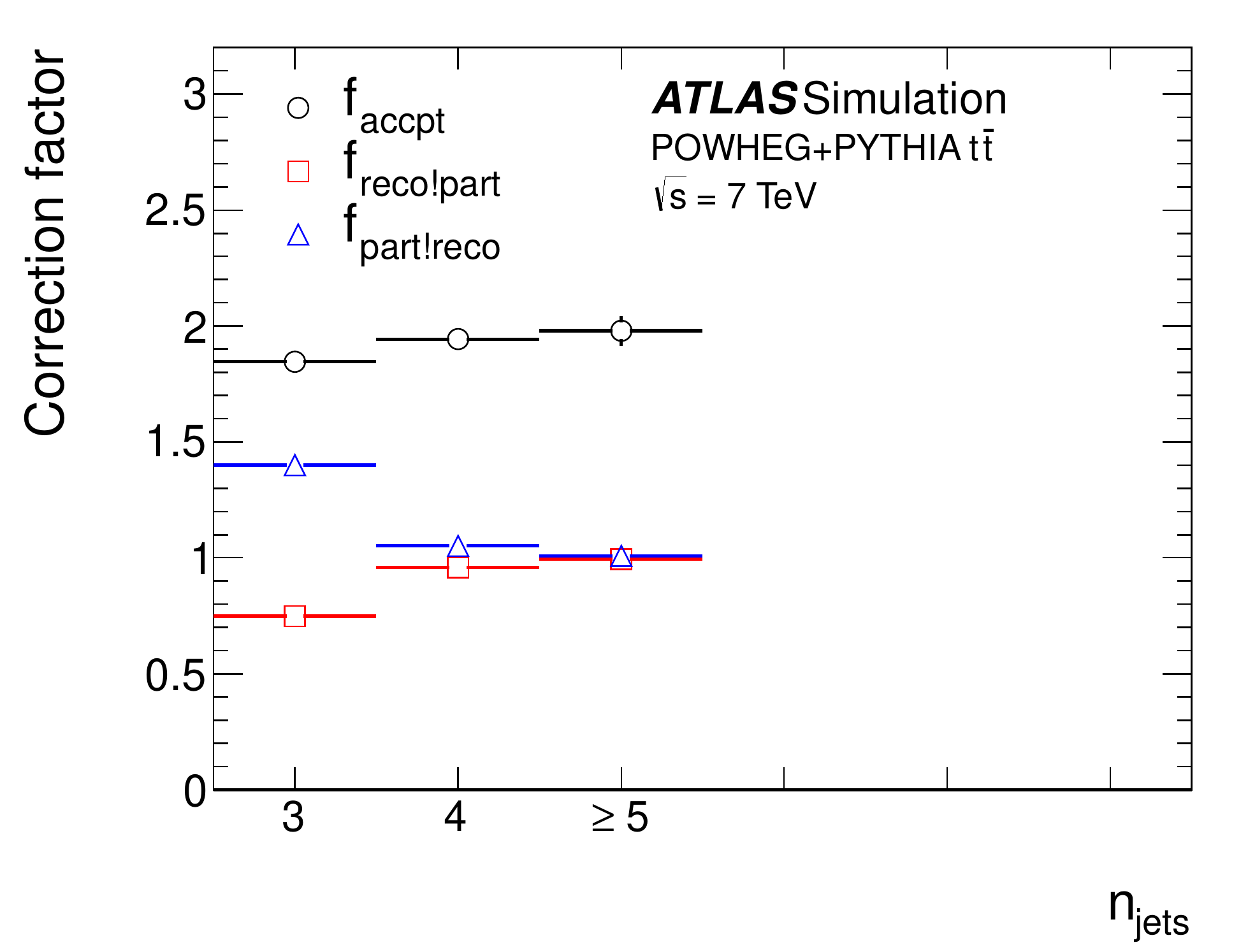}}
\subfigure[\label{fig:closure_mu_80GeV}]{\includegraphics[width=0.49\textwidth]{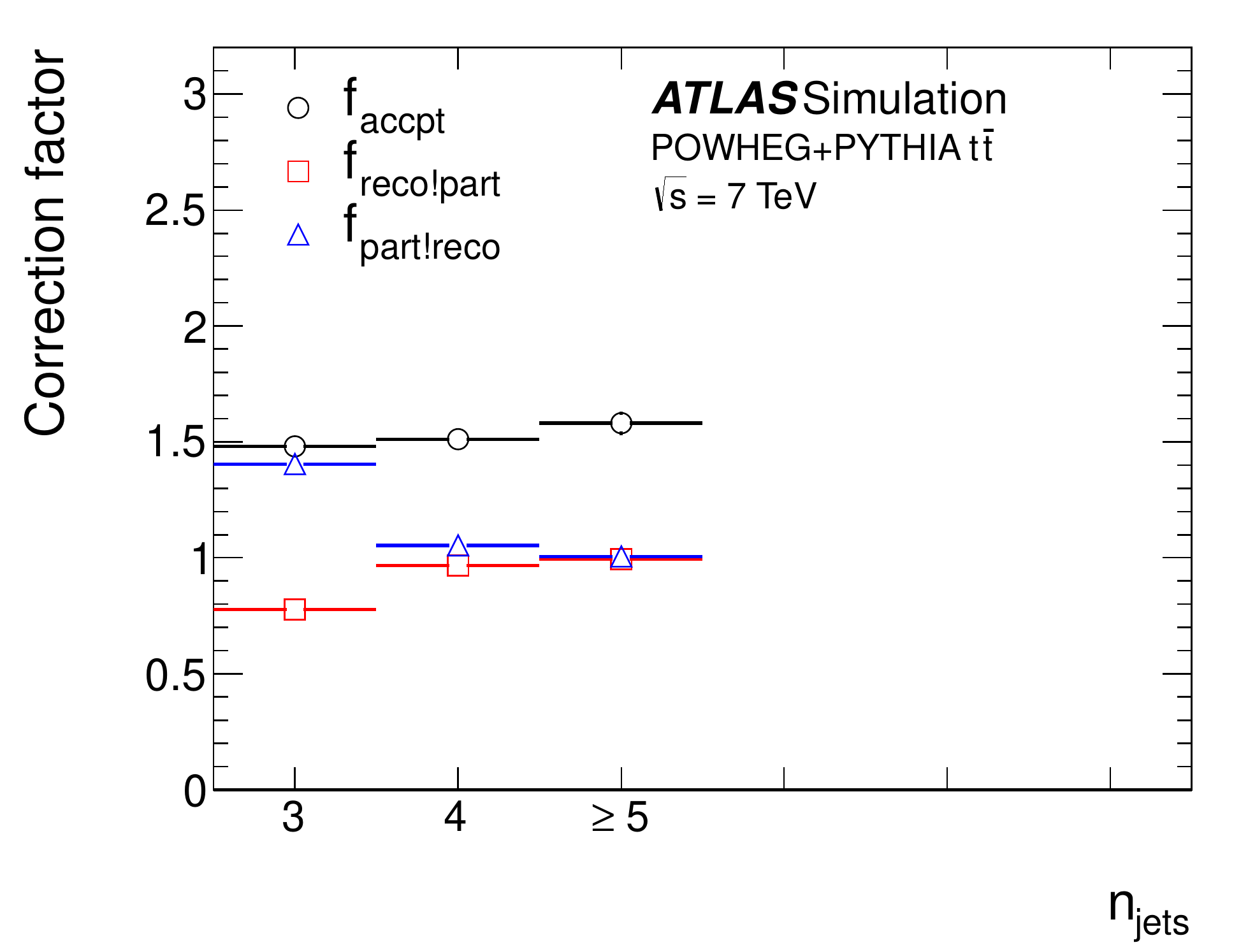}}
\caption{Global correction factors used in the unfolding of jets with $\pT>80$~\GeV\ to particle level in the electron (a) and muon (b) channel as described in the text and in eq.~(\ref{eqn:corrections}).   
The axis label $n_\mathrm{jets}$ refers to the number of particle-level jets for 
$f_\mathrm{accpt}$  and $f_\mathrm{part!reco}$ and to the number of reconstructed jets in the case of $f_\mathrm{reco!part}$.
 \label{fig:correctionfactorsA3}}
\end{figure}

%% file: appendixSystematicTables.tex
\section{Tables of results with systematic uncertainties}\label{sec:appendixsystematics}

\input{njetsPart_ljets_pT25}
\newpage
\input{njetsPart_ljets_pT40}
\newpage
\input{njetsPart_ljets_pT60}
\newpage
\input{njetsPart_ljets_pT80}
\newpage
\input{jetPtN0Part_ljets_pT25}
\newpage
\input{jetPtN1Part_ljets_pT25}
\newpage
\input{jetPtN2Part_ljets_pT25}
\newpage
\input{jetPtN3Part_ljets_pT25}
\newpage
\input{jetPtN4Part_ljets_pT25}
\newpage

%% file: njetsPart_ljets_pT25.tex
\begin{table}[tbhp]
\centering
\scriptsize
\tabcolsep=3mm
\begin{tabular}{ l c c c c c c } \hline
$\frac{d \sigma}{d n_\mathrm{jets}}$ [\%] / $n_\mathrm{jets}$& 3 & 4 & 5 & 6 & 7 & $\geq$ 8  \\ \hline \hline 
MC statistics & 0.5 & 0.5 & 0.9 & 1.8 & 2.0 & 3.4 \\ 
PDF & 1.2 & 0.8 & 2.4 & 3.8 & 5.0 & 5.2 \\ 
MC generator & 0.9 & 0.5 & 0.7 & 1.6 & 1.5 & 2.2 \\ 
Fragmentation & 2.5 & 0.3 & 0.6 & 2.5 & 6.5 & 8.1 \\ 
ISR/FSR & 2.9 & 2.8 & 1.4 & 2.6 & 3.5 & 5.5 \\ 
Colour reconnection & 1.4 & 0.2 & 0.2 & 3.5 & 4.0 & 4.4 \\ 
$\ell$ resolution \& efficiency & 0.3 & 0.3 & 0.4 & 0.4 & 0.3 & 0.7 \\ 
\met\ cell-out & 0.2 & 0.2 & 0.2 & 0.1 & 0.4 & 0.4 \\ 
$b$-quark tagging efficiency & 4.3 & 3.8 & 3.5 & 3.6 & 3.4 & 3.6 \\ 
Additional interactions & 0.1 & 0.1 & 0.1 & 0.1 & 0.4 & 0.2 \\ 
Jet reconstruction efficiency & 0.0 & 0.0 & 0.0 & 0.2 & 0.1 & 0.3 \\ 
Jet energy resolution & 1.2 & 0.7 & 1.2 & 2.9 & 4.0 & 8.8 \\ 
$b$-quark jets (JES) & 0.1 & 1.0 & 1.3 & 1.4 & 1.4 & 1.7 \\ 
Close by jets (JES) & 2.4 & 1.2 & 3.9 & 5.9 & 9.6 & 14.5 \\ 
Effective detector NP set 1 (JES) & 0.6 & 0.5 & 1.3 & 1.7 & 2.3 & 3.2 \\ 
Effective detector NP set 2 (JES) & 0.0 & 0.0 & 0.1 & 0.2 & 0.2 & 0.5 \\ 
Effective mixed NP set 1 (JES) & 0.0 & 0.0 & 0.1 & 0.2 & 0.2 & 0.4 \\ 
Effective mixed NP set 2 (JES) & 0.3 & 0.2 & 0.5 & 0.8 & 0.8 & 1.7 \\ 
Effective model NP set 1 (JES) & 1.7 & 1.1 & 3.2 & 4.3 & 5.9 & 9.4 \\ 
Effective model NP set 2 (JES) & 0.1 & 0.0 & 0.1 & 0.2 & 0.2 & 0.5 \\ 
Effective model NP set 3 (JES) & 0.1 & 0.0 & 0.1 & 0.1 & 0.0 & 0.3 \\ 
Effective model NP set 4 (JES) & 0.0 & 0.0 & 0.1 & 0.1 & 0.1 & 0.4 \\ 
Effective stat. NP set 1 (JES) & 1.0 & 0.6 & 1.8 & 2.4 & 3.3 & 5.7 \\ 
Effective stat. NP set 2 (JES) & 0.0 & 0.0 & 0.0 & 0.0 & 0.0 & 0.0 \\ 
Effective stat. NP set 3 (JES) & 0.1 & 0.1 & 0.2 & 0.3 & 0.3 & 1.0 \\ 
$\eta$-intercalibration (JES) & 1.2 & 0.8 & 2.2 & 3.1 & 4.3 & 6.4 \\ 
$\eta$-intercalibration statistics (JES) & 0.4 & 0.3 & 0.7 & 1.1 & 1.4 & 2.3 \\ 
Flavour composition  (JES) & 0.6 & 0.8 & 1.0 & 1.8 & 2.9 & 2.2 \\ 
Flavour response (JES) & 1.9 & 0.3 & 2.0 & 5.7 & 10.3 & 7.2 \\ 
Additional interactions $\mu$ (JES) & 0.1 & 0.3 & 0.3 & 0.3 & 0.4 & 1.4 \\ 
Additional interactions $N_\mathrm{PV}$ (JES) & 0.2 & 0.1 & 0.2 & 0.5 & 0.8 & 2.4 \\ 
Relative non-closure (JES) & 0.3 & 0.2 & 0.6 & 0.8 & 0.9 & 1.6 \\ 
Single particle high-\pT\ (JES) & 0.0 & 0.0 & 0.0 & 0.0 & 0.0 & 0.0 \\ 
Jet vertex fraction efficiency & 0.4 & 0.3 & 0.3 & 0.4 & 0.7 & 1.5 \\ 
$W$+jets normalisation & 5.2 & 1.9 & 1.6 & 0.9 & 0.6 & 0.9 \\ 
$W$+jets heavy/light flavour & 9.7 & 2.5 & 0.5 & 0.2 & 0.6 & 0.7 \\ 
Multijet normalisation & 1.4 & 0.4 & 0.3 & 0.6 & 0.5 & 2.3 \\ 
Multijet shape & 0.2 & 0.3 & 0.3 & 0.3 & 0.2 & 0.4 \\ 
Small backgrounds & 6.0 & 3.0 & 3.1 & 1.5 & 2.3 & 4.3 \\ 
Luminosity & 1.8 & 1.8 & 1.8 & 1.8 & 1.8 & 1.8 \\ 
\hline \hline 
Statistical uncertainty & 1.4 & 1.3 & 2.2 & 4.0 & 4.4 & 10.2 \\ 
Total uncertainty & 14.8 & 7.4 & 9.3 & 14.2 & 21.0 & 29.1 \\ 
\hline 
Cross-section [pb] & 4.34e+00 & 3.76e+00 & 1.72e+00 & 6.11e-01 & 1.61e-01 & 4.25e-02 \\ 
\hline 
\end{tabular}
\caption{Relative uncertainties on the final differential cross-section after the $e$/$\mu$ channel combination, for the jet multiplicity using a 25~\GeV\ jet \pt\ threshold.  The uncertainties are shown as a percentage of the expected \ttbar\ signal.  The energy scale uncertainty (JES) is shown for each JES nuisance parameter (NP). The effective NP are obtained by combining   a total of 54 detector, detector and model (``mixed''), modelling and statistical NPs.  An uncertainty value of 0.0 implies that the uncertainty is below 0.05.}
\label{tab:tables-paper/jet-mult-single-lepton/njetsPart_ljets_pT25}
\end{table}

%% file: njetsPart_ljets_pT40.tex
\begin{table}[tbhp]
\centering
\scriptsize
\tabcolsep=3mm
\begin{tabular}{ l c c c c c } \hline
$\frac{d \sigma}{d n_\mathrm{jets}}$ [\%] / $n_\mathrm{jets}$& 3 & 4 & 5 & 6 & $\geq$ 7  \\ \hline \hline 
MC statistics & 0.5 & 0.9 & 2.0 & 2.2 & 4.8 \\ 
PDF & 0.2 & 1.9 & 1.6 & 3.1 & 2.0 \\ 
MC generator & 1.1 & 0.7 & 0.2 & 0.9 & 5.6 \\ 
Fragmentation & 1.3 & 0.9 & 0.2 & 2.2 & 2.8 \\ 
ISR/FSR & 2.9 & 4.1 & 0.3 & 6.3 & 6.0 \\ 
Colour reconnection & 0.6 & 0.5 & 1.1 & 1.5 & 2.0 \\ 
$\ell$ resolution \& efficiency & 0.3 & 0.4 & 0.5 & 0.5 & 0.9 \\ 
\met\ cell-out & 0.2 & 0.2 & 0.4 & 0.1 & 0.9 \\ 
$b$-quark tagging efficiency & 3.9 & 3.7 & 3.8 & 3.1 & 4.2 \\ 
Additional interactions & 0.1 & 0.1 & 0.2 & 0.1 & 0.6 \\ 
Jet reconstruction efficiency & 0.0 & 0.0 & 0.0 & 0.0 & 0.1 \\ 
Jet energy resolution & 0.7 & 0.1 & 1.1 & 3.4 & 2.8 \\ 
$b$-quark jets (JES) & 1.4 & 2.5 & 2.5 & 2.9 & 3.0 \\ 
Close by jets (JES) & 0.7 & 4.6 & 7.0 & 10.8 & 14.8 \\ 
Effective detector NP set 1 (JES) & 0.4 & 1.7 & 2.3 & 3.1 & 3.5 \\ 
Effective detector NP set 2 (JES) & 0.1 & 0.1 & 0.1 & 0.3 & 0.7 \\ 
Effective mixed NP set 1 (JES) & 0.1 & 0.1 & 0.1 & 0.3 & 0.7 \\ 
Effective mixed NP set 2 (JES) & 0.1 & 0.3 & 0.4 & 0.7 & 0.7 \\ 
Effective model NP set 1 (JES) & 0.4 & 1.9 & 2.7 & 3.6 & 4.8 \\ 
Effective model NP set 2 (JES) & 0.1 & 0.3 & 0.5 & 0.7 & 1.0 \\ 
Effective model NP set 3 (JES) & 0.2 & 0.5 & 0.7 & 1.0 & 1.2 \\ 
Effective model NP set 4 (JES) & 0.1 & 0.1 & 0.2 & 0.4 & 0.7 \\ 
Effective stat. NP set 1 (JES) & 0.2 & 0.7 & 1.1 & 1.8 & 2.1 \\ 
Effective stat. NP set 2 (JES) & 0.1 & 0.1 & 0.2 & 0.3 & 0.8 \\ 
Effective stat. NP set 3 (JES) & 0.1 & 0.4 & 0.6 & 0.9 & 1.0 \\ 
$\eta$-intercalibration (JES) & 0.5 & 2.1 & 3.4 & 5.1 & 4.5 \\ 
$\eta$-intercalibration statistics (JES) & 0.1 & 0.5 & 0.8 & 1.1 & 1.4 \\ 
Flavour composition  (JES) & 0.2 & 1.0 & 1.6 & 2.0 & 1.0 \\ 
Flavour response (JES) & 0.5 & 1.2 & 3.9 & 5.5 & 2.7 \\ 
Additional interactions $\mu$ (JES) & 0.1 & 0.1 & 0.3 & 0.3 & 1.0 \\ 
Additional interactions $N_\mathrm{PV}$ (JES) & 0.0 & 0.2 & 0.5 & 0.8 & 0.8 \\ 
Relative non-closure (JES) & 0.1 & 0.2 & 0.6 & 0.7 & 1.2 \\ 
Single particle high-\pT\ (JES) & 0.0 & 0.0 & 0.0 & 0.0 & 0.0 \\ 
Jet vertex fraction efficiency & 0.2 & 0.3 & 0.5 & 1.3 & 1.8 \\ 
$W$+jets normalisation & 2.4 & 1.5 & 1.1 & 0.4 & 1.1 \\ 
$W$+jets heavy/light flavour & 4.0 & 1.2 & 0.6 & 0.4 & 0.6 \\ 
Multijet normalisation & 0.5 & 0.4 & 0.6 & 0.3 & 1.0 \\ 
Multijet shape & 0.1 & 0.3 & 0.4 & 0.4 & 0.9 \\ 
Small backgrounds & 3.2 & 2.2 & 2.1 & 2.2 & 5.2 \\ 
Luminosity & 1.8 & 1.8 & 1.8 & 1.8 & 1.8 \\ 
\hline \hline 
Statistical uncertainty & 1.2 & 2.2 & 4.5 & 5.1 & 13.1 \\ 
Total uncertainty & 8.3 & 9.8 & 12.4 & 18.3 & 25.5 \\ 
\hline 
Cross-section [pb] & 4.31e+00 & 2.00e+00 & 5.29e-01 & 1.26e-01 & 2.36e-02 \\ 
\hline 
\end{tabular}
\caption{Relative uncertainties on the final differential cross-section after the $e$/$\mu$ channel combination, for the jet multiplicity using a 40~\GeV\ jet \pt\ threshold.  The uncertainties are shown as a percentage of the expected \ttbar\ signal.  The energy scale uncertainty (JES) is shown for each JES nuisance parameter (NP). The effective NP are obtained by combining   a total of 54 detector, detector and model (``mixed''), modelling and statistical NPs.  An uncertainty value of 0.0 implies that the uncertainty is below 0.05.}
\label{tab:tables-paper/jet-mult-single-lepton/njetsPart_ljets_pT40}
\end{table}

%% file: njetsPart_ljets_pT60.tex
\begin{table}[tbhp]
\centering
\scriptsize
\tabcolsep=3mm
\begin{tabular}{ l c c c c } \hline
$\frac{d \sigma}{d n_\mathrm{jets}}$ [\%] / $n_\mathrm{jets}$& 3 & 4 & 5 & $\geq$ 6  \\ \hline \hline 
MC statistics & 0.8 & 1.2 & 2.8 & 5.9 \\ 
PDF & 0.6 & 2.1 & 0.8 & 0.8 \\ 
MC generator & 1.1 & 0.7 & 0.5 & 4.4 \\ 
Fragmentation & 1.2 & 1.1 & 0.4 & 4.1 \\ 
ISR/FSR & 4.9 & 5.3 & 7.4 & 14.6 \\ 
Colour reconnection & 0.4 & 1.4 & 4.2 & 3.1 \\ 
$\ell$ resolution \& efficiency & 0.3 & 0.4 & 0.3 & 0.5 \\ 
\met\ cell-out & 0.1 & 0.3 & 0.3 & 0.4 \\ 
$b$-quark tagging efficiency & 4.3 & 4.3 & 4.3 & 5.1 \\ 
Additional interactions & 0.1 & 0.2 & 0.3 & 0.6 \\ 
Jet reconstruction efficiency & 0.0 & 0.0 & 0.0 & 0.2 \\ 
Jet energy resolution & 0.1 & 0.9 & 3.0 & 0.8 \\ 
$b$-quark jets (JES) & 2.9 & 3.5 & 3.0 & 4.5 \\ 
Close by jets (JES) & 3.0 & 6.0 & 7.4 & 10.5 \\ 
Effective detector NP set 1 (JES) & 1.9 & 3.1 & 3.5 & 5.1 \\ 
Effective detector NP set 2 (JES) & 0.1 & 0.2 & 0.3 & 0.4 \\ 
Effective mixed NP set 1 (JES) & 0.1 & 0.2 & 0.4 & 0.4 \\ 
Effective mixed NP set 2 (JES) & 0.2 & 0.3 & 0.4 & 0.7 \\ 
Effective model NP set 1 (JES) & 1.2 & 1.8 & 2.0 & 4.0 \\ 
Effective model NP set 2 (JES) & 0.5 & 0.8 & 1.0 & 1.5 \\ 
Effective model NP set 3 (JES) & 0.7 & 1.0 & 1.3 & 1.8 \\ 
Effective model NP set 4 (JES) & 0.1 & 0.2 & 0.3 & 0.4 \\ 
Effective stat. NP set 1 (JES) & 0.2 & 0.4 & 0.4 & 1.1 \\ 
Effective stat. NP set 2 (JES) & 0.2 & 0.2 & 0.4 & 0.7 \\ 
Effective stat. NP set 3 (JES) & 0.5 & 0.7 & 0.9 & 1.5 \\ 
$\eta$-intercalibration (JES) & 2.0 & 3.2 & 4.4 & 5.9 \\ 
$\eta$-intercalibration statistics (JES) & 0.4 & 0.6 & 0.7 & 1.3 \\ 
Flavour composition  (JES) & 0.8 & 1.4 & 1.8 & 1.7 \\ 
Flavour response (JES) & 0.4 & 2.4 & 3.7 & 2.7 \\ 
Additional interactions $\mu$ (JES) & 0.1 & 0.3 & 0.4 & 1.8 \\ 
Additional interactions $N_\mathrm{PV}$ (JES) & 0.2 & 0.3 & 0.7 & 1.0 \\ 
Relative non-closure (JES) & 0.3 & 0.5 & 0.8 & 1.1 \\ 
Single particle high-\pT\ (JES) & 0.0 & 0.0 & 0.0 & 0.0 \\ 
Jet vertex fraction efficiency & 0.1 & 0.7 & 0.6 & 1.0 \\ 
$W$+jets normalisation & 2.3 & 1.2 & 0.3 & 0.3 \\ 
$W$+jets heavy/light flavour & 3.0 & 1.3 & 0.0 & 0.0 \\ 
Multijet normalisation & 0.4 & 0.6 & 0.4 & 0.1 \\ 
Multijet shape & 0.2 & 0.3 & 0.2 & 0.4 \\ 
Small backgrounds & 3.0 & 3.4 & 2.6 & 4.0 \\ 
Luminosity & 1.8 & 1.8 & 1.8 & 1.8 \\ 
\hline \hline 
Statistical uncertainty & 1.8 & 2.8 & 6.3 & 14.5 \\ 
Total uncertainty & 10.2 & 12.9 & 16.8 & 28.0 \\ 
\hline 
Cross-section [pb] & 1.99e+00 & 4.95e-01 & 1.04e-01 & 1.72e-02 \\ 
\hline 
\end{tabular}
\caption{Relative uncertainties on the final differential cross-section after the $e$/$\mu$ channel combination, for the jet multiplicity using a 60~\GeV\ jet \pt\ threshold.  The uncertainties are shown as a percentage of the expected \ttbar\ signal.  The energy scale uncertainty (JES) is shown for each JES nuisance parameter (NP). The effective NP are obtained by combining   a total of 54 detector, detector and model (``mixed''), modelling and statistical NPs.  An uncertainty value of 0.0 implies that the uncertainty is below 0.05.}
\label{tab:tables-paper/jet-mult-single-lepton/njetsPart_ljets_pT60}
\end{table}

%% file: njetsPart_ljets_pT80.tex
\begin{table}[tbhp]
\centering
\scriptsize
\tabcolsep=3mm
\begin{tabular}{ l c c c } \hline
$\frac{d \sigma}{d n_\mathrm{jets}}$ [\%] / $n_\mathrm{jets}$& 3 & 4 & $\geq$ 5  \\ \hline \hline 
MC statistics & 0.9 & 2.3 & 4.7 \\ 
PDF & 0.7 & 2.4 & 2.2 \\ 
MC generator & 1.3 & 1.2 & 8.7 \\ 
Fragmentation & 2.2 & 2.7 & 7.8 \\ 
ISR/FSR & 6.5 & 7.1 & 16.3 \\ 
Colour reconnection & 1.0 & 2.5 & 3.4 \\ 
$\ell$ resolution \& efficiency & 0.3 & 0.3 & 0.8 \\ 
\met\ cell-out & 0.1 & 0.3 & 0.6 \\ 
$b$-quark tagging efficiency & 5.2 & 4.9 & 6.2 \\ 
Additional interactions & 0.1 & 0.2 & 0.4 \\ 
Jet reconstruction efficiency & 0.0 & 0.0 & 0.1 \\ 
Jet energy resolution & 1.3 & 0.6 & 1.6 \\ 
$b$-quark jets (JES) & 3.8 & 4.2 & 3.8 \\ 
Close by jets (JES) & 2.8 & 4.3 & 6.5 \\ 
Effective detector NP set 1 (JES) & 2.5 & 3.5 & 4.8 \\ 
Effective detector NP set 2 (JES) & 0.2 & 0.4 & 0.4 \\ 
Effective mixed NP set 1 (JES) & 0.2 & 0.4 & 0.7 \\ 
Effective mixed NP set 2 (JES) & 0.2 & 0.5 & 0.6 \\ 
Effective model NP set 1 (JES) & 1.1 & 1.8 & 2.4 \\ 
Effective model NP set 2 (JES) & 0.8 & 1.4 & 1.7 \\ 
Effective model NP set 3 (JES) & 0.9 & 1.5 & 1.8 \\ 
Effective model NP set 4 (JES) & 0.1 & 0.2 & 0.2 \\ 
Effective stat. NP set 1 (JES) & 0.1 & 0.3 & 0.3 \\ 
Effective stat. NP set 2 (JES) & 0.2 & 0.4 & 0.6 \\ 
Effective stat. NP set 3 (JES) & 0.6 & 1.1 & 1.4 \\ 
$\eta$-intercalibration (JES) & 2.4 & 3.1 & 4.2 \\ 
$\eta$-intercalibration statistics (JES) & 0.4 & 0.8 & 1.0 \\ 
Flavour composition  (JES) & 1.0 & 1.4 & 1.6 \\ 
Flavour response (JES) & 1.0 & 2.1 & 3.1 \\ 
Additional interactions $\mu$ (JES) & 0.3 & 0.2 & 1.2 \\ 
Additional interactions $N_\mathrm{PV}$ (JES) & 0.3 & 0.4 & 0.4 \\ 
Relative non-closure (JES) & 0.2 & 0.6 & 0.6 \\ 
Single particle high-\pT\ (JES) & 0.0 & 0.0 & 0.0 \\ 
Jet vertex fraction efficiency & 0.5 & 0.6 & 1.7 \\ 
$W$+jets normalisation & 2.1 & 1.0 & 0.4 \\ 
$W$+jets heavy/light flavour & 3.4 & 1.1 & 0.6 \\ 
Multijet normalisation & 0.6 & 0.6 & 0.1 \\ 
Multijet shape & 0.2 & 0.2 & 0.7 \\ 
Small backgrounds & 4.4 & 4.2 & 4.7 \\ 
Luminosity & 1.8 & 1.8 & 1.8 \\ 
\hline \hline 
Statistical uncertainty & 2.2 & 5.3 & 12.4 \\ 
Total uncertainty & 12.8 & 15.1 & 28.2 \\ 
\hline 
Cross-section [pb] & 7.55e-01 & 1.49e-01 & 2.46e-02 \\ 
\hline 
\end{tabular}
\caption{Relative uncertainties on the final differential cross-section after the $e$/$\mu$ channel combination, for the jet multiplicity using a 80~\GeV\ jet \pt\ threshold.  The uncertainties are shown as a percentage of the expected \ttbar\ signal.  The energy scale uncertainty (JES) is shown for each JES nuisance parameter (NP). The effective NP are obtained by combining   a total of 54 detector, detector and model (``mixed''), modelling and statistical NPs.  An uncertainty value of 0.0 implies that the uncertainty is below 0.05.}
\label{tab:tables-paper/jet-mult-single-lepton/njetsPart_ljets_pT80}
\end{table}

%% file: jetPtN0Part_ljets_pT25.tex
\begin{table}[tbhp]
\centering
\scriptsize
\tabcolsep=1mm
\begin{tabular}{ l c c c c c c c c } \hline
$\frac{d \sigma}{d p_\mathrm{T,jet}}$ [\%] / $p_\mathrm{T,jet}$ [GeV]& [50-70] & [70-90] & [90-120] & [120-155] & [155-195] & [195-235] & [235-280] & [280-1030] \\ \hline \hline 
MC statistics & 0.9 & 0.6 & 0.6 & 0.7 & 1.0 & 1.6 & 2.1 & 2.7 \\ 
PDF & 0.3 & 0.1 & 0.1 & 0.1 & 0.1 & 0.3 & 0.1 & 0.5 \\ 
MC generator & 1.4 & 0.5 & 0.3 & 1.3 & 0.5 & 0.6 & 1.0 & 0.8 \\ 
Fragmentation & 0.9 & 0.8 & 0.7 & 1.0 & 0.7 & 0.3 & 0.8 & 1.2 \\ 
ISR/FSR & 1.2 & 1.5 & 3.0 & 3.6 & 5.4 & 2.3 & 3.4 & 3.1 \\ 
Colour reconnection & 0.8 & 0.1 & 0.1 & 0.3 & 0.6 & 1.1 & 0.9 & 1.6 \\ 
$\ell$ resolution \& efficiency & 1.3 & 1.4 & 1.4 & 1.3 & 1.3 & 1.3 & 1.3 & 1.3 \\ 
\met\ cell-out & 0.3 & 0.2 & 0.1 & 0.1 & 0.1 & 0.1 & 0.1 & 0.2 \\ 
$b$-quark tagging efficiency & 3.7 & 3.7 & 4.4 & 5.6 & 7.4 & 7.1 & 6.2 & 5.6 \\ 
Additional interactions & 0.2 & 0.1 & 0.1 & 0.1 & 0.1 & 0.1 & 0.1 & 0.1 \\ 
Jet reconstruction efficiency & 0.0 & 0.0 & 0.0 & 0.0 & 0.0 & 0.1 & 0.0 & 0.0 \\ 
Jet energy resolution & 2.3 & 1.6 & 0.6 & 1.2 & 1.9 & 1.3 & 2.7 & 1.3 \\ 
$b$-quark jets (JES) & 4.6 & 0.7 & 2.1 & 3.5 & 3.8 & 3.2 & 3.1 & 2.0 \\ 
Close by jets (JES) & 2.9 & 1.7 & 2.0 & 1.4 & 1.4 & 1.6 & 1.7 & 1.5 \\ 
Effective detector NP set 1 (JES) & 3.0 & 0.6 & 0.9 & 1.7 & 2.2 & 2.9 & 3.3 & 2.9 \\ 
Effective detector NP set 2 (JES) & 0.2 & 0.0 & 0.2 & 0.1 & 0.1 & 0.1 & 0.3 & 0.1 \\ 
Effective mixed NP set 1 (JES) & 0.2 & 0.1 & 0.1 & 0.1 & 0.2 & 0.3 & 0.8 & 1.1 \\ 
Effective mixed NP set 2 (JES) & 0.1 & 0.1 & 0.2 & 0.2 & 0.1 & 0.1 & 0.2 & 0.2 \\ 
Effective model NP set 1 (JES) & 0.7 & 0.7 & 1.2 & 1.1 & 0.9 & 0.7 & 0.6 & 0.2 \\ 
Effective model NP set 2 (JES) & 1.1 & 0.3 & 0.2 & 0.4 & 0.6 & 0.8 & 1.1 & 0.8 \\ 
Effective model NP set 3 (JES) & 1.2 & 0.3 & 0.3 & 0.6 & 0.8 & 0.6 & 0.4 & 0.1 \\ 
Effective model NP set 4 (JES) & 0.0 & 0.0 & 0.1 & 0.0 & 0.3 & 0.4 & 0.5 & 0.2 \\ 
Effective stat. NP set 1 (JES) & 0.7 & 0.6 & 0.4 & 0.3 & 0.1 & 0.1 & 0.1 & 0.2 \\ 
Effective stat. NP set 2 (JES) & 0.3 & 0.0 & 0.1 & 0.1 & 0.2 & 0.2 & 0.4 & 0.3 \\ 
Effective stat. NP set 3 (JES) & 0.8 & 0.1 & 0.2 & 0.3 & 0.6 & 0.6 & 0.8 & 0.4 \\ 
$\eta$-intercalibration (JES) & 2.4 & 0.2 & 1.1 & 1.4 & 1.9 & 2.4 & 2.5 & 1.5 \\ 
$\eta$-intercalibration statistics (JES) & 0.4 & 0.1 & 0.3 & 0.3 & 0.4 & 0.3 & 0.4 & 0.2 \\ 
Flavour composition  (JES) & 0.7 & 0.3 & 0.6 & 0.7 & 0.6 & 0.9 & 1.1 & 1.2 \\ 
Flavour response (JES) & 0.5 & 0.0 & 0.3 & 0.4 & 0.5 & 0.9 & 1.0 & 1.3 \\ 
Additional interactions $\mu$ (JES) & 0.4 & 0.1 & 0.1 & 0.1 & 0.2 & 0.3 & 0.5 & 0.2 \\ 
Additional interactions $N_\mathrm{PV}$ (JES) & 0.3 & 0.0 & 0.2 & 0.1 & 0.2 & 0.4 & 0.2 & 0.1 \\ 
Relative non-closure (JES) & 0.2 & 0.1 & 0.1 & 0.2 & 0.5 & 0.7 & 0.5 & 0.2 \\ 
Single particle high-\pT\ (JES) & 0.0 & 0.0 & 0.0 & 0.0 & 0.0 & 0.0 & 0.0 & 0.0 \\ 
Jet vertex fraction efficiency & 1.4 & 1.5 & 1.5 & 1.6 & 1.7 & 1.8 & 2.0 & 2.2 \\ 
$W$+jets normalisation & 3.2 & 2.1 & 1.8 & 1.9 & 2.6 & 3.3 & 3.8 & 4.7 \\ 
$W$+jets heavy/light flavour & 6.1 & 3.6 & 3.1 & 3.7 & 5.1 & 5.9 & 6.9 & 7.0 \\ 
Multijet normalisation & 1.5 & 0.6 & 0.4 & 0.8 & 1.1 & 1.0 & 1.7 & 1.6 \\ 
Multijet shape & 0.5 & 0.1 & 0.1 & 0.1 & 0.1 & 0.1 & 0.3 & 0.4 \\ 
Small backgrounds & 2.0 & 1.3 & 1.1 & 1.4 & 1.7 & 2.2 & 3.0 & 2.6 \\ 
Luminosity & 1.8 & 1.8 & 1.8 & 1.8 & 1.8 & 1.8 & 1.8 & 1.8 \\ 
\hline \hline 
Statistical uncertainty & 2.2 & 1.4 & 1.4 & 1.5 & 2.4 & 3.7 & 5.6 & 5.8 \\ 
Total uncertainty & 11.9 & 7.3 & 8.2 & 10.1 & 13.0 & 12.9 & 14.7 & 14.3 \\ 
\hline 
Cross-section [pb/GeV] & 3.93e-03 & 6.44e-03 & 3.35e-03 & 1.46e-03 & 4.89e-04 & 2.18e-04 & 8.07e-05 & 2.69e-07 \\ 
\hline 
\end{tabular}
\caption{Relative uncertainties on the final differential cross-section after the $e$/$\mu$ channel combination, for the leading jet.  The uncertainties are shown as a percentage of the expected \ttbar\ signal.  The energy scale uncertainty (JES) is shown for each JES nuisance parameter (NP). The effective NP are obtained by combining   a total of 54 detector, detector and model (``mixed''), modelling and statistical NPs.  An uncertainty value of 0.0 implies that the uncertainty is below 0.05.}
\label{tab:tables-paper/jet-pt-single-lepton/jetPtN0Part_ljets_pT25}
\end{table}

%% file: jetPtN1Part_ljets_pT25.tex
\begin{table}[tbhp]
\centering
\scriptsize
\tabcolsep=1mm
\begin{tabular}{ l c c c c c c c } \hline
$\frac{d \sigma}{d p_\mathrm{T,jet}}$ [\%] / $p_\mathrm{T,jet}$ [GeV]& [35-50] & [50-70] & [70-90] & [90-120] & [120-155] & [155-195] & [195-830] \\ \hline \hline 
MC statistics & 0.9 & 0.5 & 0.8 & 0.8 & 1.1 & 2.0 & 2.7 \\ 
PDF & 0.4 & 0.0 & 0.2 & 0.3 & 0.3 & 0.5 & 0.3 \\ 
MC generator & 1.1 & 0.1 & 0.7 & 0.5 & 1.0 & 0.6 & 1.4 \\ 
Fragmentation & 0.9 & 1.0 & 0.7 & 0.0 & 1.5 & 1.7 & 1.0 \\ 
ISR/FSR & 0.9 & 2.9 & 2.8 & 3.8 & 4.4 & 2.5 & 2.3 \\ 
Colour reconnection & 0.5 & 0.2 & 0.4 & 0.3 & 0.4 & 0.4 & 0.9 \\ 
$\ell$ resolution \& efficiency & 1.3 & 1.4 & 1.5 & 1.3 & 1.3 & 1.3 & 1.3 \\ 
\met\ cell-out & 0.2 & 0.2 & 0.2 & 0.1 & 0.2 & 0.1 & 0.4 \\ 
$b$-quark tagging efficiency & 4.4 & 4.1 & 4.3 & 5.4 & 7.5 & 9.3 & 6.9 \\ 
Additional interactions & 0.1 & 0.1 & 0.1 & 0.0 & 0.1 & 0.1 & 0.3 \\ 
Jet reconstruction efficiency & 0.1 & 0.0 & 0.0 & 0.0 & 0.0 & 0.0 & 0.0 \\ 
Jet energy resolution & 1.2 & 1.0 & 0.1 & 0.7 & 0.4 & 2.5 & 1.4 \\ 
$b$-quark jets (JES) & 3.3 & 0.2 & 2.6 & 3.3 & 4.2 & 4.9 & 5.0 \\ 
Close by jets (JES) & 5.1 & 1.6 & 3.6 & 2.2 & 1.2 & 1.9 & 2.0 \\ 
Effective detector NP set 1 (JES) & 2.3 & 0.4 & 1.5 & 2.0 & 2.4 & 3.4 & 3.8 \\ 
Effective detector NP set 2 (JES) & 0.2 & 0.0 & 0.1 & 0.2 & 0.1 & 0.1 & 0.2 \\ 
Effective mixed NP set 1 (JES) & 0.1 & 0.0 & 0.1 & 0.1 & 0.1 & 0.4 & 0.8 \\ 
Effective mixed NP set 2 (JES) & 0.1 & 0.1 & 0.2 & 0.2 & 0.2 & 0.1 & 0.1 \\ 
Effective model NP set 1 (JES) & 0.6 & 0.8 & 1.2 & 1.2 & 0.9 & 1.0 & 0.2 \\ 
Effective model NP set 2 (JES) & 0.9 & 0.3 & 0.4 & 0.6 & 0.7 & 1.1 & 1.0 \\ 
Effective model NP set 3 (JES) & 1.1 & 0.3 & 0.5 & 0.7 & 0.9 & 1.2 & 0.4 \\ 
Effective model NP set 4 (JES) & 0.2 & 0.1 & 0.1 & 0.0 & 0.1 & 0.5 & 0.4 \\ 
Effective stat. NP set 1 (JES) & 0.5 & 0.6 & 0.3 & 0.2 & 0.1 & 0.2 & 0.3 \\ 
Effective stat. NP set 2 (JES) & 0.3 & 0.0 & 0.1 & 0.2 & 0.2 & 0.4 & 0.3 \\ 
Effective stat. NP set 3 (JES) & 0.7 & 0.1 & 0.4 & 0.5 & 0.5 & 1.1 & 0.6 \\ 
$\eta$-intercalibration (JES) & 2.4 & 0.4 & 1.8 & 1.8 & 2.1 & 2.6 & 2.3 \\ 
$\eta$-intercalibration statistics (JES) & 0.3 & 0.2 & 0.3 & 0.4 & 0.4 & 0.5 & 0.2 \\ 
Flavour composition  (JES) & 0.7 & 0.3 & 0.7 & 0.7 & 0.7 & 0.9 & 0.8 \\ 
Flavour response (JES) & 0.8 & 0.1 & 0.4 & 0.6 & 0.7 & 1.1 & 1.1 \\ 
Additional interactions $\mu$ (JES) & 0.2 & 0.2 & 0.1 & 0.1 & 0.2 & 0.4 & 0.1 \\ 
Additional interactions $N_\mathrm{PV}$ (JES) & 0.1 & 0.0 & 0.1 & 0.2 & 0.3 & 0.5 & 0.1 \\ 
Relative non-closure (JES) & 0.3 & 0.1 & 0.3 & 0.2 & 0.4 & 0.7 & 0.4 \\ 
Single particle high-\pT\ (JES) & 0.0 & 0.0 & 0.0 & 0.0 & 0.0 & 0.0 & 0.0 \\ 
Jet vertex fraction efficiency & 1.6 & 1.4 & 1.6 & 1.7 & 1.7 & 2.0 & 2.3 \\ 
$W$+jets normalisation & 4.3 & 2.0 & 1.7 & 1.9 & 2.4 & 3.3 & 6.2 \\ 
$W$+jets heavy/light flavour & 8.1 & 3.7 & 2.6 & 3.2 & 4.3 & 6.3 & 8.3 \\ 
Multijet normalisation & 1.6 & 0.6 & 0.3 & 0.5 & 1.0 & 1.3 & 1.7 \\ 
Multijet shape & 0.6 & 0.0 & 0.2 & 0.1 & 0.1 & 0.2 & 0.4 \\ 
Small backgrounds & 2.5 & 1.3 & 1.1 & 1.4 & 1.9 & 2.7 & 3.8 \\ 
Luminosity & 1.8 & 1.8 & 1.8 & 1.8 & 1.8 & 1.8 & 1.8 \\ 
\hline \hline 
Statistical uncertainty & 2.2 & 1.1 & 1.6 & 1.8 & 2.8 & 5.0 & 6.8 \\ 
Total uncertainty & 13.6 & 7.8 & 8.8 & 9.9 & 12.6 & 16.0 & 17.5 \\ 
\hline 
Cross-section [pb/GeV] & 7.61e-03 & 9.43e-03 & 6.13e-03 & 1.71e-03 & 4.92e-04 & 1.33e-04 & 3.16e-07 \\ 
\hline 
\end{tabular}
\caption{Relative uncertainties on the final differential cross-section after the $e$/$\mu$ channel combination, for the 2nd jet.  The uncertainties are shown as a percentage of the expected \ttbar\ signal.  The energy scale uncertainty (JES) is shown for each JES nuisance parameter (NP). The effective NP are obtained by combining   a total of 54 detector, detector and model (``mixed''), modelling and statistical NPs.  An uncertainty value of 0.0 implies that the uncertainty is below 0.05.}
\label{tab:tables-paper/jet-pt-single-lepton/jetPtN1Part_ljets_pT25}
\end{table}

%% file: jetPtN2Part_ljets_pT25.tex
\begin{table}[tbhp]
\centering
\scriptsize
\tabcolsep=1mm
\begin{tabular}{ l c c c c c c } \hline
$\frac{d \sigma}{d p_\mathrm{T,jet}}$ [\%] / $p_\mathrm{T,jet}$ [GeV]& [25-35] & [35-50] & [50-70] & [70-90] & [90-120] & [120-530] \\ \hline \hline 
MC statistics & 0.8 & 0.4 & 0.8 & 1.0 & 1.6 & 2.2 \\ 
PDF & 0.4 & 0.1 & 0.3 & 0.5 & 0.2 & 0.4 \\ 
MC generator & 1.2 & 0.5 & 0.2 & 0.6 & 0.3 & 1.0 \\ 
Fragmentation & 1.3 & 0.8 & 0.6 & 0.1 & 0.1 & 0.4 \\ 
ISR/FSR & 2.5 & 3.1 & 3.2 & 2.1 & 4.8 & 4.8 \\ 
Colour reconnection & 0.5 & 0.6 & 0.0 & 0.5 & 0.8 & 0.9 \\ 
$\ell$ resolution \& efficiency & 1.3 & 1.3 & 1.7 & 1.4 & 1.4 & 1.4 \\ 
\met\ cell-out & 0.2 & 0.2 & 0.1 & 0.2 & 0.1 & 0.3 \\ 
$b$-quark tagging efficiency & 5.0 & 4.4 & 4.6 & 5.3 & 6.5 & 8.3 \\ 
Additional interactions & 0.1 & 0.1 & 0.1 & 0.1 & 0.1 & 0.2 \\ 
Jet reconstruction efficiency & 0.1 & 0.0 & 0.0 & 0.0 & 0.0 & 0.1 \\ 
Jet energy resolution & 0.2 & 0.9 & 0.4 & 0.7 & 1.5 & 0.8 \\ 
$b$-quark jets (JES) & 1.6 & 0.2 & 2.5 & 3.3 & 3.8 & 4.9 \\ 
Close by jets (JES) & 4.5 & 0.4 & 4.5 & 4.6 & 2.2 & 2.0 \\ 
Effective detector NP set 1 (JES) & 1.3 & 0.4 & 1.3 & 2.6 & 2.7 & 4.3 \\ 
Effective detector NP set 2 (JES) & 0.1 & 0.0 & 0.1 & 0.2 & 0.2 & 0.1 \\ 
Effective mixed NP set 1 (JES) & 0.1 & 0.0 & 0.1 & 0.2 & 0.2 & 0.5 \\ 
Effective mixed NP set 2 (JES) & 0.1 & 0.1 & 0.2 & 0.3 & 0.2 & 0.2 \\ 
Effective model NP set 1 (JES) & 0.8 & 0.7 & 1.4 & 1.4 & 1.2 & 1.1 \\ 
Effective model NP set 2 (JES) & 0.5 & 0.3 & 0.3 & 0.8 & 0.9 & 1.3 \\ 
Effective model NP set 3 (JES) & 0.6 & 0.3 & 0.5 & 1.0 & 1.0 & 1.3 \\ 
Effective model NP set 4 (JES) & 0.2 & 0.0 & 0.2 & 0.1 & 0.0 & 0.5 \\ 
Effective stat. NP set 1 (JES) & 0.1 & 0.6 & 0.5 & 0.2 & 0.1 & 0.1 \\ 
Effective stat. NP set 2 (JES) & 0.2 & 0.1 & 0.2 & 0.3 & 0.2 & 0.4 \\ 
Effective stat. NP set 3 (JES) & 0.3 & 0.2 & 0.4 & 0.6 & 0.7 & 1.0 \\ 
$\eta$-intercalibration (JES) & 1.8 & 0.1 & 2.0 & 2.7 & 2.4 & 2.9 \\ 
$\eta$-intercalibration statistics (JES) & 0.3 & 0.1 & 0.4 & 0.5 & 0.5 & 0.7 \\ 
Flavour composition  (JES) & 0.6 & 0.1 & 0.7 & 1.2 & 1.1 & 1.4 \\ 
Flavour response (JES) & 1.5 & 0.1 & 0.7 & 1.0 & 1.1 & 1.7 \\ 
Additional interactions $\mu$ (JES) & 0.1 & 0.2 & 0.0 & 0.0 & 0.2 & 0.3 \\ 
Additional interactions $N_\mathrm{PV}$ (JES) & 0.1 & 0.1 & 0.2 & 0.4 & 0.2 & 0.4 \\ 
Relative non-closure (JES) & 0.0 & 0.1 & 0.3 & 0.4 & 0.2 & 0.6 \\ 
Single particle high-\pT\ (JES) & 0.0 & 0.0 & 0.0 & 0.0 & 0.0 & 0.0 \\ 
Jet vertex fraction efficiency & 1.7 & 1.4 & 1.6 & 1.8 & 1.8 & 2.2 \\ 
$W$+jets normalisation & 5.3 & 2.1 & 1.4 & 1.4 & 2.1 & 2.4 \\ 
$W$+jets heavy/light flavour & 10.2 & 3.9 & 1.7 & 2.1 & 3.1 & 3.8 \\ 
Multijet normalisation & 1.7 & 0.8 & 0.3 & 0.4 & 0.4 & 1.4 \\ 
Multijet shape & 0.7 & 0.0 & 0.3 & 0.2 & 0.2 & 0.2 \\ 
Small backgrounds & 2.9 & 1.4 & 0.9 & 1.1 & 1.4 & 2.7 \\ 
Luminosity & 1.8 & 1.8 & 1.8 & 1.8 & 1.8 & 1.8 \\ 
\hline \hline 
Statistical uncertainty & 2.0 & 1.0 & 1.5 & 2.2 & 3.6 & 5.4 \\ 
Total uncertainty & 15.0 & 8.0 & 9.3 & 10.4 & 12.2 & 15.4 \\ 
\hline 
Cross-section [pb/GeV] & 1.92e-02 & 1.73e-02 & 7.49e-03 & 2.47e-03 & 4.90e-04 & 8.72e-07 \\ 
\hline 
\end{tabular}
\caption{Relative uncertainties on the final differential cross-section after the $e$/$\mu$ channel combination, for the 3rd jet.  The uncertainties are shown as a percentage of the expected \ttbar\ signal.  The energy scale uncertainty (JES) is shown for each JES nuisance parameter (NP). The effective NP are obtained by combining   a total of 54 detector, detector and model (``mixed''), modelling and statistical NPs.  An uncertainty value of 0.0 implies that the uncertainty is below 0.05.}
\label{tab:tables-paper/jet-pt-single-lepton/jetPtN2Part_ljets_pT25}
\end{table}

%% file: jetPtN3Part_ljets_pT25.tex
\begin{table}[tbhp]
\centering
\scriptsize
\tabcolsep=1mm
\begin{tabular}{ l c c c c c } \hline
$\frac{d \sigma}{d p_\mathrm{T,jet}}$ [\%] / $p_\mathrm{T,jet}$ [GeV]& [25-35] & [35-50] & [50-70] & [70-90] & [90-280] \\ \hline \hline 
MC statistics & 0.6 & 0.7 & 1.0 & 1.8 & 3.7 \\ 
PDF & 0.1 & 0.1 & 0.5 & 0.7 & 0.4 \\ 
MC generator & 1.0 & 0.3 & 0.8 & 1.5 & 0.8 \\ 
Fragmentation & 0.5 & 0.2 & 0.2 & 0.4 & 1.8 \\ 
ISR/FSR & 1.6 & 3.7 & 2.0 & 6.3 & 5.7 \\ 
Colour reconnection & 0.2 & 0.7 & 0.3 & 1.2 & 2.0 \\ 
$\ell$ resolution \& efficiency & 1.3 & 1.5 & 1.3 & 1.4 & 1.5 \\ 
\met\ cell-out & 0.2 & 0.2 & 0.1 & 0.3 & 0.1 \\ 
$b$-quark tagging efficiency & 4.6 & 4.4 & 4.8 & 5.4 & 7.0 \\ 
Additional interactions & 0.1 & 0.1 & 0.0 & 0.2 & 0.1 \\ 
Jet reconstruction efficiency & 0.1 & 0.0 & 0.0 & 0.0 & 0.0 \\ 
Jet energy resolution & 0.2 & 0.0 & 0.5 & 1.4 & 0.6 \\ 
$b$-quark jets (JES) & 0.1 & 1.6 & 3.2 & 3.8 & 4.5 \\ 
Close by jets (JES) & 0.9 & 4.2 & 7.5 & 5.5 & 2.9 \\ 
Effective detector NP set 1 (JES) & 0.2 & 1.0 & 2.5 & 3.6 & 3.7 \\ 
Effective detector NP set 2 (JES) & 0.0 & 0.1 & 0.3 & 0.3 & 0.4 \\ 
Effective mixed NP set 1 (JES) & 0.0 & 0.1 & 0.2 & 0.3 & 0.5 \\ 
Effective mixed NP set 2 (JES) & 0.3 & 0.4 & 0.3 & 0.3 & 0.3 \\ 
Effective model NP set 1 (JES) & 1.5 & 2.2 & 2.1 & 1.7 & 1.5 \\ 
Effective model NP set 2 (JES) & 0.4 & 0.1 & 0.7 & 1.2 & 1.3 \\ 
Effective model NP set 3 (JES) & 0.5 & 0.2 & 0.9 & 1.4 & 1.5 \\ 
Effective model NP set 4 (JES) & 0.1 & 0.1 & 0.3 & 0.1 & 0.2 \\ 
Effective stat. NP set 1 (JES) & 1.1 & 1.2 & 0.6 & 0.1 & 0.2 \\ 
Effective stat. NP set 2 (JES) & 0.2 & 0.1 & 0.3 & 0.3 & 0.5 \\ 
Effective stat. NP set 3 (JES) & 0.2 & 0.2 & 0.7 & 1.0 & 1.0 \\ 
$\eta$-intercalibration (JES) & 0.3 & 1.8 & 3.2 & 3.6 & 3.0 \\ 
$\eta$-intercalibration statistics (JES) & 0.3 & 0.5 & 0.7 & 0.7 & 0.7 \\ 
Flavour composition  (JES) & 0.7 & 0.8 & 1.4 & 1.4 & 1.6 \\ 
Flavour response (JES) & 0.7 & 1.8 & 2.4 & 2.1 & 2.5 \\ 
Additional interactions $\mu$ (JES) & 0.3 & 0.0 & 0.1 & 0.5 & 0.2 \\ 
Additional interactions $N_\mathrm{PV}$ (JES) & 0.1 & 0.3 & 0.3 & 0.5 & 0.7 \\ 
Relative non-closure (JES) & 0.4 & 0.2 & 0.6 & 0.5 & 0.5 \\ 
Single particle high-\pT\ (JES) & 0.0 & 0.0 & 0.0 & 0.0 & 0.0 \\ 
Jet vertex fraction efficiency & 1.6 & 1.6 & 1.9 & 2.1 & 2.2 \\ 
$W$+jets normalisation & 2.1 & 1.3 & 0.8 & 1.1 & 1.0 \\ 
$W$+jets heavy/light flavour & 3.2 & 1.4 & 1.0 & 1.2 & 1.2 \\ 
Multijet normalisation & 0.7 & 0.2 & 0.3 & 0.6 & 1.2 \\ 
Multijet shape & 0.2 & 0.3 & 0.3 & 0.3 & 0.5 \\ 
Small backgrounds & 1.4 & 0.8 & 1.0 & 1.3 & 2.4 \\ 
Luminosity & 1.8 & 1.8 & 1.8 & 1.8 & 1.8 \\ 
\hline \hline 
Statistical uncertainty & 1.3 & 1.4 & 2.2 & 4.2 & 6.5 \\ 
Total uncertainty & 7.8 & 9.2 & 12.1 & 14.1 & 15.4 \\ 
\hline 
Cross-section [pb/GeV] & 2.62e-02 & 1.10e-02 & 2.46e-03 & 5.51e-04 & 2.77e-06 \\ 
\hline 
\end{tabular}
\caption{Relative uncertainties on the final differential cross-section after the $e$/$\mu$ channel combination, for the 4th jet.  The uncertainties are shown as a percentage of the expected \ttbar\ signal.  The energy scale uncertainty (JES) is shown for each JES nuisance parameter (NP). The effective NP are obtained by combining   a total of 54 detector, detector and model (``mixed''), modelling and statistical NPs.  An uncertainty value of 0.0 implies that the uncertainty is below 0.05.}
\label{tab:tables-paper/jet-pt-single-lepton/jetPtN3Part_ljets_pT25}
\end{table}

%% file: jetPtN4Part_ljets_pT25.tex
\begin{table}[tbhp]
\centering
\scriptsize
\tabcolsep=1mm
\begin{tabular}{ l c c c c } \hline
$\frac{d \sigma}{d p_\mathrm{T,jet}}$ [\%] / $p_\mathrm{T,jet}$ [GeV]& [25-35] & [35-50] & [50-70] & [70-195] \\ \hline \hline 
MC statistics & 1.0 & 1.0 & 2.1 & 3.8 \\ 
PDF & 0.1 & 0.4 & 0.6 & 0.4 \\ 
MC generator & 0.4 & 0.3 & 1.1 & 1.8 \\ 
Fragmentation & 0.1 & 2.1 & 0.2 & 0.7 \\ 
ISR/FSR & 2.9 & 3.9 & 1.6 & 6.5 \\ 
Colour reconnection & 0.7 & 1.4 & 1.5 & 4.0 \\ 
$\ell$ resolution \& efficiency & 1.6 & 1.3 & 1.3 & 1.4 \\ 
\met\ cell-out & 0.1 & 0.2 & 0.4 & 0.4 \\ 
$b$-quark tagging efficiency & 4.5 & 4.5 & 4.4 & 6.2 \\ 
Additional interactions & 0.1 & 0.1 & 0.2 & 0.2 \\ 
Jet reconstruction efficiency & 0.1 & 0.0 & 0.0 & 0.1 \\ 
Jet energy resolution & 1.8 & 0.6 & 1.8 & 1.4 \\ 
$b$-quark jets (JES) & 0.6 & 2.2 & 3.3 & 3.8 \\ 
Close by jets (JES) & 2.3 & 7.2 & 9.7 & 5.1 \\ 
Effective detector NP set 1 (JES) & 0.7 & 2.0 & 3.6 & 4.1 \\ 
Effective detector NP set 2 (JES) & 0.1 & 0.2 & 0.3 & 0.3 \\ 
Effective mixed NP set 1 (JES) & 0.1 & 0.2 & 0.2 & 0.3 \\ 
Effective mixed NP set 2 (JES) & 0.5 & 0.6 & 0.4 & 0.3 \\ 
Effective model NP set 1 (JES) & 3.5 & 3.0 & 2.6 & 1.9 \\ 
Effective model NP set 2 (JES) & 0.5 & 0.3 & 1.0 & 1.4 \\ 
Effective model NP set 3 (JES) & 0.5 & 0.5 & 1.2 & 1.7 \\ 
Effective model NP set 4 (JES) & 0.0 & 0.3 & 0.3 & 0.1 \\ 
Effective stat. NP set 1 (JES) & 2.2 & 1.5 & 0.6 & 0.5 \\ 
Effective stat. NP set 2 (JES) & 0.1 & 0.2 & 0.3 & 0.4 \\ 
Effective stat. NP set 3 (JES) & 0.1 & 0.5 & 1.0 & 0.9 \\ 
$\eta$-intercalibration (JES) & 1.5 & 3.2 & 4.2 & 4.1 \\ 
$\eta$-intercalibration statistics (JES) & 0.7 & 0.8 & 0.8 & 0.5 \\ 
Flavour composition  (JES) & 0.9 & 1.5 & 1.6 & 1.6 \\ 
Flavour response (JES) & 2.5 & 4.3 & 3.7 & 2.9 \\ 
Additional interactions $\mu$ (JES) & 0.3 & 0.1 & 0.5 & 0.5 \\ 
Additional interactions $N_\mathrm{PV}$ (JES) & 0.3 & 0.3 & 0.5 & 0.9 \\ 
Relative non-closure (JES) & 0.6 & 0.3 & 0.8 & 0.4 \\ 
Single particle high-\pT\ (JES) & 0.0 & 0.0 & 0.0 & 0.0 \\ 
Jet vertex fraction efficiency & 1.8 & 1.9 & 2.2 & 2.4 \\ 
$W$+jets normalisation & 1.9 & 0.7 & 0.4 & 0.5 \\ 
$W$+jets heavy/light flavour & 1.5 & 0.9 & 0.3 & 0.4 \\ 
Multijet normalisation & 0.3 & 0.3 & 0.4 & 0.3 \\ 
Multijet shape & 0.3 & 0.3 & 0.3 & 0.5 \\ 
Small backgrounds & 0.8 & 0.9 & 1.1 & 1.7 \\ 
Luminosity & 1.8 & 1.8 & 1.8 & 1.8 \\ 
\hline \hline 
Statistical uncertainty & 1.8 & 2.1 & 4.3 & 9.4 \\ 
Total uncertainty & 9.3 & 13.1 & 15.2 & 17.9 \\ 
\hline 
Cross-section [pb/GeV] & 1.51e-02 & 3.59e-03 & 6.10e-04 & 3.20e-06 \\ 
\hline 
\end{tabular}
\caption{Relative uncertainties on the final differential cross-section after the $e$/$\mu$ channel combination, for the 5th jet.  The uncertainties are shown as a percentage of the expected \ttbar\ signal.  The energy scale uncertainty (JES) is shown for each JES nuisance parameter (NP). The effective NP are obtained by combining   a total of 54 detector, detector and model (``mixed''), modelling and statistical NPs.  An uncertainty value of 0.0 implies that the uncertainty is below 0.05.}
\label{tab:tables-paper/jet-pt-single-lepton/jetPtN4Part_ljets_pT25}
\end{table}

%% file: atlas_authlist.tex
\begin{flushleft}
{\Large The ATLAS Collaboration}

\bigskip

G.~Aad$^{\rm 84}$,
B.~Abbott$^{\rm 112}$,
J.~Abdallah$^{\rm 152}$,
S.~Abdel~Khalek$^{\rm 116}$,
O.~Abdinov$^{\rm 11}$,
R.~Aben$^{\rm 106}$,
B.~Abi$^{\rm 113}$,
M.~Abolins$^{\rm 89}$,
O.S.~AbouZeid$^{\rm 159}$,
H.~Abramowicz$^{\rm 154}$
% ...
\bigskip
\\
% ...
$^{11}$ Institute of Physics, Azerbaijan Academy of Sciences, Baku, Azerbaijan\\
% ...
$^{84}$ CPPM, Aix-Marseille Universit{\'e} and CNRS/IN2P3, Marseille, France\\
% ...
$^{89}$ Department of Physics and Astronomy, Michigan State University, East Lansing MI, United States of America\\
% ...
$^{106}$ Nikhef National Institute for Subatomic Physics and University of Amsterdam, Amsterdam, Netherlands\\
% ...
$^{112}$ Homer L. Dodge Department of Physics and Astronomy, University of Oklahoma, Norman OK, United States of America\\
$^{113}$ Department of Physics, Oklahoma State University, Stillwater OK, United States of America\\
% ...
$^{116}$ LAL, Universit{\'e} Paris-Sud and CNRS/IN2P3, Orsay, France\\
% ...
$^{152}$ Institute of Physics, Academia Sinica, Taipei, Taiwan\\
% ...
$^{154}$ Raymond and Beverly Sackler School of Physics and Astronomy, Tel Aviv University, Tel Aviv, Israel\\
% ...
$^{159}$ Department of Physics, University of Toronto, Toronto ON, Canada\\
% ...
\end{flushleft}
